\documentclass[a4paper,11pt]{article}
\pdfoutput=1 

\usepackage{jheppub} 

\usepackage[T1]{fontenc} 
\usepackage{comment}
\usepackage{color,latexsym,array,multirow,verbatim,enumerate,cancel}
\usepackage{pifont}% http://ctan.org/pkg/pifont
\usepackage[table,xcdraw]{xcolor}
\usepackage{booktabs}

\title{\boldmath 
Boosted top tagging and its interpretation using Shapley values }

\author[a]{Biplob Bhattacherjee}
\author[a]{, Camellia Bose}
\author[b]{, Amit Chakraborty}
\author[a,c]{, and Rhitaja Sengupta}

\affiliation[a]{Center for High Energy Physics, Indian Institute of Science, Bengaluru, \\Karnataka, India}
\affiliation[b]{Department of Physics, School of Engineering and Sciences,\\ SRM University AP, Amaravati, Mangalagiri 522240, India}
\affiliation[c]{Bethe Center for Theoretical Physics and Physikalisches Institut der Universit\"at Bonn, \\
Nu{\ss}allee 12, 53115 Bonn, Germany}

\emailAdd{biplob@iisc.ac.in}
\emailAdd{camelliabose@iisc.ac.in}
\emailAdd{amit.c@srmap.edu.in}
\emailAdd{rsengupt@uni-bonn.de}

\abstract{Top tagging has emerged as a fast-evolving subject due to the top quark's significant role in probing physics beyond the standard model. For the reconstruction of top jets, machine learning models have shown a substantial improvement in the classification performance compared to the previous methods. In this work, we build top taggers using $N$-Subjettiness ratios and several Energy Correlation observables as
input features to train the eXtreme Gradient BOOSTed decision tree (\texttt{XGBOOST}). The study finds that tighter parton-level matching lead to more accurate tagging. However, in real experimental data, where the parton level data are unknown, this matching cannot be done. We train the \texttt{XGBOOST} models without performing this matching and show that this difference impacts the taggers' effectiveness. Additionally, we test the tagger under different simulation conditions, including changes in center-of-mass energy, parton distribution functions (PDFs), and pileup effects, demonstrating its robustness with performance deviations of less than 1\%. Furthermore, we use the SHapley Additive exPlanation (SHAP) framework to calculate the importance of the features of the trained models. It helps us to estimate how much each feature of the data contributed to the model's prediction and what regions are of more importance for each input variable. Finally, we combine all the tagger variables to form a hybrid tagger and interpret the results using the Shapley values.}

\begin{document} 
\maketitle
\flushbottom

\section{Introduction}
\label{sec:intro}

The Standard Model (SM) top quark is an essential probe of new physics.
Models of beyond the SM physics (BSM) trying to alleviate the hierarchy problem\,\cite{tHooft:1979rat}, like the supersymmetric extensions of the SM\,\cite{Gildener:1976ih,Susskind:1978ms} or the little Higgs models\,\cite{Arkani-Hamed:2001nha,Arkani-Hamed:2002ikv}, include top partners which naturally decay to top quarks.
To dig out any new physics from top quarks, we require ways to correctly identify and fully reconstruct them at collider experiments, like the Large Hadron Collider (LHC) experiments. While its leptonic and semi-leptonic decay modes are less challenging to identify in the busy environment of LHC detectors, these have lower branching fractions than the hadronic decay mode, and part of the final state is invisible due to the neutrinos.
However, there are still ways to reconstruct the top quark mass from the leptonic and semi-leptonic decay of the top quark, see Refs.\,\cite{CDF:2009mbf,CMS-PAS-TOP-15-008,CMS:2016ixg,ATLAS:2017dhr,ATLAS:2018fwq,ATLAS:2022jbw}.
The hadronic decay mode of top quarks can be fully reconstructed. However, the high jet multiplicity in the final state makes the combinatorics challenging. This problem is resolved when we are in the regime of boosted top quarks, where they have very high momenta, which makes their decay products collimated. In the collider detectors, this is then reconstructed as a single object $-$ a fat jet consisting of all the decay products of the top quark. 

Traditionally, algorithms for finding jet substructure were used for tagging jets from the hadronisation of top quarks from a background of jets associated with light quarks or gluons.
Top taggers have evolved significantly over the years. They range from taggers based on explicitly reconstructing the subjets inside the fat jet, like Johns Hopkins\,\cite{Kaplan:2008ie} and HepTopTagger\,\cite{Plehn:2009rk, Plehn:2010st, Plehn:2011sj}, to approaches utilising variables that naturally capture the correct pronged structure of a top jet, such as $N$-Subjettiness\,\cite{Thaler:2010tr} and energy correlation functions\,\cite{Larkoski:2013eya, Moult:2016cvt, Larkoski:2014zma, Larkoski:2014gra, Larkoski:2015kga}. Additionally, there are methods employing machine learning (ML) techniques like image recognition on jet images using convolutional neural networks (CNNs), as well as networks trained on the low-level information of a jet, such as the transverse momenta and pseudorapidity of the jet constituents, or high-level variables constructed out of them.
There's a plethora of research works going on in this direction to improve the tagging efficiencies of top quarks\,\cite{Shmakov:2021qdz,Erdmann:2013rxa,Erdmann:2017hra,Almeida:2015jua,Kasieczka:2017nvn,Butter:2017cot,Macaluso:2018tck,Moore:2018lsr,Qu:2019gqs,Kasieczka:2019dbj,Roy:2019jae,Diefenbacher:2019ezd,Chakraborty:2020yfc,Bhattacharya:2020aid,Lim:2020igi,Dreyer:2020brq,Aguilar-Saavedra:2021rjk,Andrews:2021ejw,Qu:2022mxj,Dreyer:2022yom,Ahmed:2022hct,Munoz:2022gjq,Choi:2023slq,He:2023cfc,Bogatskiy:2023nnw,Shen:2023ofd,Isildak:2023dnf,Sahu:2023uwb,Baron:2023yhw,Bogatskiy:2023fug,Liu:2023dio,Batson:2023ohn,Furuichi:2023vdx,Ngairangbam:2023cps} in experiments to better our chances of observing hints of new physics. Both the ATLAS and CMS experimental collaborations have also included many such sophisticated taggers in their searches which has improved their limits\,\cite{ATLAS:2020lks,ATL-PHYS-PUB-2020-017,ATL-PHYS-PUB-2022-039,CMS:2012bti,CMS:2017ucf,CMS:2021beq,Keicher:2023mer}. Since ML algorithms can sometimes be black boxes, it is essential to understand the physics behind what it has learnt and how it can distinguish top jets from others. Several studies, therefore, attempt to explore the interpretability of these algorithms and their results, such as Refs.\,\cite{deOliveira:2015xxd, Chang:2017kvc,Diefenbacher:2019ezd, Agarwal:2020fpt,Romero:2021qlf,Collins:2021pld,Mokhtar:2021bkf,Anzalone:2022hrt,Grojean:2020ech,Bradshaw:2022qev,Grojean:2022mef,Khot:2022aky,Roy:2022gge,Das:2022cjl,Mengel:2023mnw,Ngairangbam:2023cps}, which include but are not limited to top quark tagging. 

Although being very well-explored, there are still some subtleties that can affect the performance of top tagging algorithms. In our present work, we pose the following questions and try to answer them. Most ML-based studies on top tagging work with a subset of jets where the decay products of the top quark at the parton level are contained within a certain distance from the axis of the fat jet. This ensures that the reconstructed jet contains the total decay of the top quark and is called the ``matching'' condition\,\cite{Kasieczka:2017nvn,Macaluso:2018tck,Kasieczka:2019dbj}. The choice of this distance from the jet axis used for matching is ad hoc, and more importantly, this criterion cannot be used in data. This leads us to question the impact of this matching condition on the performance of the various taggers. Moreover, tagging efficiencies depend on the momentum of the initial top quark\,\cite{Kaplan:2008ie}, which is again difficult to control in the data, where we only have the momentum of the reconstructed fat jet. 
% In contrast, 
A large angle final-state radiation (FSR) or the presence of neutrinos and muons from the decay of $B$ mesons might reduce the energy of the final jet with respect to the initially produced top quark. It is interesting to explore how the migration along different energy bins of the jet reconstructed at the detector compared to the parton-level top quark affects the performance of the various top taggers. 

We begin our present study addressing these questions by training \texttt{XGBOOST}\,\cite{DBLP:journals/corr/ChenG16} networks, which is an ML algorithm using the gradient boosting framework. We use various sets of $N$-Subjettiness variables and energy correlation functions as inputs to the \texttt{XGBOOST} networks and find out which set of variables among them have better performance as top taggers. Furthermore, we train our \texttt{XGBOOST} tagger using the complete combination of the $N$-Subjettiness variables and Energy Correlation Functions and study the impact of reducing these input variables for the training on the performance of the tagger. We present an interpretation of our results from the \texttt{XGBOOST} training, making use of the \texttt{SHAP} (SHapley Additive exPlanations)\,\cite{DBLP:journals/corr/LundbergL17, DBLP:journals/corr/abs-1802-03888, DBLP:journals/corr/abs-1905-04610, Shapley+2016+307+318} framework, which is based on Shapley values, defined later in Section\,\ref{sec:SHAP}. SHAP has been widely used to interpret machine learning output in collider studies \cite{Cornell:2021gut, Alvestad:2021sje, Grojean:2020ech, CMS:2021qzz, Grojean:2022mef, Adhikary:2022jfp}. To summarise, we ask the following questions in our study.

\begin{enumerate}
\item How does the matching condition affect how well top taggers work?

\item How do we understand the model's decision and extract information from the interpretation?

\item How does the choice of QCD background (quark-initiated jets versus gluon-initiated jets) influence tagging performance?

\item How is the tagging affected under varying simulation conditions?

\item How does the tagger perform in a different energy bin, and how do energy migration of both top jets and QCD jets affect top tagger performance?

\end{enumerate}

The rest of the paper is organised as follows- in Section\,\ref{sec:dataset}, we discuss the simulation, matching criterion, and the energy bin of the data set of top jets used in this work. In Section\,\ref{sec:standard_top_taggers_and_xgb}, we first present a brief review of the $N$-Subjettiness variables and the energy correlation functions used for top tagging and then present results of our \texttt{XGBOOST} tagger for each of these sets. In Section\,\ref{sec:SHAP}, we introduce the SHAP framework of interpreting results from the \texttt{XGBOOST}-based taggers, construct a hybrid tagger using variables across the different sets of standard top tagging variables, and interpret the results from this hybrid tagger. Section \ref{sec:xgb_quark} discusses the differences that arise in top tagging when background is composed of light quark jets instead of gluon jets. We compare results from different event generators in Section \ref{sec:herwig}. We include the results that we obtain with some test cases that contain jets generated at a different center-of-mass energy, a different PDF, and with Pileup. We also compare the results for jets in different energy bins in Section \ref{sec:1000GeV}. The background jets with higher transverse momenta, that lose energy to fake as jets in the energy bins of our interest are studied in Section \ref{sec:high_pt_gluon}. We finally conclude in Section\,\ref{sec:conclusion}.

\section{Data set of top jets}
\label{sec:dataset}

In this section, we discuss the data set of jets associated with top quarks used by us in the present work. We first briefly outline our simulation details and then describe the matching of the parton-level information to the final reconstructed top jet. This includes ensuring the top jet consists of the full hadronic activity of the three partons coming from its decay and whether the energy of the top jet lies in the same bin as the initially generated top quark.

\subsection{Simulation details}
\label{ssec:simulation}

\begin{figure}[htb!]
    \centering
    \includegraphics[width=0.35\textwidth]{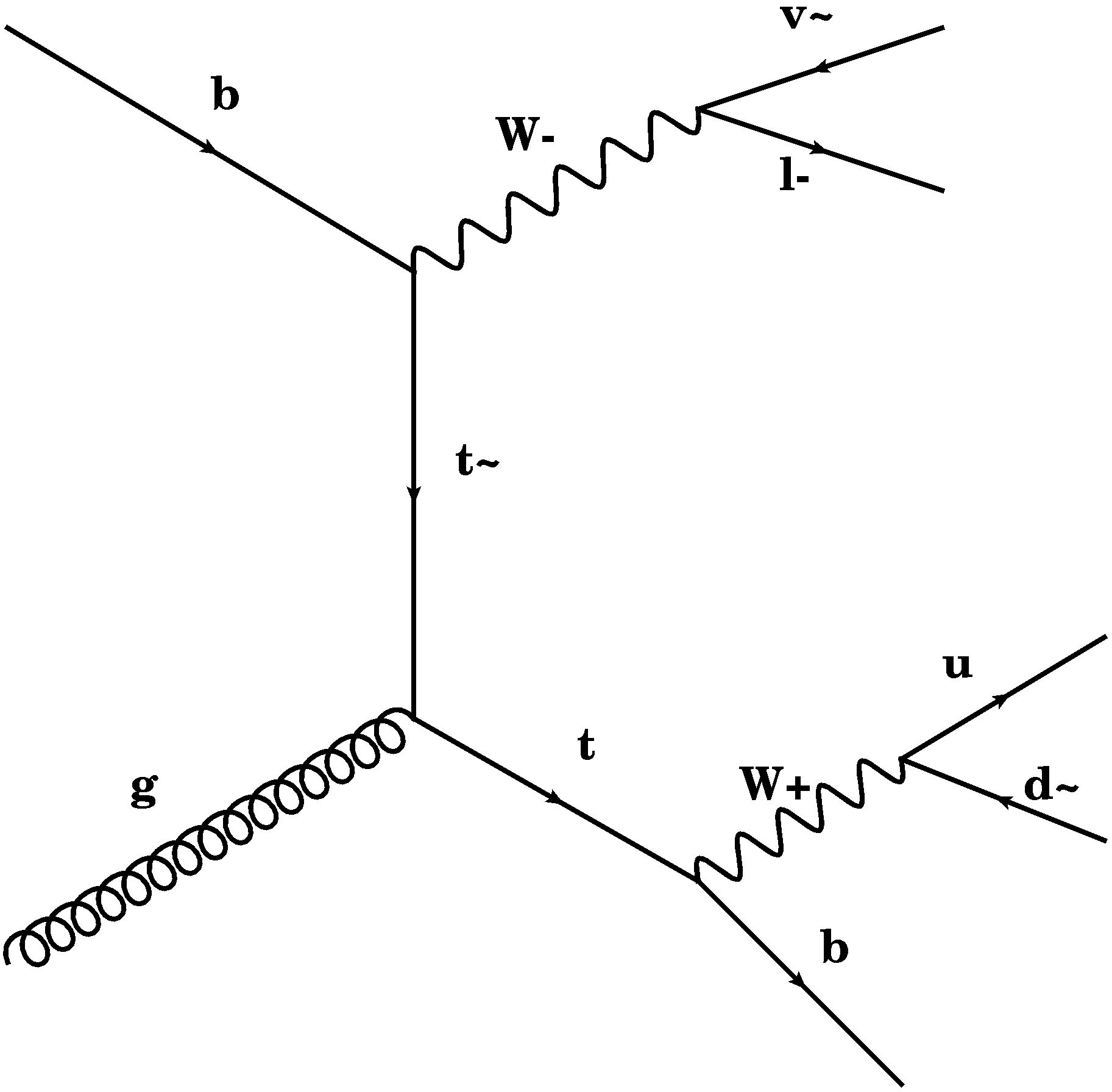}\hspace{15mm}
    \includegraphics[width=0.3\textwidth]{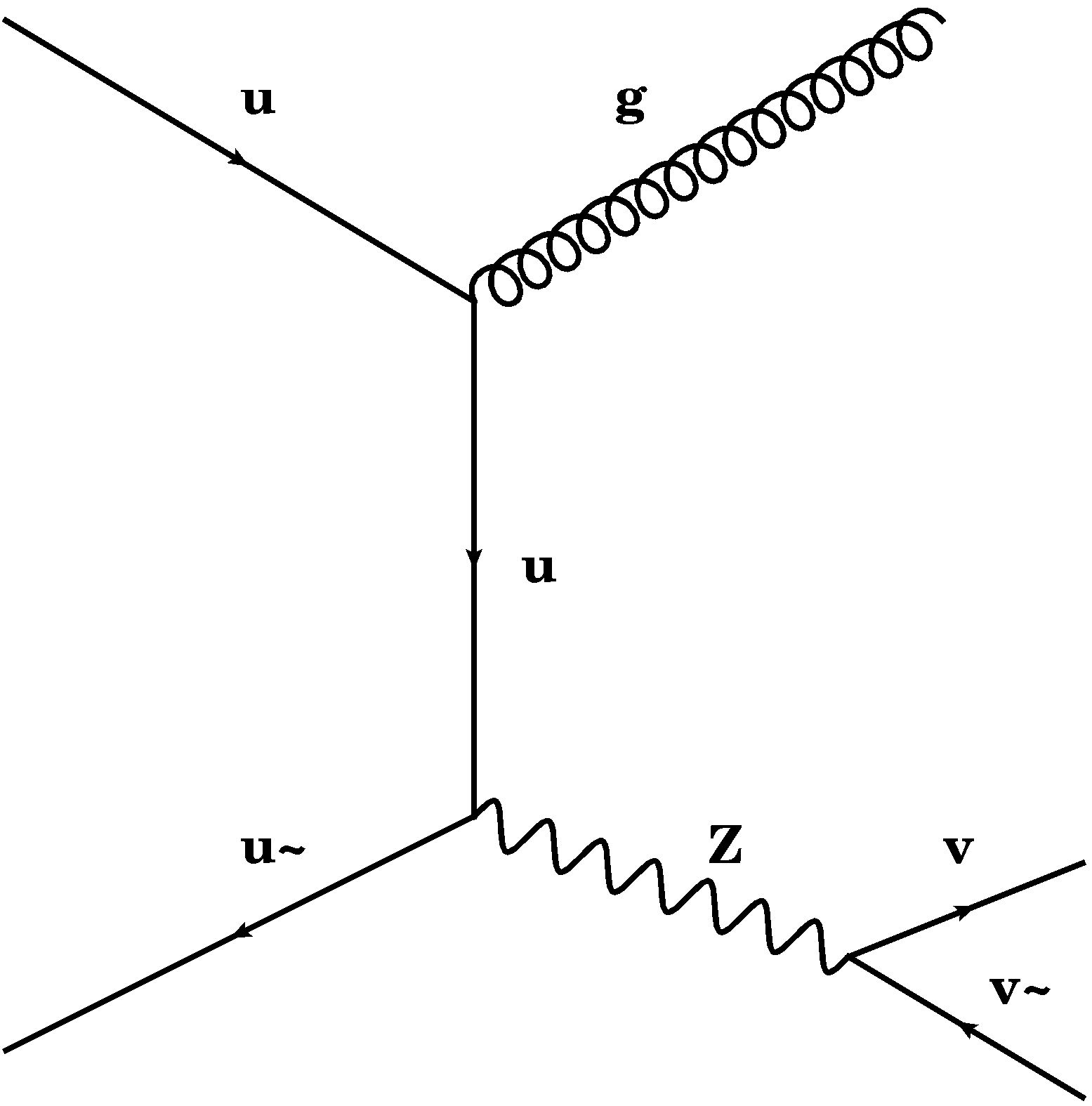}
    \caption{Feynman diagrams of the \textit{t-}channel signal \textit{(left)} and the background \textit{(right)} processes.}
    \label{fig:feyn}
\end{figure}

We generate single top jets using the following process
$$p p \rightarrow t W,\ ~~ t \to b W.$$ at a center-of-mass energy of $\sqrt{s}$=14 TeV.
The $W$-boson coming from the top quark decay is allowed to decay hadronically only, while the other $W$-boson decays leptonically (we choose a muonic mode only). Gluon and light quark (u, d)-initiated jets are also analysed to estimate the mistagging rates. For the background, we simulate the QCD-jets using the following processes
$$p p \rightarrow Z g$$ (gluon jets) and $$p p \rightarrow Z q$$ (light quark jets) with $Z$ decaying invisibly to neutrinos. Figure \ref{fig:feyn} shows the Feynman diagrams of the signal and the background processes. 

We put a generation level cut on the transverse momenta $p_T^t$ (signal) and $p_T^{g/q}$ (background) of 450 GeV. The hard collision events are generated using MadGraph (\texttt{MG5\_aMC\_v2\_8\_2}) \cite{Alwall:2011uj, Alwall:2014hca}, and then passed to \texttt{PYTHIA 8} \cite{Sjostrand:2007gs} for hadronization and parton shower. Jets are generated with initial-state radiation (ISR), final-state radiation (FSR), and multiple parton interaction (MPI) effects. \texttt{Delphes-3.5.0 \cite{deFavereau:2013fsa}} is used to simulate detector effects where we use the default ATLAS card.

We use the \texttt{cteq6l1} PDF set from \texttt{LHAPDF6} \cite{Buckley:2014ana} along with the ATLAS UE Tune \texttt{AU2-CTEQ6L1} for the event generation. The simulated events are reconstructed as jets using Cambridge-Aachen (C/A) jet clustering algorithm \cite{Dokshitzer:1997in} and a fixed jet radius parameter R = 1.0. We change the segmentation of the Hadronic Calorimeter (HCAL) to $\sim 0.1\times0.1$ in $\eta-\phi$ plane with $|\eta|$ extended up to 3.0. Jets are clustered using the calorimeter tower elements. We construct separate datasets for two different ranges of transverse momentum, $p_T^J = [500, 600)$ GeV, and $p_T^J = [1000, 1100)$ GeV. In each of the $p_T$ bins, we consider the leading jet (jet with the highest $p_T$) with $|\eta| < 2.5$ and store its constituents to construct different observables, which are then used as inputs to the various top taggers, as discussed in the next section. 

We study the contribution of background with higher $p_T$ that can fake as top jets with $p_T^J = [500, 600)$ GeV. We generate gluon jets with a cut on the generation level $p_T^g$ of 550, 650, 750, 850, and 950 GeV and select the jets that lose energy and acquire the $p_T$ ranges of our interest. Similarly, for the $p_T^J = [1000, 1100)$ GeV case study, we generate gluon jets with higher $p_T^g$ of 1050, 1150, 1250, 1350, and 1450 GeV.

To study the generator dependence of our results, we also generate top jets and QCD jets using \texttt{HERWIG 7} \cite{Bahr:2008pv, Bellm:2015jjp} for hadronization and parton shower, in the range of transverse momentum, $p_T^J = [500, 600)$ GeV. The remaining simulation details stay unchanged. To study the dependence of the classifiers on various collider conditions, we generate some datasets that contain events simulated at $\sqrt{s}$ = 27 TeV, events generated with the \texttt{NNPDF2.3NNLO} PDF, and simulation with pileup events which are multiple proton-proton collisions in the same or nearby bunch crossings during a single event. These conditions are studied in Section \ref{sec:herwig}.

\subsection{Matching of the partons with the top jet}
\label{ssec:match_energybinmigration}

The top ($t$) quark decays to a $b$ quark and a $W$ boson, and the $W$ further decays to two light quarks, which we shall refer to from now on as $u$ and $d$. Thus, the final state contains three quarks. In the left (right) panel of Figure \ref{fig:ptb_DRb}, we plot the 2-D distribution of the fraction of the top's $p_T$ carried away by a \textit{b (u)} quark and the distance of the \textit{b (u)}  quark from \textit{t} in $\eta-\phi$ plane. The distance between two objects in the $\eta-\phi$ plane is given by $\Delta R = \sqrt{(\Delta\eta_{ij})^2+(\Delta\phi_{ij})^2}$.  We see that the softer the parton is, the farther it is from \textit{t}. Since \textit{u} and \textit{d} decay from a \textit{W}, they are softer than \textit{b} and have a wider spread in $p_T$ as can be seen in Figure \ref{fig:ptb_DRb}.  

\begin{figure}[htb!]
    \centering
    \includegraphics[width=0.5\textwidth]{./PT_500_Fig1a.pdf}~
    \includegraphics[width=0.5\textwidth]{./PT_500_Fig1b.pdf}
    \caption{\textit{Left:} 2-D histogram of $p_T$ fraction carried by a \textit{b} parton and the $\Delta R$ between \textit{b} and \textit{t} at the truth level. \textit{Right:} 2-D histogram of $p_T$ fraction carried by a \textit{u} parton and the $\Delta R$ between \textit{u} and \textit{t} at the truth level, where \textit{u} represents one of the decay products of \textit{W}. The colour bar shows the density of a bin.}
    \label{fig:ptb_DRb}
\end{figure}

It is a common practice to process top jets before using them to train any machine learning (ML) classifier by selecting only those top jets which contain the initial partons \textit{b,u} and \textit{d} matched within a radius $\Delta R_m$ from the jet axis. The matching radius $\Delta R_m$ is typically the same as the jet radius. We select three matching radii for our analysis, $\Delta R_m$ = 0.6, 0.8, 1.0. We also consider top jets without being subjected to any matching procedure, which we shall refer to as unmatched criteria from now on. Table \ref{tab:match_500_600} shows the fraction of top jets belonging to each matching criteria and the fraction of top jets where one or more quarks lie outside $\Delta R_m$. All jets have $p_T^J\in[500,600)$ GeV. 
%%%%%%%%%%%%%%%%%%%%%%%%%%%%%%%%%%%%%%%%%%%%%%%%%%%%%%%%%%%%%%%%%%%%%%%%%%%%%%%%
\begin{table}[htb!]
\centering
\begin{tabular}{|ccc|}
\hline
\multicolumn{2}{|c|}{MG5 + Py8, $p_T^{J} \in [500,600)$ GeV}                           
\\ \hline
\multicolumn{1}{|c|}{\begin{tabular}[c]{@{}c@{}}Matching radius\\ $\Delta R_m$\end{tabular}} & \multicolumn{1}{c|}{\begin{tabular}[c]{@{}c@{}}Fraction of top jets \\ fulfilling matching \\ criteria\end{tabular}} \\ \hline
\multicolumn{1}{|c|}{0.6} & \multicolumn{1}{c|}{0.32}   \\ \hline
\multicolumn{1}{|c|}{0.8} & \multicolumn{1}{c|}{0.55}   \\ \hline
\multicolumn{1}{|c|}{1.0} & \multicolumn{1}{c|}{0.68}   \\ \hline
\end{tabular}
\caption{Fraction of top jets that pass the matching criteria with four different matching criteria $\Delta R_m$ = 0.6, 0.8, and 1.0.}
\label{tab:match_500_600}
\end{table}
%%%%%%%%%%%%%%%%%%%%%%%%%%%%%%%%%%%%%%%%%%%%%%%%%%%%%%%%%%%%%%%%%%%%%%%%%%%%%%%%%

Before going into the taggers themselves, we consider the jet mass variable $m_{jet}$ for the different categories of top jets. Figure \ref{fig:mjet_DRm} shows the normalised $m_{jet}$ distribution for four cases of $\Delta R_m$ versus the QCD background of gluon jets and light quark jets.
As our matching criteria tightens, we see that the jet mass variable reconstructs the top quark mass more accurately. For unmatched jets, i.e., when no matching criterion is applied, the $m_{jet}$ distribution becomes flatter, and a peak forms at \textit{W} mass, $m_W = 80.3~GeV$, populated by events where one or more quarks lies outside the jet cone radius. 

\begin{figure}[htb!]
    \centering
    \includegraphics[width=0.5\textwidth]{./PT_500_mjet.pdf}~
    \includegraphics[width=0.5\textwidth]{./PT_500_mjet_q.pdf}
    \caption{Normalised $m_{jet}$ distribution of top jets with $\Delta R_m$ = 0.6, 0.8, 1.0, unmatched top jets. The distribution for top jets is compared with gluon jets \textit{(left)} and quark jets \textit{(right)}.}
    \label{fig:mjet_DRm}
\end{figure}

We calculate the Wasserstein-1 distance \cite{arjovsky2017wasserstein} $W_1(T, G)$ and $W_1(T, Q)$ between the distributions of top jets (T) and gluon jets (G), and top jets (T) and quark jets (Q). $W_1(T, G(Q))$ is a measure of the separation between two distributions T and G(Q). It can be calculated as the integral of the absolute difference between the cumulative distribution functions of the two distributions T and G(Q). Appendix \ref{app:wass} discusses the Wasserstein-1 distance in detail. 
We see that the top jets are better separated from the light quark jets than from the gluon jets in terms of the jet mass variable.

As mentioned, we generate truth-level events by applying a cut on $p_T^t$ > 450 GeV because we primarily study jets with $p_T^J\in [500,600)$ GeV for our analysis. Smearing of the jet energy in the detectors and the inclusion of an initial-state radiation (ISR) jet increase the $p_T^J$  from a lower $p_T^t$ of the original top quark. The cross-section at generation is 8.6 fb, and 25$\%$ of the generated events occupy the $p_T^J\in [500,600)$ GeV bin. A top quark can have an initial $p_T^t$ > 600 GeV, which then loses energy and migrates to the $p_T^J \in [500,600)$  bin. We generate signal events with a generation level cut of $p_T^t$ > 600 GeV, which yield a cross-section of 1.6 fb and again 25$\%$ of such events lie in the $p_T^J \in [500,600)$  bin. Figure \ref{fig:ptt_ptj} shows the correlation between $p_T$ of the top quark and the final top jet for the two $p_T$ cuts at the generation level. From the left panel, we see that while most events carry their initial energy from parton to jets, there are also events that undergo bin migration. The right panel shows that such events that migrate from higher to lower $p_T$ mainly populate the upper half of the $p_T^J\in [500,600)$ GeV bin, as expected. 

\begin{figure}[htb!]
    \centering
    \includegraphics[width=0.5\textwidth]{./PT_500_ptt_ptJ_um_450.pdf}~
    \includegraphics[width=0.5\textwidth]{./PT_600_ptt_ptJ_um_600.pdf}
    \caption{2-D matrix of truth level top quark $p_T$ and selection level top jet $p_T$ when generation $p_T$ cut is 450 GeV \textit{(left)} and 600 GeV \textit{(right)}. Each bin shows the percentage of unmatched jets that migrated from $p_T^t$ to $p_T^J$. }
    \label{fig:ptt_ptj}
\end{figure}

The phenomenon of bin migration affects variables like jet mass. Figure \ref{fig:mjet_600_500} shows the $m_{jet}$ distribution of unmatched jets. Jets that lose energy to have $p_T^J\in [500,600)$ GeV have lower $m_{jet}$ values, resulting in their distribution to shift towards lower $m_{jet}$.

\begin{figure}[htb!]
    \centering
    \includegraphics[width=0.5\textwidth]{./PT_600_mjet_bins.pdf}
    \caption{Normalised $m_{jet}$ distribution of top jets for unmatched top jets with the two different generation level $p_T$ cuts.}
    \label{fig:mjet_600_500}
\end{figure}

Later in Section \ref{sec:high_pt_gluon}, we also discuss the migration of the QCD jets from a higher $p_T$ range to the $p_T$ range of our study. We show how the phenomenon in the background jets affects the signal versus background classification. 

\section{\texttt{XGBOOST} analysis with jet substructure observables for 500 GeV jets}
\label{sec:standard_top_taggers_and_xgb}

This section briefly discusses the jet substructure observables that we use for top vs. gluon jet classification and the results when those observables are fed to a decision tree model.

\subsection{Jet substructure observables}
We use the \texttt{fastjet-contrib} library, which is an extension of the \texttt{Fastjet} \cite{Cacciari:2005hq,Cacciari:2011ma} package, to construct the jet substructure variables, namely $N$-Subjettiness \cite{Thaler:2010tr} and Energy Correlation Functions (ECFs) \cite{Larkoski:2013eya} and their ratios. These variables are well documented in existing literature where their proficiency in top tagging has been extensively studied and implemented \cite{Thaler:2010tr, Larkoski:2013eya, Larkoski:2014zma, Larkoski:2014gra, Larkoski:2015kga, Moult:2016cvt}. Appendix \ref{app:standard_top_taggers} provides a detailed discussion of the $N$-Subjettiness and the ECF observables used in our analysis.

Previous studies have explored top taggers that demonstrated improved performance through machine learning techniques that use a combination of high-level variables, jet images, and the four-momentum of jet constituents
 \cite{
%Grojean:2022mef,Kasieczka_2019, ANN_2015,Chakraborty:2020yfc, Diefenbacher_2020, Butter:2017cot, Kasieczka:2017nvn, Macaluso:2018tck, Bhattacharya:2020aid, Lim:2020igi, Dreyer:2020brq, Aguilar-Saavedra:2021rjk, Andrews:2021ejw, Dreyer:2022yom, Ahmed:2022hct, Munoz:2022gjq
Almeida:2015jua, Kasieczka:2017nvn, Butter:2017cot, Moore:2018lsr, Macaluso:2018tck, Kasieczka:2019dbj, Roy:2019jae, Diefenbacher:2019ezd, Chakraborty:2020yfc, Bhattacharya:2020aid, Lim:2020igi, Dreyer:2020brq, Aguilar-Saavedra:2021rjk, Andrews:2021ejw, Dreyer:2022yom, Ahmed:2022hct, Munoz:2022gjq}. Any machine learning-based top-tagger that uses kinematic variables has the same basic underlying strategy: it takes the jet substructure variables as input features for each event, trains on numerous signal and background events, and produces a probability score for each event in the test data set. For binary classification, this would correspond to the probability that the jet is a top jet.

Among the jet substructure observables, $N$-Subjettiness ($\tau_N^{(\beta)}$) quantifies the alignment of particles within a jet to $N$ subjet directions. Energy Correlation Functions (ECFs), denoted as $e_n^{(\beta)}$, measure multi-particle correlations within a jet and are categorised into five series for detailed analysis. The C-Series and the D-Series uses complex ratios to distinguish between jets with different prong structures. The U-Series generalises correlators by considering the smallest angular separations, simplifying the calculation of angular scales and was initially designed to distinguish between quark and gluon jets. The M-Series and N-Series ($N_i$) offer boost-invariant ratios of the generalised correlators as effective discriminants.

We categorise the $N$-Subjettiness ratios and the C-Series, D-Series, M-Series, N-Series, and U-Series of ECF ratios into six sets. Each set also includes the jet mass variable: $m_{jet}$. Table \ref{tab:xgb_features} summarises the variables used in the different categories of top taggers.

\begin{table}[hbt!]
\centering
\begin{tabular}{|c|l|}
\hline
Variable set          & \multicolumn{1}{c|}{Jet substructure variables used}                                                                                                                                                                                                                                                                                                   \\ \hline
$N$-Subjettiness series \cite{Thaler:2010tr} & \begin{tabular}[c]{@{}l@{}}$\tau_{21}^{(0.5)},~\tau_{32}^{(0.5)},~\tau_{43}^{(0.5)},~\tau_{54}^{(0.5)},~\tau_{65}^{(0.5)}$,\\ $\tau_{21}^{(1.0)},~\tau_{32}^{(1.0)},~\tau_{43}^{(1.0)},~\tau_{54}^{(1.0)},~\tau_{65}^{(1.0)}$,\\ $\tau_{21}^{(2.0)},~\tau_{32}^{(2.0)},~\tau_{43}^{(2.0)},~\tau_{54}^{(2.0)},~\tau_{65}^{(2.0)},~m_{jet}$\end{tabular} \\ \hline
C-Series \cite{Larkoski:2013eya}             & \begin{tabular}[c]{@{}l@{}}$C_1^{(1.0)},~C_2^{(1.0)},~C_3^{(1.0)}$,\\ $C_1^{(2.0)},~C_2^{(2.0)},~C_3^{(2.0)},~m_{jet}$\end{tabular}                                                                                                                                                                                                                    \\ \hline
D-Series \cite{Larkoski:2014zma, Larkoski:2014gra,Larkoski:2015kga}             & \begin{tabular}[c]{@{}l@{}}$D_2^{(1.0,1.0)},~D_2^{(1.0,2.0)},~D_2^{(2.0,1.0)},~D_2^{(2.0,2.0)}$,\\ $D_{3,a},~D_{3,b},~D_{3,c},~D_3^{(2,0.8,0.6)},~m_{jet}$\end{tabular}                                                                                                                                                                                \\ \hline
U-Series \cite{Moult:2016cvt}             & \begin{tabular}[c]{@{}l@{}}$U_1^{(0.5)},~U_2^{(0.5)},~U_3^{(0.5)}$,\\ $U_1^{(1.0)},~U_2^{(1.0)},~U_3^{(1.0)}$,\\ $U_1^{(2.0)},~U_2^{(2.0)},~U_3^{(2.0)},~m_{jet}$\end{tabular}                                                                                                                                                                         \\ \hline
M-Series \cite{Moult:2016cvt}            & \begin{tabular}[c]{@{}l@{}}$M_2^{(1.0)},~M_3^{(1.0)}$,\\ $M_2^{(2.0)},~M_3^{(2.0)},~m_{jet}$\end{tabular}                                                                                                                                                                                                                                              \\ \hline
N-Series \cite{Moult:2016cvt}             & \begin{tabular}[c]{@{}l@{}}$N_2^{(1.0)},~N_3^{(1.0)}$,\\ $N_2^{(2.0)},~N_3^{(2.0)},~m_{jet}$\end{tabular}                                                                                                                                                                                                                                              \\ \hline
\end{tabular}
\caption{Summary of the variables used as input features in the different categories of top taggers.}
\label{tab:xgb_features}
\end{table}

Our work uses a decision tree-based algorithm: the eXtreme Gradient BOOSTed decision tree \cite{DBLP:journals/corr/ChenG16}. \texttt{XGBOOST} implements gradient-boosted decision trees designed for speed and performance. The decision tree uses \texttt{binary:logloss} as the objective loss function. The hyperparameters \texttt{eta, gamma, max\_depth, min\_child\_weight, subsample,} and \texttt{colsample\_by\_tree} were optimised using the \texttt{HYPEROPT} package. The definitions of the hyperparameters and details of the optimization are presented in Table \ref{tab:hyperparams}, in Appendix \ref{app:hyperopt}. We evaluate the model's accuracy (ACC) at a threshold of 0.5 probability for classifying the jet as a top jet. To prevent overfitting, we use early stopping, which tells the model to stop training if validation performance continues to degrade after 5 rounds. The Receiving Operator Characteristics (ROC) curve of a model depicts the trade-off between true and false positive rates for different classification thresholds. The Area Under the Curve or AUC measures the effectiveness of a model. AUC values range from 0 to 1, where higher values indicate superior model discrimination, while an AUC of 0.5 represents random classification. We use AUC as a metric to evaluate the performance of our model during training and validation. 

Figure \ref{fig:tau32b1_DRm} shows the kinematic distributions of the variables $\tau_{32}^{(1.0)},~C_1^{1.0)},~D_3^{(2.0,0.8,0.6)}$ and $N_3^{(1.0)}$ with an additional cut applied on the jet mass, $m_{jet}$, such that $145<m_{jet}<205$ GeV. The cut $C_2^{(1.0)}$ > 0.1 in the distribution of $C_3^{(1.0)}$ ensures the infrared safety of the variable \cite{Larkoski:2013eya}. 

We can see from Figure \ref{fig:tau32b1_DRm} that higher $\Delta R_m$ values result in a lower $W_1(T,G)$ for all four variables, meaning that the overlap between the distribution of top jets and gluon jets increases. Smaller $\Delta R_m$ shifts the distributions of the variables to lower values by making the jets more three-prong-like. From Figure \ref{fig:ptb_DRb}, we know that the $\Delta R$ between the top and its decay products is inversely proportional to the $p_T$ of the latter. By tightening $\Delta R_m$, it is also ensured that the jet constituents along the prongs have higher $p_T$. Appendix \ref{app:mass_cut} shows the distributions without a jet mass cut. 

\begin{figure}[hbt!]
    \centering
    \includegraphics[width=0.45\textwidth]{./PT_500_tau32b1m.pdf}~
    \includegraphics[width=0.45\textwidth]{./PT_500_C3b1mc2.pdf}\\
    \includegraphics[width=0.45\textwidth]{./PT_500_D3m.pdf}~
    \includegraphics[width=0.45\textwidth]{./PT_500_N3b1m.pdf}
    \caption{Normalised distributions of jet substructure observables $\tau_{32}^{(1.0)}$ (\textit{top left}), $C_1^{1.0)}$ (\textit{top right}), $D_3^{(2.0,0.8,0.6)}$ (\textit{bottom left}), and $N_3^{(1.0)}$ (\textit{bottom right}) of top jets with $\Delta R_m$ = 0.6, 0.8, 1.0, unmatched top jets, and gluon jets.}
    \label{fig:tau32b1_DRm}
\end{figure}

Similar distributions of the observables for light quark jets have been presented in the Appendix \ref{app:quark_dist}. Clearly the observables $D_3$ and $N_3$ show greater separation between top jets and quark jets compared to when the background consists of gluon jets.
To evaluate the effect of both types of background, we train and test the models separately using each dataset. We start with gluon jets, which are harder to distinguish from top jets due to their higher multiplicity. In Section \ref{sec:xgb_quark}, we then analyse the classification between top jets and light quark jets.

\begin{figure}[hbt!]
    \centering
    \includegraphics[width=0.45\textwidth]{./PT_500_Fig5b.pdf}~
    \includegraphics[width=0.45\textwidth]{./PT_500_Fig6b.pdf}\\
    \includegraphics[width=0.45\textwidth]{./PT_500_Fig7b.pdf}~
    \includegraphics[width=0.45\textwidth]{./PT_500_Fig8b.pdf}
    \caption{2-D distribution of $f_{p_T}^{min}$ and $\tau_{32}^{(1.0)}$ (\textit{top left}), $C_1^{1.0)}$ (\textit{top right}), $D_3^{(2.0,0.8,0.6)}$ (\textit{bottom left}), and $N_3^{(1.0)}$ (\textit{bottom right}) of unmatched top jets.}
    \label{fig:fptmin}
\end{figure}

From the truth level information of top jets, we compute $p_T$ fractions of partons ($p_{T,p}/p_{T,t},~p=~b,~u,~d)$ and select the minimum, $f_{p_T}^{min} = \min (\frac{p_{T,b}}{p_{T,t}},\frac{p_{T,u}}{p_{T,t}},\frac{p_{T,d}}{p_{T,t}})$. 
$f_{p_T}^{min}$ thus refers to the softest $p_T$ fraction carried by a top decay product. The higher the $p_T$ fraction of a parton, the smaller is the distance of the parton from the top quark. In other words, top jets with high $f_{p_T}^{min}$ have more tightly constrained partons with less energy lost at the boundary. It turns out that lower values of $f_{p_T}^{min}$ correspond to higher values of the variables, as can be seen in Figure \ref{fig:fptmin}. When the softest parton of the three is too soft, the jet loses its three-prong structure, resulting in the higher values of the variables.

No cut is employed before training the \texttt{XGBOOST} models; instead, we include $m_{jet}$ and $C_2^{(1.0)}$ as input features in our top taggers, as mentioned above. This is done to enable the models to learn the intrinsic characteristics of the variables.

We also construct the jet substructure variables listed in Table \ref{tab:xgb_features} on pruned jets. Pruning \cite{Ellis:2009me} is a technique by which soft particles are removed from a jet if the following conditions hold while recombining two particles $i$ and $j$:

\begin{equation}
    \frac{min(p_T^i,p_T^j)}{p_T^i + p_T^j} < z_{cut} ~and~ \Delta R_{ij} > r_{cut}
\label{eq:prune}
\end{equation}

The parameters $z_{cut}$ and $r_{cut}$ are defined before the pruning procedure begins to control the extent of pruning. If the condition in Equation \ref{eq:prune} is satisfied, the softer particle is discarded, and the jet algorithm reconstructs a pruned jet. We set the two parameters $z_{cut}$ and $r_{cut}$ to be 0.1 and $\frac{m_{jet}}{p_{T^J}}$, respectively. Finally, we use the observables in Table \ref{tab:xgb_features} formulated on jets before pruning (unpruned features) and after pruning (pruned features) together.

%Table \ref{tab:NS_results} shows the test accuracy and AUC for unpruned, pruned, and combined cases. The results depend on the matching radius $\Delta R_m$. Both test accuracy and AUC decrease with an increase in $\Delta R_m$. This can be understood from the distribution of $\tau_{32}^{(1.0)}$ in Figure \ref{fig:tau32b1_DRm}, which shows more separation between the top and gluon jets with smaller matching radii. The set with pruned $N$-Subjettiness ratios perform worse than the unpruned case, and combining both improves the performance of the unpruned tagger only slightly.

%%%%%%%%%%%%%%%%%%%%%%%%%%%%%%%%%%%%%%%%%%%%%%%%%%%%%%%%%%%%%%%%%%%%%%%%%%%%%%%%%%%%%%%%%%%%%%%%

To summarise, we have six taggers with unpruned features, six with pruned features, and six with unpruned and pruned features combined. Furthermore, each tagger is trained on top and gluon jets generated with the four categories of truth level matching process used earlier. Each matching category has 600k top jets and 600k gluon jets for training, 200k top jets and 200k gluon jets for validation, and 200k top jets and 200k gluon jets for testing. Details of generation and simulation are listed in Section \ref{ssec:simulation}. The results from the different taggers are discussed in the next section.

\subsection{Results of the \texttt{XGBOOST} analysis}

The results of our \texttt{XGBOOST} top taggers are listed in Table \ref{tab:results_p0}, Table \ref{tab:results_p1}, and Table \ref{tab:results_p0p1} for the cases of unpruned jets, pruned jets, and a combination of pruned and unpruned respectively. 

\begin{table}[hbt!]
\centering
\begin{tabular}{|c|cccccccc|}
\hline
\multirow{3}{*}{\begin{tabular}[c]{@{}c@{}}Taggers with\\ features before jet\\ pruning\end{tabular}} & \multicolumn{8}{c|}{$p_T^J\in$[500,600) GeV}                                                                                                                                                                            \\ \cline{2-9} 
                              & \multicolumn{2}{c|}{$\Delta R_m$=0.6}                                   & \multicolumn{2}{c|}{$\Delta R_m$=0.8}                                   & \multicolumn{2}{c|}{$\Delta R_m$=1.0}                                   & \multicolumn{2}{c|}{unmatched}        \\ \cline{2-9} 
                              & \multicolumn{1}{c|}{ACC}   & \multicolumn{1}{c|}{AUC} & \multicolumn{1}{c|}{ACC}   & \multicolumn{1}{c|}{AUC} & \multicolumn{1}{c|}{ACC}   & \multicolumn{1}{c|}{AUC} & \multicolumn{1}{c|}{ACC}   & AUC \\ \hline
$N$-Subjettiness        & \multicolumn{1}{c|}{0.929} & \multicolumn{1}{c|}{\textbf{0.981}}    & \multicolumn{1}{c|}{\textbf{0.917}} & \multicolumn{1}{c|}{\textbf{0.974}}    & \multicolumn{1}{c|}{\textbf{0.905}} & \multicolumn{1}{c|}{\textbf{0.967}}    & \multicolumn{1}{c|}{\textbf{0.854}} & \textbf{0.928}    \\ \hline
C-Series                      & \multicolumn{1}{c|}{0.925} & \multicolumn{1}{c|}{0.978}    & \multicolumn{1}{c|}{0.910} & \multicolumn{1}{c|}{0.969}    & \multicolumn{1}{c|}{0.896} & \multicolumn{1}{c|}{0.961}    & \multicolumn{1}{c|}{0.842} & 0.917    \\ \hline
D-Series                      & \multicolumn{1}{c|}{0.928} & \multicolumn{1}{c|}{0.980}    & \multicolumn{1}{c|}{0.913} & \multicolumn{1}{c|}{0.972}    & \multicolumn{1}{c|}{0.899} & \multicolumn{1}{c|}{0.963}    & \multicolumn{1}{c|}{0.847} & 0.922    \\ \hline
U-Series                      & \multicolumn{1}{c|}{\textbf{0.930}} & \multicolumn{1}{c|}{\textbf{0.981}}    & \multicolumn{1}{c|}{0.916} & \multicolumn{1}{c|}{\textbf{0.974}}    & \multicolumn{1}{c|}{0.901} & \multicolumn{1}{c|}{0.964}    & \multicolumn{1}{c|}{0.851} & 0.926    \\ \hline
M-Series                      & \multicolumn{1}{c|}{0.910} & \multicolumn{1}{c|}{0.973}    & \multicolumn{1}{c|}{0.900} & \multicolumn{1}{c|}{0.967}    & \multicolumn{1}{c|}{0.886} & \multicolumn{1}{c|}{0.957}    & \multicolumn{1}{c|}{0.819} & 0.903    \\ \hline
N-Series                      & \multicolumn{1}{c|}{0.924} & \multicolumn{1}{c|}{0.978}    & \multicolumn{1}{c|}{0.906} & \multicolumn{1}{c|}{0.967}    & \multicolumn{1}{c|}{0.890} & \multicolumn{1}{c|}{0.956}    & \multicolumn{1}{c|}{0.839} & 0.915    \\ \hline
%Combined set                  & \multicolumn{1}{c|}{}      & \multicolumn{1}{c|}{}         & \multicolumn{1}{c|}{}      & \multicolumn{1}{c|}{}         & \multicolumn{1}{c|}{}      & \multicolumn{1}{c|}{}         & \multicolumn{1}{c|}{}      &          \\ \hline
\end{tabular}
\caption{Test Accuracy at 50$\%$ threshold probability score (ACC) and Area Under the Curve (AUC) for the \texttt{XGBOOST} based taggers using $N$-Subjettiness, C, D, U, M, and N-Series of jet substructure observables with varying matching radius ($\Delta R_m$) of top jets. The features are constructed on unpruned top and gluon jets.}
\label{tab:results_p0}
\end{table}

The top taggers' performance in all the cases increases with a decreasing $\Delta R_m$. This is evident from the distributions of the variables in Figure \ref{fig:tau32b1_DRm} showing a better separation between top and gluon jets with smaller matching radii. In the case of variables formed from unpruned jets, the $N$-Subjettiness tagger has the highest AUC among all the taggers for all $\Delta R_m$, outperforming the C-Series, D-Series, M-Series, and N-Series. In Ref. \cite{Larkoski:2014zma}, it was shown that the variable $D_3^{(2.0,0.8,0.6)}$ has the best discrimination power compared to $\tau_{32}^{(1.0)}$ and $C_3^{(1.0)}$. Nevertheless, when the D-Series variables are combined in a multivariate tagger, its performance is inferior to that of the $N$-Subjettiness tagger by $\sim 0.1-0.6\%$, although it shows improvement over the C-Series tagger by $\sim 0.2-0.5\%$. Although the U-Series of variables were designed as quark/gluon discriminants, they also have a considerable discriminating power between top and QCD jets similar to $N$-Subjettiness. The M-Series tagger exhibits the lowest performance among the six sets.

\begin{table}[hbt!]
\centering
\begin{tabular}{|c|cccccccc|}
\hline
\multirow{3}{*}{\begin{tabular}[c]{@{}c@{}}Taggers with\\ features after jet\\ pruning\end{tabular}} & \multicolumn{8}{c|}{$p_T^J\in$[500,600) GeV}                                                                                                                                                                \\ \cline{2-9} 
& \multicolumn{2}{c|}{$\Delta R_m$=0.6}                                & \multicolumn{2}{c|}{$\Delta R_m$=0.8}                                & \multicolumn{2}{c|}{$\Delta R_m$=1.0}                                & \multicolumn{2}{c|}{unmatched}     \\ \cline{2-9} 
 & \multicolumn{1}{c|}{ACC}   & \multicolumn{1}{c|}{AUC}   & \multicolumn{1}{c|}{ACC}   & \multicolumn{1}{c|}{AUC}   & \multicolumn{1}{c|}{ACC}   & \multicolumn{1}{c|}{AUC}   & \multicolumn{1}{c|}{ACC}   & AUC   \\ \hline
$N$-Subjettiness                                                                                & \multicolumn{1}{c|}{0.922} & \multicolumn{1}{c|}{\textbf{0.978}} & \multicolumn{1}{c|}{0.896} & \multicolumn{1}{c|}{\textbf{0.964}} & \multicolumn{1}{c|}{\textbf{0.877}} & \multicolumn{1}{c|}{\textbf{0.950}} & \multicolumn{1}{c|}{\textbf{0.843}} & \textbf{0.917} \\ \hline
C-Series                                                                                            & \multicolumn{1}{c|}{0.922} & \multicolumn{1}{c|}{\textbf{0.978}} & \multicolumn{1}{c|}{0.895} & \multicolumn{1}{c|}{0.962} & \multicolumn{1}{c|}{\textbf{0.877}} & \multicolumn{1}{c|}{\textbf{0.950}} & \multicolumn{1}{c|}{0.839} & 0.913 \\ \hline
D-Series                                                                                            & \multicolumn{1}{c|}{0.921} & \multicolumn{1}{c|}{0.977} & \multicolumn{1}{c|}{0.894} & \multicolumn{1}{c|}{0.962} & \multicolumn{1}{c|}{\textbf{0.877}} & \multicolumn{1}{c|}{\textbf{0.950}} & \multicolumn{1}{c|}{0.840} & 0.914 \\ \hline
U-Series                                                                                            & \multicolumn{1}{c|}{\textbf{0.923}} & \multicolumn{1}{c|}{\textbf{0.978}} & \multicolumn{1}{c|}{\textbf{0.897}} & \multicolumn{1}{c|}{\textbf{0.964}} & \multicolumn{1}{c|}{0.875} & \multicolumn{1}{c|}{\textbf{0.950}} & \multicolumn{1}{c|}{0.842} & 0.916 \\ \hline
M-Series                                                                                            & \multicolumn{1}{c|}{0.909} & \multicolumn{1}{c|}{0.973} & \multicolumn{1}{c|}{0.878} & \multicolumn{1}{c|}{0.955} & \multicolumn{1}{c|}{0.860} & \multicolumn{1}{c|}{0.940} & \multicolumn{1}{c|}{0.822} & 0.903 \\ \hline
N-Series                                                                                            & \multicolumn{1}{c|}{0.919} & \multicolumn{1}{c|}{0.977} & \multicolumn{1}{c|}{0.889} & \multicolumn{1}{c|}{0.959} & \multicolumn{1}{c|}{0.870} & \multicolumn{1}{c|}{0.945} & \multicolumn{1}{c|}{0.836} & 0.911 \\ \hline

\end{tabular}
\caption{Test Accuracy at 50$\%$ threshold probability score (ACC) and Area Under the Curve (AUC) for the \texttt{XGBOOST} based taggers using $N$-Subjettiness, C, D, U, M, and N-Series of jet substructure observables with varying matching radius ($\Delta R_m$) of top jets. The features are constructed on pruned top and gluon jets with the parameters set to $z_{cut}$ = 0.1 and $r_{cut}$ = $\frac{m_{jet}}{p_{T^J}}$.}
\label{tab:results_p1}
\end{table}

Figure \ref{fig:mjet_pru_DRm} shows the $m_{jet}$ distribution after pruning the jets. Although $m_{jet}$ shows better separation after pruning, comparing Tables \ref{tab:results_p0} and \ref{tab:results_p1}, we see that features derived from pruned jets yield subpar performance.  

\begin{figure}[htb!]
    \centering
    \includegraphics[width=0.5\textwidth]{./PT_500_mjet_p.pdf}
    \caption{Normalised $m_{jet}$ distribution of top jets with $\Delta R_m$ = 0.6, 0.8, 1.0, unmatched top jets, and gluon jets after pruning with the parameters $z_{cut}$ = 0.1 and $r_{cut}$ = $\frac{m_{jet}}{p_{T^J}}$.}
    \label{fig:mjet_pru_DRm}
\end{figure}

On pruning the soft emissions, the gluon jet mass peaks at lower values compared to the unpruned scenario, and it also improves the jet mass resolution. However, pruned top jets give rise to prominent peaks at lower $m_{jet}$ values. This effect increases with increasing $\Delta R_m$. The $N$-Subjettiness and the ECF ratios were designed to capture the energy distributions along the subjets and their angular correlations. The removal of soft and wide-angle radiation during pruning can lead to the loss of information embedded in the soft component. Removing energy from a soft prong might cause the top jets to lose their distinctive three-prong structure, making it two or one-prong-like. Figure \ref{fig:tau32b1_DRm_p} shows the distribution of variables after jet pruning. With a cut on the mass of the pruned jet, $\tau_{32}$ and $D_3$ show increase in $W_1(T,G)$ after pruning while $C_3$ and $N_3$ show decrease in $W_1(T,G)$. However, without applying a cut on the jet mass, these variables lose their discrimination power considerably. The distributions without the jet mass cut are shown in Appendix \ref{app:mass_cut}. Observables like $C_3$ and $D_3$ peak at lower values for gluon jets after pruning.

\begin{figure}[hbt!]
    \centering
    \includegraphics[width=0.45\textwidth]{./PT_500_tau32b1m_p.pdf}~
    \includegraphics[width=0.45\textwidth]{./PT_500_C3b1mc2_p.pdf}\\
    \includegraphics[width=0.45\textwidth]{./PT_500_D3m_p.pdf}~
    \includegraphics[width=0.45\textwidth]{./PT_500_N3b1m_p.pdf}
    \caption{Normalised distributions of jet substructure observables $\tau_{32}^{(1.0)}$ (\textit{top left}), $C_1^{(1.0)}$ (\textit{top right}), $D_3^{(2.0,0.8,0.6)}$ (\textit{bottom left}), and $N_3^{(1.0)}$ (\textit{bottom right}) of top jets with $\Delta R_m$ = 0.6, 0.8, 1.0, after pruning unmatched top jets, and gluon jets.}
    \label{fig:tau32b1_DRm_p}
\end{figure}
 
% \usepackage{multirow}
\begin{table}[hbt!]
\centering
\begin{tabular}{|c|cccccccc|}
\hline
\multirow{3}{*}{\begin{tabular}[c]{@{}c@{}}Tagger with \\ pruned and \\ unpruned features\end{tabular}} & \multicolumn{8}{c|}{$p_T^J\in$[500,600) GeV}                                                                                                                                                                \\ \cline{2-9} 
& \multicolumn{2}{c|}{$\Delta R_m$=0.6}                                & \multicolumn{2}{c|}{$\Delta R_m$=0.8}                                & \multicolumn{2}{c|}{$\Delta R_m$=1.0}                                & \multicolumn{2}{c|}{unmatched}     \\ \cline{2-9} 
& \multicolumn{1}{c|}{ACC}   & \multicolumn{1}{c|}{AUC}   & \multicolumn{1}{c|}{ACC}   & \multicolumn{1}{c|}{AUC}   & \multicolumn{1}{c|}{ACC}   & \multicolumn{1}{c|}{AUC}   & \multicolumn{1}{c|}{ACC}   & AUC   \\ \hline
$N$-Subjettiness                                                                                  & \multicolumn{1}{c|}{\textbf{0.933}} & \multicolumn{1}{c|}{\textbf{0.983}} & \multicolumn{1}{c|}{\textbf{0.921}} & \multicolumn{1}{c|}{\textbf{0.976}} & \multicolumn{1}{c|}{\textbf{0.907}} & \multicolumn{1}{c|}{\textbf{0.969}} & \multicolumn{1}{c|}{\textbf{0.858}} & \textbf{0.932} \\ \hline
C-Series                                                                                                & \multicolumn{1}{c|}{0.931} & \multicolumn{1}{c|}{0.981} & \multicolumn{1}{c|}{0.917} & \multicolumn{1}{c|}{0.974} & \multicolumn{1}{c|}{0.902} & \multicolumn{1}{c|}{0.965} & \multicolumn{1}{c|}{0.851} & 0.926 \\ \hline
D-Series                                                                                                & \multicolumn{1}{c|}{0.931} & \multicolumn{1}{c|}{0.981} & \multicolumn{1}{c|}{0.917} & \multicolumn{1}{c|}{0.974} & \multicolumn{1}{c|}{0.903} & \multicolumn{1}{c|}{0.965} & \multicolumn{1}{c|}{0.852} & 0.926 \\ \hline
U-Series                                                                                                & \multicolumn{1}{c|}{0.932} & \multicolumn{1}{c|}{0.978} & \multicolumn{1}{c|}{0.918} & \multicolumn{1}{c|}{0.975} & \multicolumn{1}{c|}{0.902} & \multicolumn{1}{c|}{0.966} & \multicolumn{1}{c|}{0.854} & 0.928 \\ \hline
M-Series                                                                                                & \multicolumn{1}{c|}{0.924} & \multicolumn{1}{c|}{0.979} & \multicolumn{1}{c|}{0.911} & \multicolumn{1}{c|}{0.972} & \multicolumn{1}{c|}{0.895} & \multicolumn{1}{c|}{0.963} & \multicolumn{1}{c|}{0.838} & 0.919 \\ \hline
N-Series                                                                                                & \multicolumn{1}{c|}{0.928} & \multicolumn{1}{c|}{0.981} & \multicolumn{1}{c|}{0.914} & \multicolumn{1}{c|}{0.973} & \multicolumn{1}{c|}{0.897} & \multicolumn{1}{c|}{0.962} & \multicolumn{1}{c|}{0.850} & 0.925 \\ \hline
\end{tabular}
\caption{Test Accuracy at 50$\%$ threshold probability score (ACC) and Area Under the Curve (AUC) for the \texttt{XGBOOST} based taggers using $N$-Subjettiness, C, D, U, M, and N-Series of jet substructure observables with varying matching radius ($\Delta R_m$) of top jets. The features are constructed on unpruned and pruned top and gluon jets with the parameters set to $z_{cut}$ = 0.1 and $r_{cut}$ = $\frac{m_{jet}}{p_{T^J}}$.}
\label{tab:results_p0p1}
\end{table}

From Table \ref{tab:results_p0p1}, we see that merging input features derived from both unpruned and pruned jets results in better performance of taggers compared to using either the unpruned or pruned set of features alone. The combination captures the distinctive characteristics of top and gluon jets both before and after the pruning process. As stated above, a model with solely pruned features relies on reduced information due to the loss of soft radiation properties. However, the inclusion of features pre-pruning is able to compensate for the loss, thus forming a complimentary set. Figure \ref{fig:p0_p1} shows the 2-D histograms between unpruned and pruned versions of $m_{jet}$ and $\tau_{32}^{(1.0)}$ for top and gluon jets. They populate different regions in top jets and in gluon jets. In the 2D plane of unpruned and pruned versions of the same observable, pruning introduces a greater non-diagonality in the case of gluon jets than in the case of top jets. This extra information is passed on to the taggers when we use the two versions together as input.

\begin{figure}[hbt!]
    \centering
    \includegraphics[width=0.45\textwidth]{./PT_500_mass_top_mr10.pdf}~
    \includegraphics[width=0.45\textwidth]{./PT_500_mass_gluon.pdf}\\
    \includegraphics[width=0.45\textwidth]{./PT_500_tau32_top_mr10.pdf}~
    \includegraphics[width=0.45\textwidth]{./PT_500_tau32_gluon.pdf}
    \caption{Distribution of top and gluon jets in the two dimensional plane of pruned and unpruned versions of $m_{jet}$ (\textit{top}) and $\tau_{32}^{(1.0)}$ (\textit{bottom}).}
    \label{fig:p0_p1}
\end{figure}

 The superior performance of the $N$-Subjettiness tagger compared to other multivariate \texttt{XGBOOST} based taggers constructed from the ECF ratios, despite some of these taggers showing better performance in their original studies \cite{Larkoski:2014zma, Moult:2016cvt}, implies that the $N$-Subjettiness variables, when used together, form a uniquely effective set of discriminants for top vs. gluon classification. 

Having constructed the six small sets of taggers with \texttt{XGBOOST}, we try to interpret the results in the next section. 

\section{SHAP as a method of interpretation}
\label{sec:SHAP}

Each feature can be a valuable tool that helps the model better identify a jet, but some features play a more critical role than others. Therein lies the usefulness of \textit{feature importances}. Feature importances can be either \textit{global} or \textit{local}. Global feature importance calculates how useful a feature is throughout the entire dataset. Local feature importance is the same for a single prediction (an individual data point or event). 

Global feature importances provide an overall ranking of feature's impact on the model across the entire dataset. Features that consistently contribute to reducing the model's loss function or increasing its accuracy across multiple iterations of training are considered important. In decision tree models, features that are used more frequently to a split a node while training are also assigned a higher importance. However, global importances do not capture the variation in a feature's contribution across different subsets of the data or individual predictions. Local importances on the other hand quantify the contribution of each feature for every individual prediction. The feature attributions when aggregated across the entire dataset give us the overall ranking, which can then be used for feature selection and dimensionality reduction.

Among various methods used for interpreting a model, we choose to use SHAP \cite{DBLP:journals/corr/LundbergL17}. It is a local feature importance quantification method based on Shapley values \cite{Shapley+2016+307+318}. A Shapley value is a quantity designed in the context of game theory to calculate a player's contribution to a game of $n$ such players with a fixed outcome. This is done by calculating the outcome when one player is removed. The difference between the game's outcome with and without the player present is a measure of how good that player is. In other words, For a given feature, the Shapley value calculates the difference in the model's prediction when the feature is included versus when it is excluded, considering all possible combinations with the other features. The Shapley values are calculated using the following formula \cite{DBLP:journals/corr/LundbergL17, DBLP:journals/corr/abs-1802-03888}.

\begin{equation}
    \phi_i = \sum_{S\subseteq N\setminus \{i\}} \frac{|S|!(M-|S|-1)!}{M!} [f_x(S\cup \{i\})-f_x(S)]
\end{equation}

Here, $i$ is the feature for which SHAP value is calculated, $N$ is the set of all features, and $M$ is the total number of features. $S$ is a subset of $N$ that does not include $i$. The function $f_x$ is the model prediction. Thus, the algorithm computes a weighted sum of the differences in model prediction outcomes with and without the $i^{th}$ feature for all possible arrangements of the subset $S$.

After the decision trees are trained, we use the package \texttt{shap} to calculate the SHAP values for each feature using the test samples. To compute the SHAP values, \texttt{shap} uses \texttt{TreeExplainer}\cite{DBLP:journals/corr/abs-1905-04610}, which has been specifically designed to explain tree-based machine learning models and is a faster algorithm to estimate SHAP values compared to general SHAP explainers.
%SHAP has been widely used to interpret machine learning output in collider studies \cite{Cornell:2021gut, Alvestad:2021sje, Grojean:2020ech, CMS:2021qzz, Grojean:2022mef, Adhikary:2022jfp}. 
The base value or expected value of a model is the average model output for the entire dataset prior to training. The classification result for each event is equal to the sum of the SHAP values of all features for that event, plus the base value.

%%%%%%%%%%%%%%%%%%% Shap summary plots of different sets of variables %%%%%%%%%%%%%%%%%%%%%%%%%
\begin{figure}[htb!]
    \centering
    \includegraphics[width=0.45\textwidth]{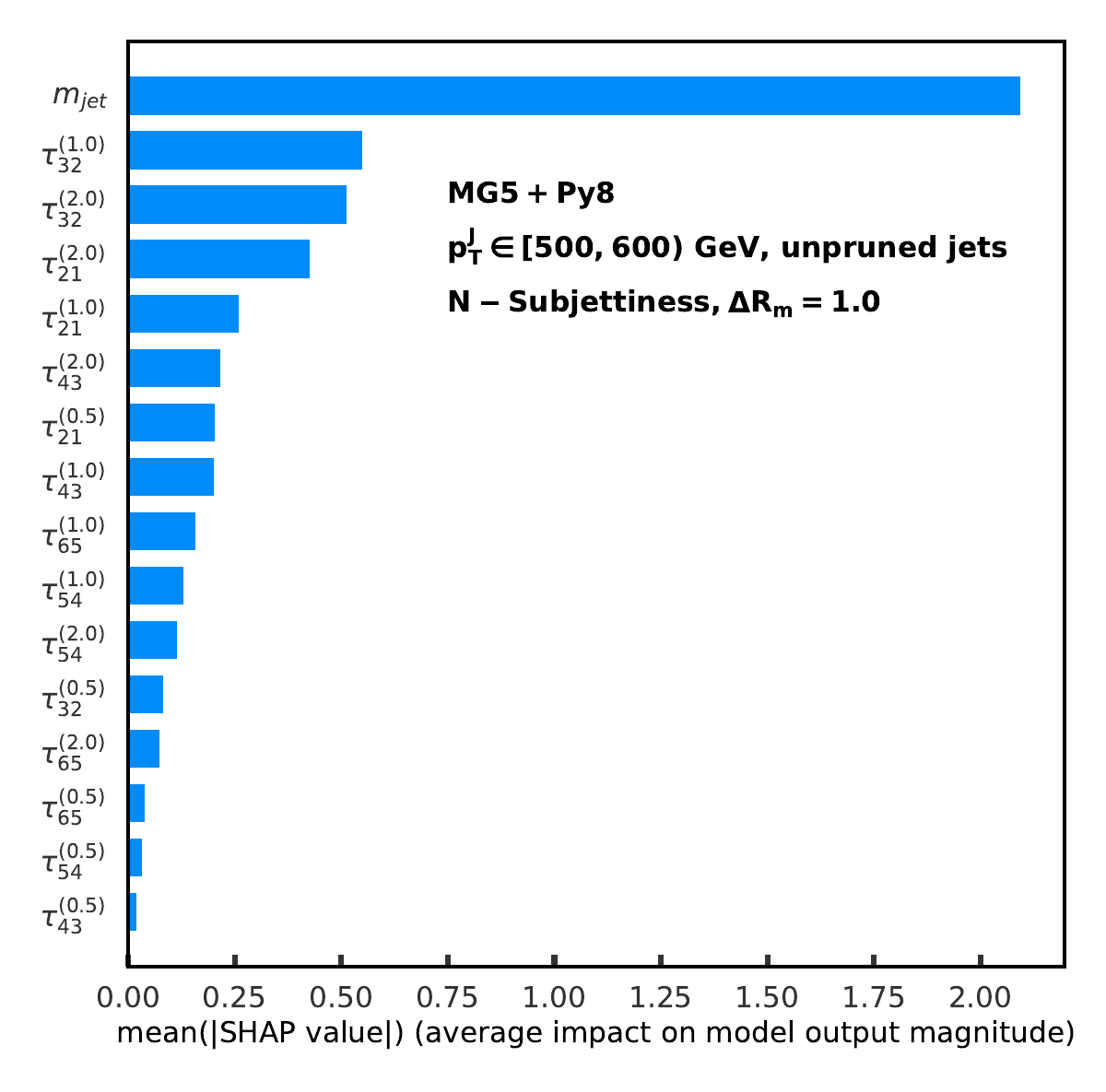}~
    \includegraphics[width=0.5\textwidth]{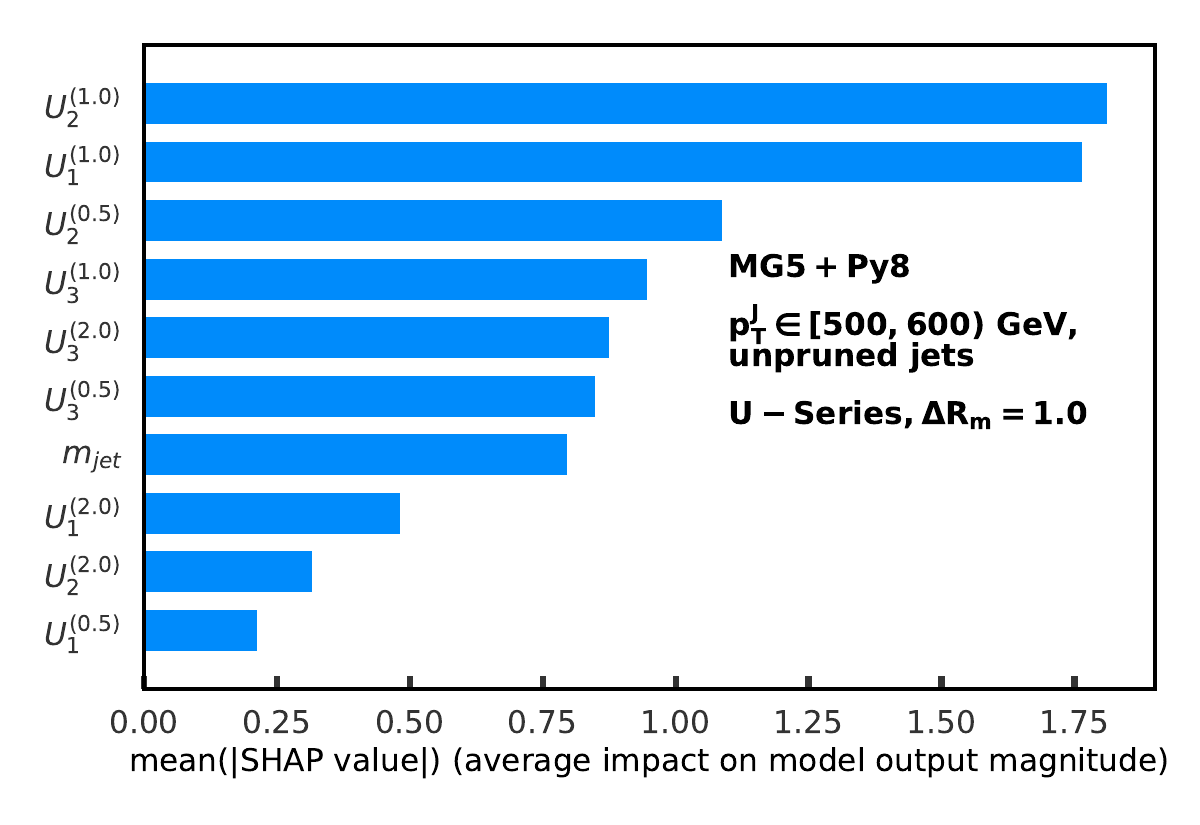}\\
    \includegraphics[width=0.45\textwidth]{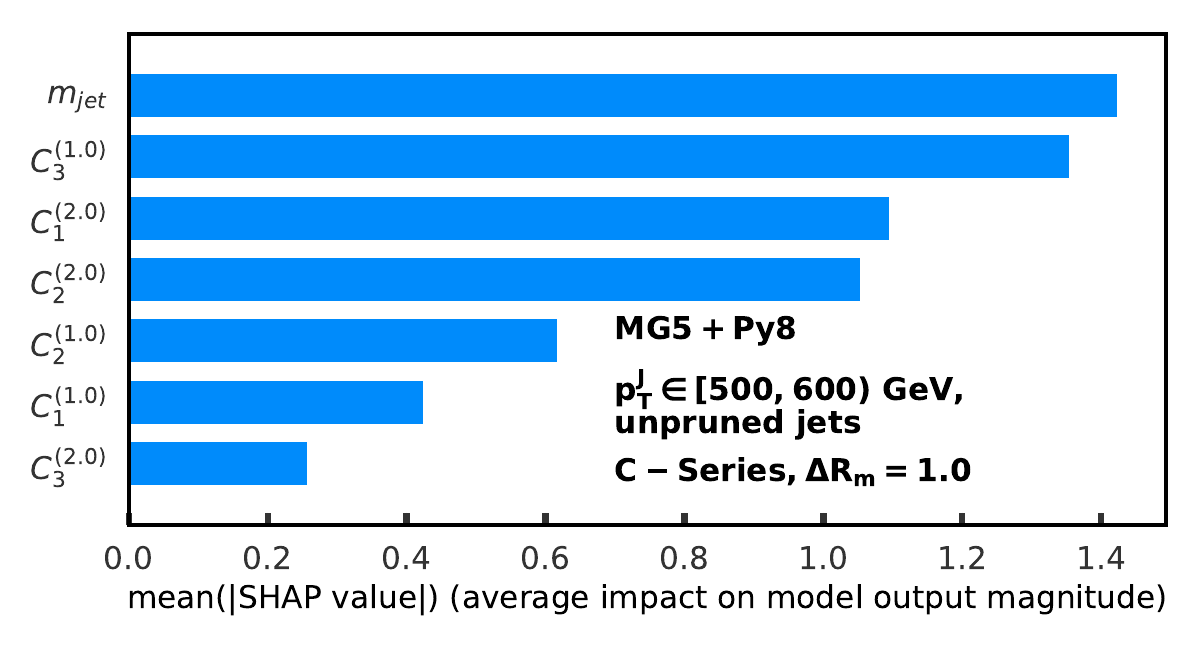}~
    \includegraphics[width=0.5\textwidth]{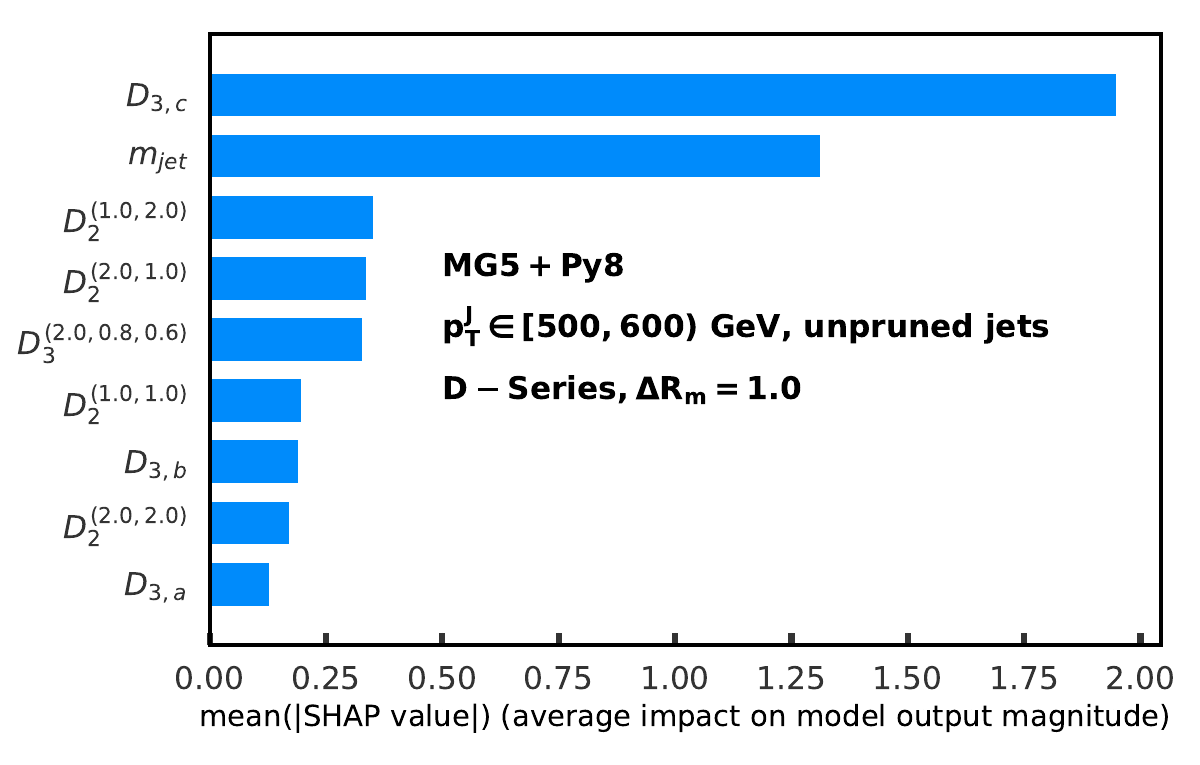}\\
    \includegraphics[width=0.45\textwidth]{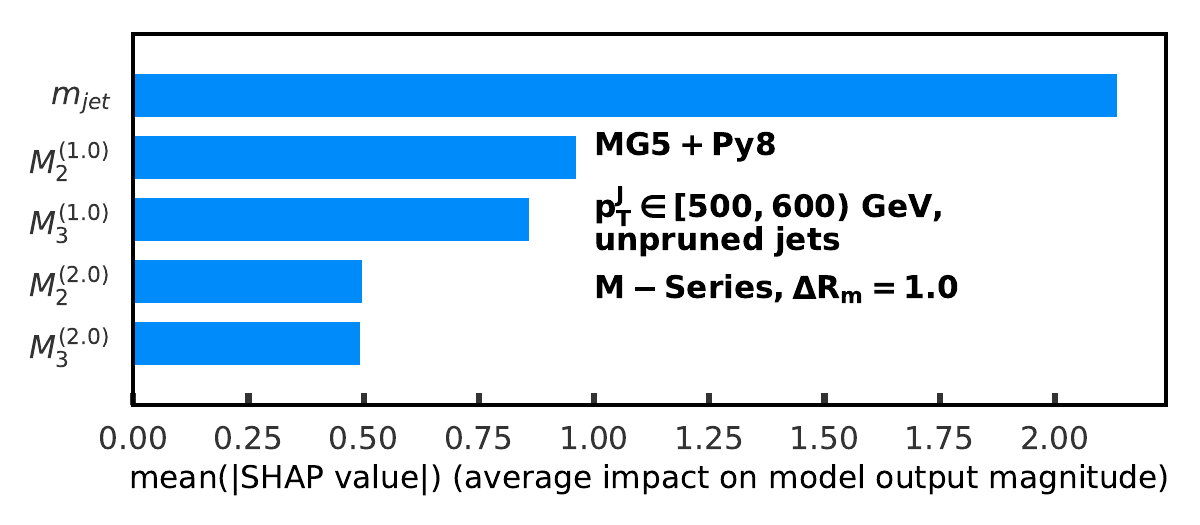}~
    \includegraphics[width=0.5\textwidth]{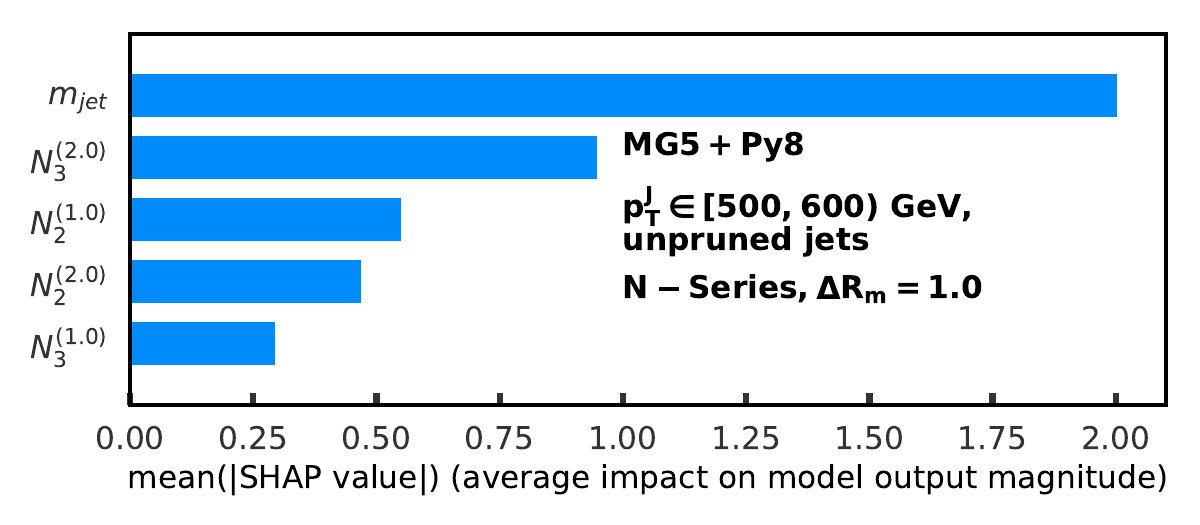}
    \caption{Bar plots depicting the hierarchy of feature importance. Each panel from the top left in the first row to the bottom right in the third row shows the feature importance calculated by SHAP for $N$-Subjettiness, U, C, D, M, and N-Series consecutively. The dataset contains unpruned top and gluon jets with $\Delta R_m$ = 1.0 for top jets.}
    \label{fig:shap_bar}
\end{figure}
%%%%%%%%%%%%%%%%%%%%%%%%%%%%%%%%%%%%%%%%%%%%%%%%%%%%%%%%%%%%%%%%%%%%%%%%%%%%%%%%%%%%%%%%%%%%%%%

Figure \ref{fig:shap_bar} shows, for every \texttt{XGBOOST} model trained on unpruned, $\Delta R_m$ = 1.0 jets with six different sets of variables, the features in decreasing order of importance based on the average of their absolute SHAP values over the entire dataset. Top jets, owing to their rich substructure, can include both three-prong and two-prong configurations, especially when energy loss at the jet boundary causes them to lose the distinct three-prong structure. The $N$-Subjetiness, C-Series, D-Series, M-Series, and N-Series taggers have the most important contribution from the observables that were designed to identify both three-prong and two-prong structures. This suggests that the model corectly uses the relevant observables associated with top tagging to identify differences in jet substructures

The model trained with the $N$-Subjettiness series of features has $m_{jet}$ as the most important feature. In other words, $m_{jet}$ contributes the most to the model's output score, followed by $\tau_{32}^{(1.0)}$, $\tau_{32}^{(2.0)}$, $\tau_{21}^{(2.0)}$, and $\tau_{21}^{(1.0)}$. Higher $N$-Subjettiness ratios are less important in classifying three-pronged jets from QCD. Apart from the U-Series, $m_{jet}$ acquires a high SHAP score in all the models. For the D-Series, the term that has the highest feature importance is $D_{3,c}$,

\begin{equation}
    \frac{e_{4}^{(0.6)}(e_{2}^{(2.0)})^{\frac{2\times 0.8}{0.6}-\frac{0.6}{2.0}}}{(e_{3}^{(0.8)})^{2}}
\end{equation}

This term corresponds to the three-prong phase space where a soft emission has a $p_T$ fraction parametrically much smaller than the opening angle. Surprisingly, it outperforms the $D_3$ variable, which includes this term.  

In the case of the U-Series, the variable $U_2^{(1.0)}$ has the highest contribution, which is the generalised ECF,

\begin{equation}
    U_2^{(1.0)} = {}_{1}e_{3}^{(1.0)} = \sum_{i_{1}<i_{2}<i_{3}\in J}z_{i_{1}}z_{i_{2}}z_{i_{3}} \min \left\lbrace\Delta R_{i_{1}i_{2}},~\Delta R_{i_{2}i_{3}},~\Delta R_{i_{1}i_{3}}\right\rbrace
\end{equation}

where $i_n$ is the $n^{th}$ constituent of jet $J$, $z_{i_n} = p_{T}^{i_n}/p_{T}^J$, and $R_{ij} = \sqrt{\Delta \eta_{ij}^2 + \Delta \phi_{ij}^2}$ is the distance between two constituents $i$ and $j$ in the $\eta-\phi$ plane.
Variables in the U-Series are highly correlated among each other as they differ only by the powers of the ECFs, and also with and $m_{jet}$, which is why $m_{jet}$ is ranked lower. Decision trees handle correlations between two features by prioritising one over the other during the splitting of a node. It is interesting to note that the observables that were proposed in \cite{Moult:2016cvt} for quark vs. gluon discrimination also work well for top vs. gluon discrimination.

\subsection{Constructing a hybrid tagger}
\label{ssec:hybrid}
We now construct a tagger that combines input features from the six taggers. We first calculate the correlation between all the variables and identify the ones more than 95$\%$ correlated. 
\begin{figure}
    \centering
    \includegraphics[width=0.95\textwidth]{./PT_500_corr_um_500GeV.pdf}
    \caption{Correlation matrix between  $m_{jet}$, $C_1^{(1.0)}, ~C_1^{(2.0)}, ~U_1^{(0.5)}, ~U_1^{(1.0)}, ~U_1^{(2.0)}, ~U_2^{(1.0)},  ~U_3^{(0.5)}$, and $U_3^{(1.0)}$ for unmatched jets.}
    \label{fig:corr_py8}
\end{figure}

These include the variables $C_1^{(1.0)}, ~C_1^{(2.0)}, ~U_1^{(0.5)}, ~U_1^{(1.0)}, ~U_1^{(2.0)}, ~U_2^{(1.0)},  ~U_3^{(0.5)}$, and $U_3^{(1.0)}$. Figure \ref{fig:corr_py8} shows the correlation between all the above variables and $m_{jet}$ for unmatched jets. We remove all the variables from the upper diagonal that show more than 95$\%$ correlation. The highly correlated set of features remains the same for all $\Delta R_m$. After removing the highly correlated features from all six sets, we combine them into an extensive set of 39 features, which we call the combined set of features, and use them for training and testing our decision tree on unpruned jets with $\Delta R_m$ = 0.6, 0.8, and 1.0 top jets.  We also train and test on unmatched jets. We select the 15 highest-ranked features for each of the $\Delta R_m$ categories using SHAP to form a reduced hybrid set of tagger variables. 

Table \ref{tab:best15_deltaRm} contains the list of the most important observables for the different $\Delta R_m$ categories. Out of these, 11 features are common across all the four $\Delta R_m$ sets. They are:

\begin{equation}
\begin{aligned}
    & m_{jet}, \tau_{21}^{(0.5)}, \tau_{32}^{(2.0)}, \tau_{43}^{(2.0)}, D_2^{(1.0,2.0)},
    & D_{3,c}, D_3^{(2,0.8,0.6)}, N_3^{(1.0)}, M_3^{(1.0)},M_3^{(2.0)},		 U_2^{(0.5)}
    \label{eq:commom_11}
\end{aligned}
\end{equation}

% \usepackage{multirow}
\begin{table}[htb!]
\centering
\begin{tabular}{|c|cccc|}
\hline
\multirow{2}{*}{\begin{tabular}[c]{@{}c@{}}Features in the \\ reduced set tagger\end{tabular}} & \multicolumn{4}{c|}{$p_T^J\in$[500,600) GeV} \\ \cline{2-5} 
 & $\Delta R_m$=0.6 & $\Delta R_m$=0.8 & $\Delta R_m$=1.0 & unmatched \\ \hline
$m_{jet}$ & $\checkmark$ & $\checkmark$ & $\checkmark$ & $\checkmark$ \\
$\tau_{21}^{(0.5)}$ & $\checkmark$ & $\checkmark$ & $\checkmark$ & $\checkmark$ \\
$\tau_{21}^{(1.0)}$ & $\times$ & $\checkmark$ & $\checkmark$ & $\times$ \\
$\tau_{32}^{(0.5)}$ & $\times$ & $\checkmark$ & $\times$ & $\times$ \\
$\tau_{32}^{(1.0)}$ & $\times$ & $\checkmark$ & $\checkmark$ & $\checkmark$ \\
$\tau_{32}^{(2.0)}$ & $\checkmark$ & $\checkmark$ & $\checkmark$ & $\checkmark$ \\
$\tau_{43}^{(1.0)}$ & $\times$ & $\times$ & $\times$ & $\checkmark$ \\
$\tau_{43}^{(2.0)}$ & $\checkmark$ & $\checkmark$ & $\checkmark$ & $\checkmark$ \\
$\tau_{65}^{(1.0)}$ & $\times$ & $\times$ & $\checkmark$ & $\checkmark$ \\
$C_2^{(1.0)}$ & $\checkmark$ & $\times$ & $\times$ & $\times$ \\
$D_2^{(1.0,1.0)}$ & $\checkmark$ & $\times$ & $\times$ & $\times$ \\
$D_2^{(1.0,2.0)}$ & $\checkmark$ & $\checkmark$ & $\checkmark$ & $\checkmark$ \\
$D_{3,c}$ & $\checkmark$ & $\checkmark$ & $\checkmark$ & $\checkmark$ \\
$D_3^{(2,0.8,0.6)}$ & $\checkmark$ & $\checkmark$ & $\checkmark$ & $\checkmark$ \\
$N_2^{(1.0)}$ & $\times$ & $\times$ & $\checkmark$ & $\checkmark$ \\
$N_3^{(1.0)}$ & $\checkmark$ & $\checkmark$ & $\checkmark$ & $\checkmark$ \\
$N_3^{(2.0)}$ & $\checkmark$ & $\times$ & $\times$ & $\times$ \\
$M_3^{(1.0)}$ & $\checkmark$ & $\checkmark$ & $\checkmark$ & $\checkmark$ \\
$M_3^{(2.0)}$ & $\checkmark$ & $\checkmark$ & $\checkmark$ & $\checkmark$ \\
$U_2^{(0.5)}$ & $\checkmark$ & $\checkmark$ & $\checkmark$ & $\checkmark$ \\
$U_3^{(2.0)}$ & $\checkmark$ & $\checkmark$ & $\times$ & $\times$ \\ \hline
\end{tabular}
\caption{Features appearing in the reduced set with 15 highest ranked features for unpruned jets with $\Delta R_m = 0.6,~0.8,~1.0$ and unmatched matching criteria.}
\label{tab:best15_deltaRm}
\end{table}

\begin{figure}[htb!]
    \centering
    \includegraphics[width=0.5\textwidth]{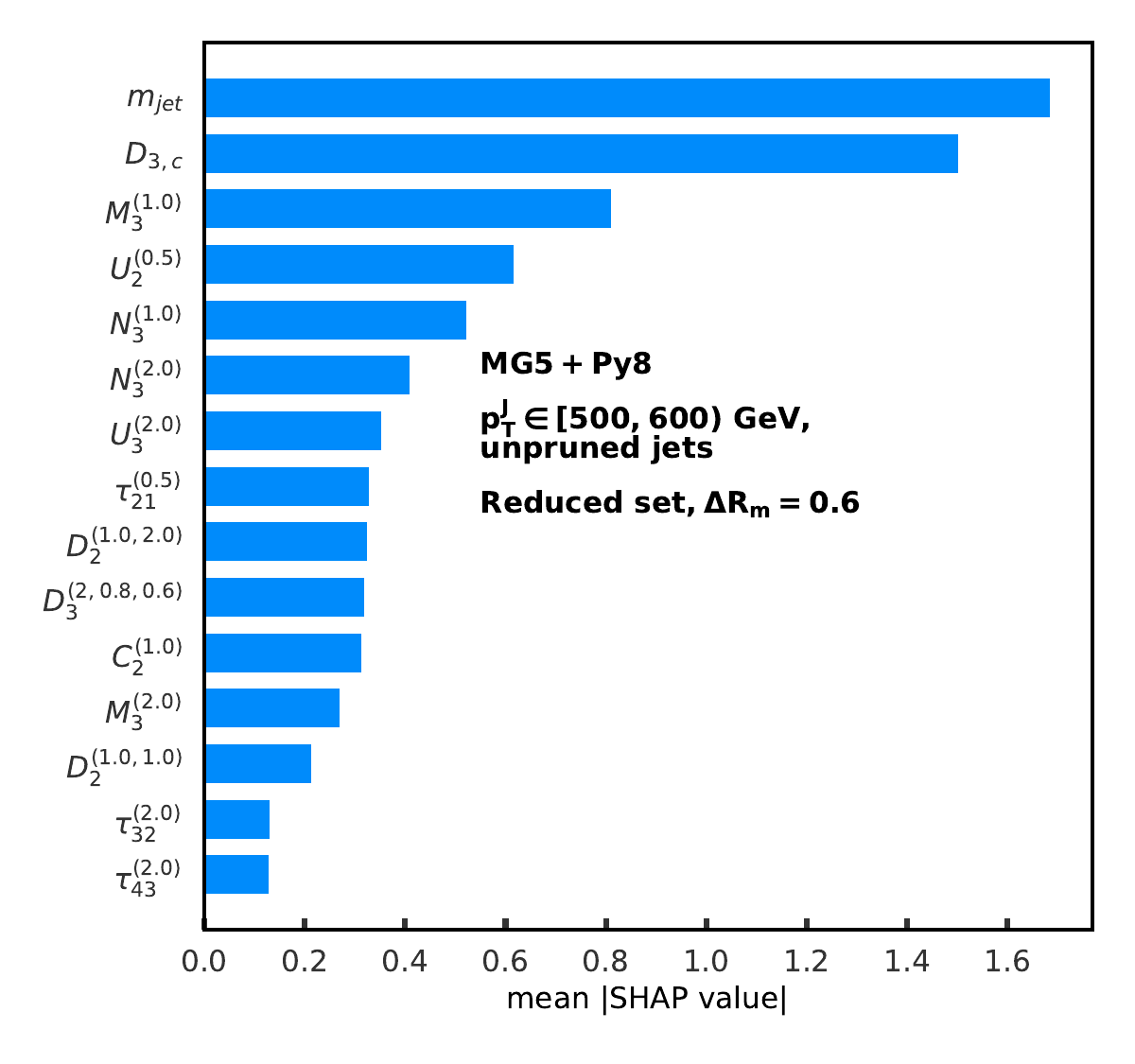}~
    \includegraphics[width=0.5\textwidth]{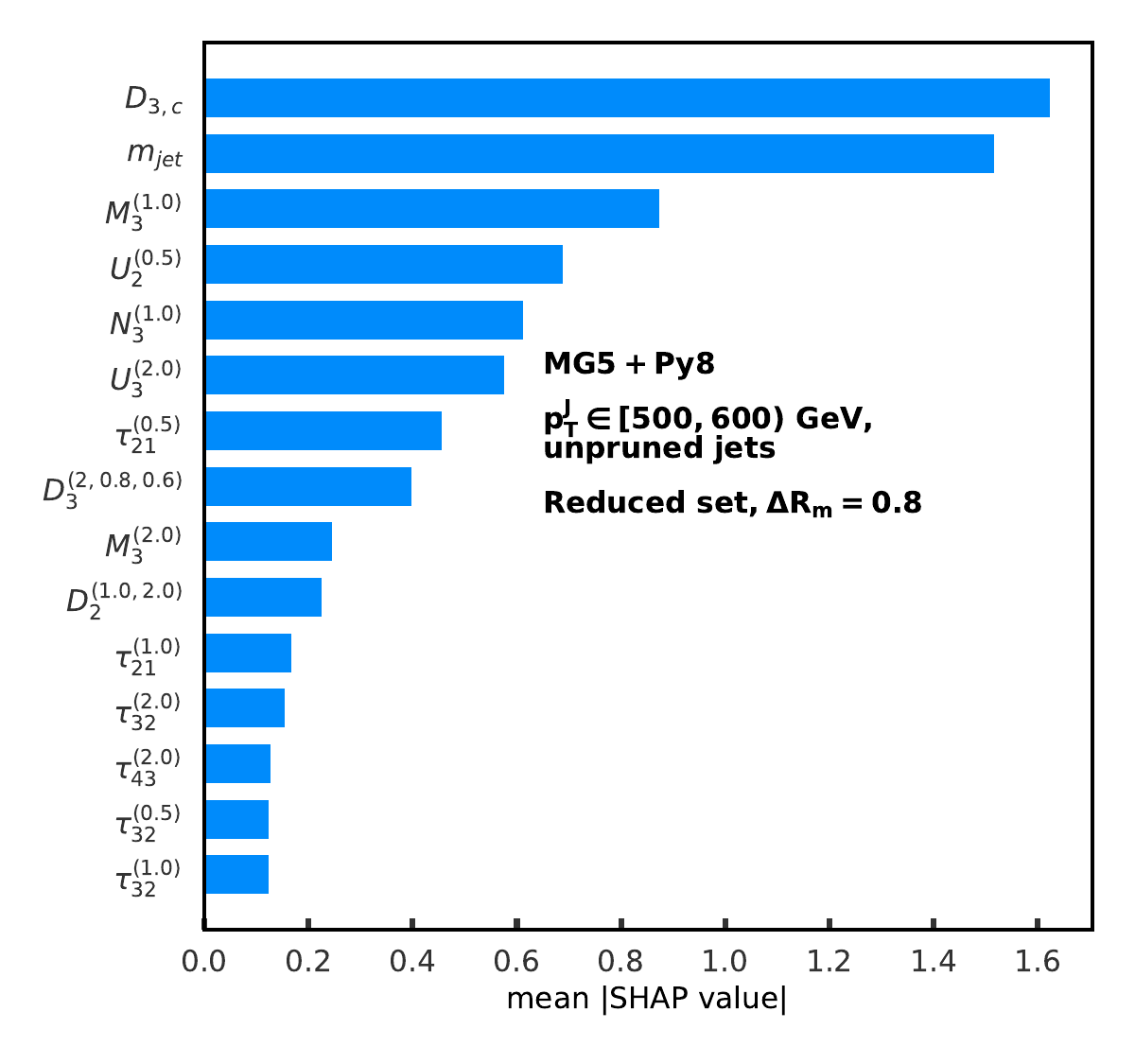}
    \includegraphics[width=0.5\textwidth]{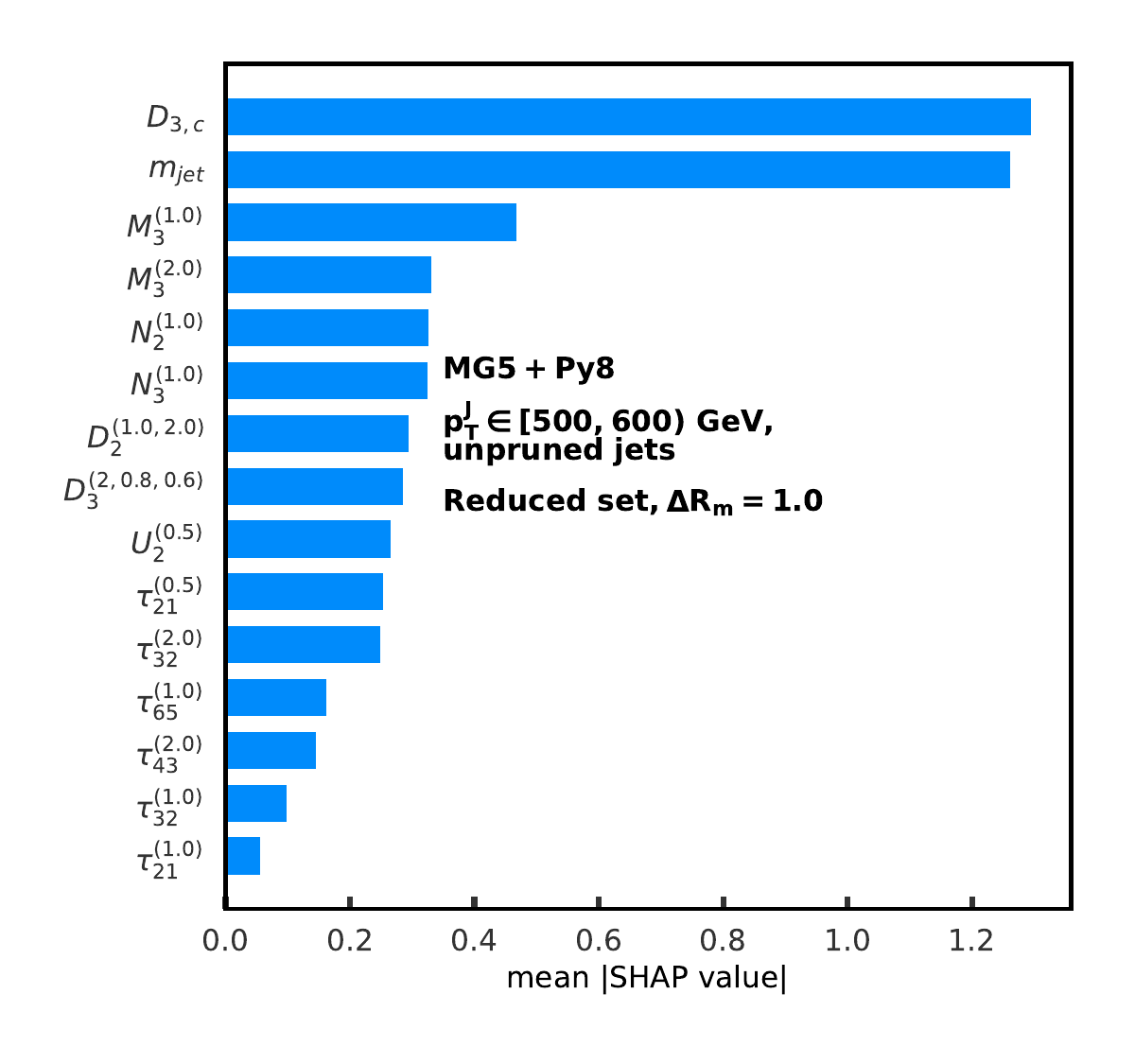}~
    \includegraphics[width=0.5\textwidth]{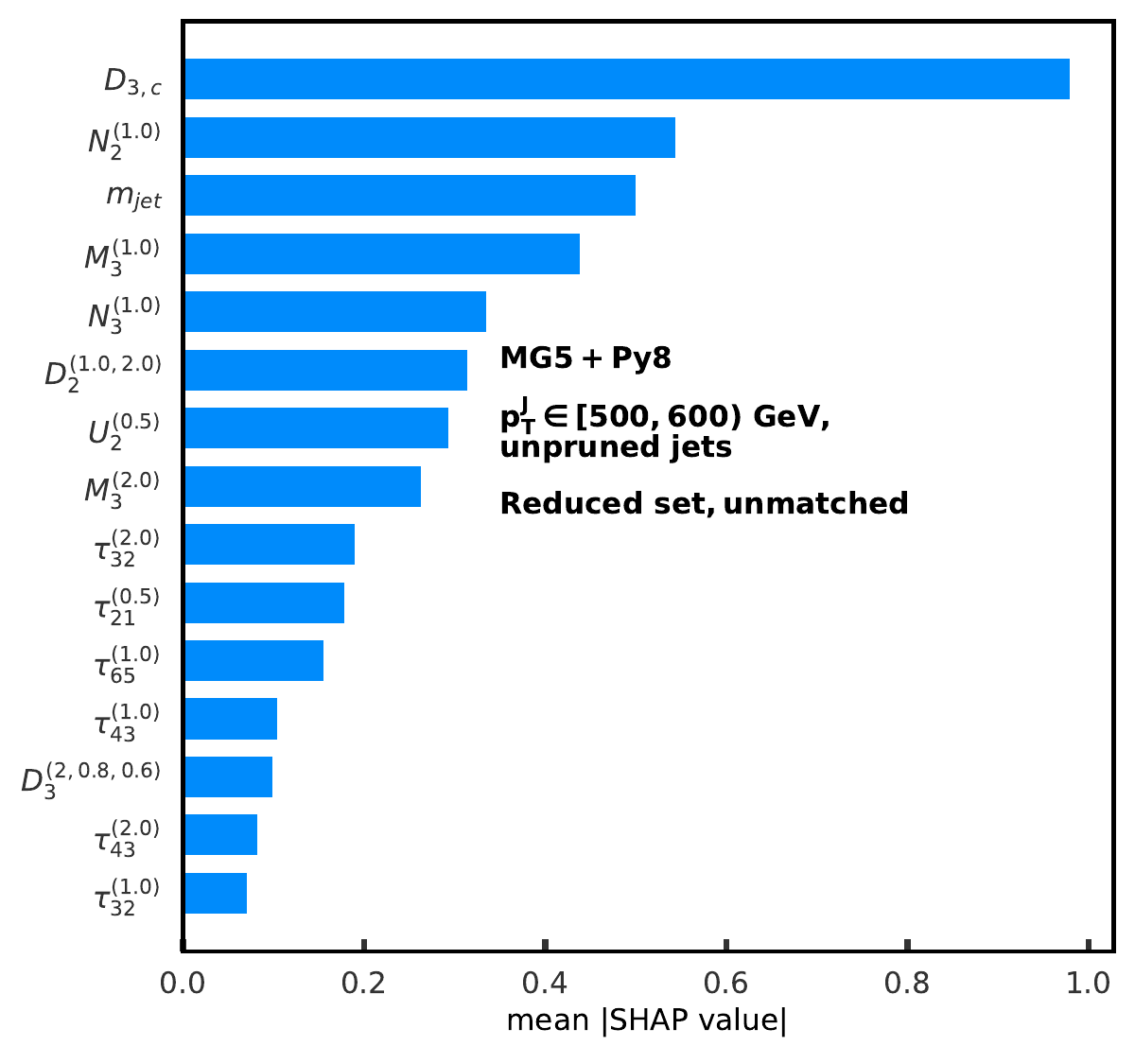}
    \caption{Bar plots depicting the hierarchy of feature importance for the hybrid tagger comprising the top 15 variables from all six sets combined. The dataset contains unpruned top and gluon jets with $\Delta R_m$ = 1.0 (\textit{left}) and unmatched (\textit{right}) top jets.}
    \label{fig:shap_bar_top15}
\end{figure}

These 15 features, in order of decreasing importance, are listed in Figure \ref{fig:shap_bar_top15}. We see that $D_{3,c}$, $m_{jet}$, and $M_3^{(1.0)}$ rank the highest in all the cases. For $\Delta R_m$ = 0.6 and 0.8, $U_2^{(0.5)}$ plays an important role, while for For $\Delta R_m$ = 1.0 and unmatched jets, $N_2^{1.0}$ is an important feature. Interestingly, the $N$-Subjettiness ratios had performed the best as a series compared to the ECF series but the different categories of ECF ratios combined together overshadow the contribution of $N$-Subjettiness observables. The top 15 features also include higher $N$-Subjettiness ratios like $\tau_{65}$ and $\tau_{43}$ with increasing $\Delta R_m$, although with low importance. At higher $\Delta R_m$, the higher $N$-Subjettiness ratios become relevant for capturing the complex substructure such as additional prong-like splittings within the jets.

The results of the taggers with the combined and the reduced set of features are listed in Table \ref{tab:results_p0_all}, along with the results of the separate taggers of Section \ref{sec:standard_top_taggers_and_xgb}. With this smaller, reduced set, we obtain a test accuracy of 0.909 and an AUC of 0.969, which is only 0.1$\%$ degradation from the combined set with $\Delta R_m = 1.0$ jets, while with unmatched jets, we obtain a test accuracy of 0.859 and an AUC of 0.932, which is 0.2$\%$ lower than the combined set results.

\begin{table}[hbt!]
\centering
\begin{tabular}{|c|cccccccc|}
\hline
\multirow{3}{*}{\begin{tabular}[c]{@{}c@{}}Taggers with\\ features before jet\\ pruning\end{tabular}} & \multicolumn{8}{c|}{$p_T^J\in$[500,600) GeV}                                                                                                                                                                            \\ \cline{2-9} 
                              & \multicolumn{2}{c|}{$\Delta R_m$=0.6}                                   & \multicolumn{2}{c|}{$\Delta R_m$=0.8}                                   & \multicolumn{2}{c|}{$\Delta R_m$=1.0}                                   & \multicolumn{2}{c|}{unmatched}        \\ \cline{2-9} 
                              & \multicolumn{1}{c|}{ACC}   & \multicolumn{1}{c|}{AUC} & \multicolumn{1}{c|}{ACC}   & \multicolumn{1}{c|}{AUC} & \multicolumn{1}{c|}{ACC}   & \multicolumn{1}{c|}{AUC} & \multicolumn{1}{c|}{ACC}   & AUC \\ \hline
$N$-Subjettiness        & \multicolumn{1}{c|}{0.929} & \multicolumn{1}{c|}{0.981}    & \multicolumn{1}{c|}{0.917} & \multicolumn{1}{c|}{0.974}    & \multicolumn{1}{c|}{0.905} & \multicolumn{1}{c|}{0.967}    & \multicolumn{1}{c|}{0.854} & 0.928    \\ \hline
C-Series                      & \multicolumn{1}{c|}{0.925} & \multicolumn{1}{c|}{0.978}    & \multicolumn{1}{c|}{0.910} & \multicolumn{1}{c|}{0.969}    & \multicolumn{1}{c|}{0.896} & \multicolumn{1}{c|}{0.961}    & \multicolumn{1}{c|}{0.842} & 0.917    \\ \hline
D-Series                      & \multicolumn{1}{c|}{0.928} & \multicolumn{1}{c|}{0.980}    & \multicolumn{1}{c|}{0.913} & \multicolumn{1}{c|}{0.972}    & \multicolumn{1}{c|}{0.899} & \multicolumn{1}{c|}{0.963}    & \multicolumn{1}{c|}{0.847} & 0.922    \\ \hline
U-Series                      & \multicolumn{1}{c|}{0.930} & \multicolumn{1}{c|}{0.981}    & \multicolumn{1}{c|}{0.916} & \multicolumn{1}{c|}{0.974}    & \multicolumn{1}{c|}{0.901} & \multicolumn{1}{c|}{0.964}    & \multicolumn{1}{c|}{0.851} & 0.926    \\ \hline
M-Series                      & \multicolumn{1}{c|}{0.910} & \multicolumn{1}{c|}{0.973}    & \multicolumn{1}{c|}{0.900} & \multicolumn{1}{c|}{0.967}    & \multicolumn{1}{c|}{0.886} & \multicolumn{1}{c|}{0.957}    & \multicolumn{1}{c|}{0.819} & 0.903    \\ \hline
N-Series                      & \multicolumn{1}{c|}{0.924} & \multicolumn{1}{c|}{0.978}    & \multicolumn{1}{c|}{0.906} & \multicolumn{1}{c|}{0.967}    & \multicolumn{1}{c|}{0.890} & \multicolumn{1}{c|}{0.956}    & \multicolumn{1}{c|}{0.839} & 0.915    \\ \hline
Combined set                  & \multicolumn{1}{c|}{\textbf{0.938}} & 
\multicolumn{1}{c|}{\textbf{0.985}}    & \multicolumn{1}{c|}{\textbf{0.924}} &
\multicolumn{1}{c|}{\textbf{0.978}}    & \multicolumn{1}{c|}{\textbf{0.910}} & \multicolumn{1}{c|}{\textbf{0.970}} & \multicolumn{1}{c|}{\textbf{0.860}}    & \textbf{0.934} \\ \hline
Reduced set                   & \multicolumn{1}{c|}{0.936} & 
\multicolumn{1}{c|}{0.984}    & \multicolumn{1}{c|}{0.923} &
\multicolumn{1}{c|}{0.977}    & \multicolumn{1}{c|}{0.909} & \multicolumn{1}{c|}{0.969}    & \multicolumn{1}{c|}{0.859} & 0.932 \\ \hline
\end{tabular}
\caption{Test Accuracy at 50$\%$ threshold probability score (ACC) and Area Under the Curve (AUC) for the \texttt{XGBOOST} based taggers using $N$-Subjettiness, C, D, U, M, N-Series of jet substructure observables (from Table \ref{tab:results_p0}), the combined set with 39 features, and the reduced set with 15 highest ranked features on unpruned jets with $\Delta R_m = ~0.6,~0.8,~1.0$ and unmatched matching criteria.}
\label{tab:results_p0_all}
\end{table}

As we observed in the previous section, $N$-Subjettiness consistently captures the essential features that distinguish top jets from gluon jets, making it a reliable standalone top tagger. The M-Series, on the other hand, falls short in comparison with all five sets. Interestingly, while the $N$-Subjettiness tagger does well on its own, the combined and reduced set taggers that integrate all the features not only outperform it, but the high ranking features contain $M_3$ instead of the $\tau$ ratios. This suggests that these ECFs, even the $M$ series ratios, collectively capture more detailed aspects of jet substructure, making $N$-Subjettiness ratios less crucial when used alongside them. 

In going from the combined set to the reduced set tagger, the performance does not degrade much. This implies that the reduced set tagger, which is trained on the best performing variables from the $N$-subjettiness ratios and the different ECFs, contains enough information to discriminate between top and gluon jets efficiently and robustly. This simple decision tree using the reduced set of features can form a reliable and effective tagger in different use cases. 

\begin{figure}[htb!]
    \centering
    \includegraphics[width=0.7\textwidth]{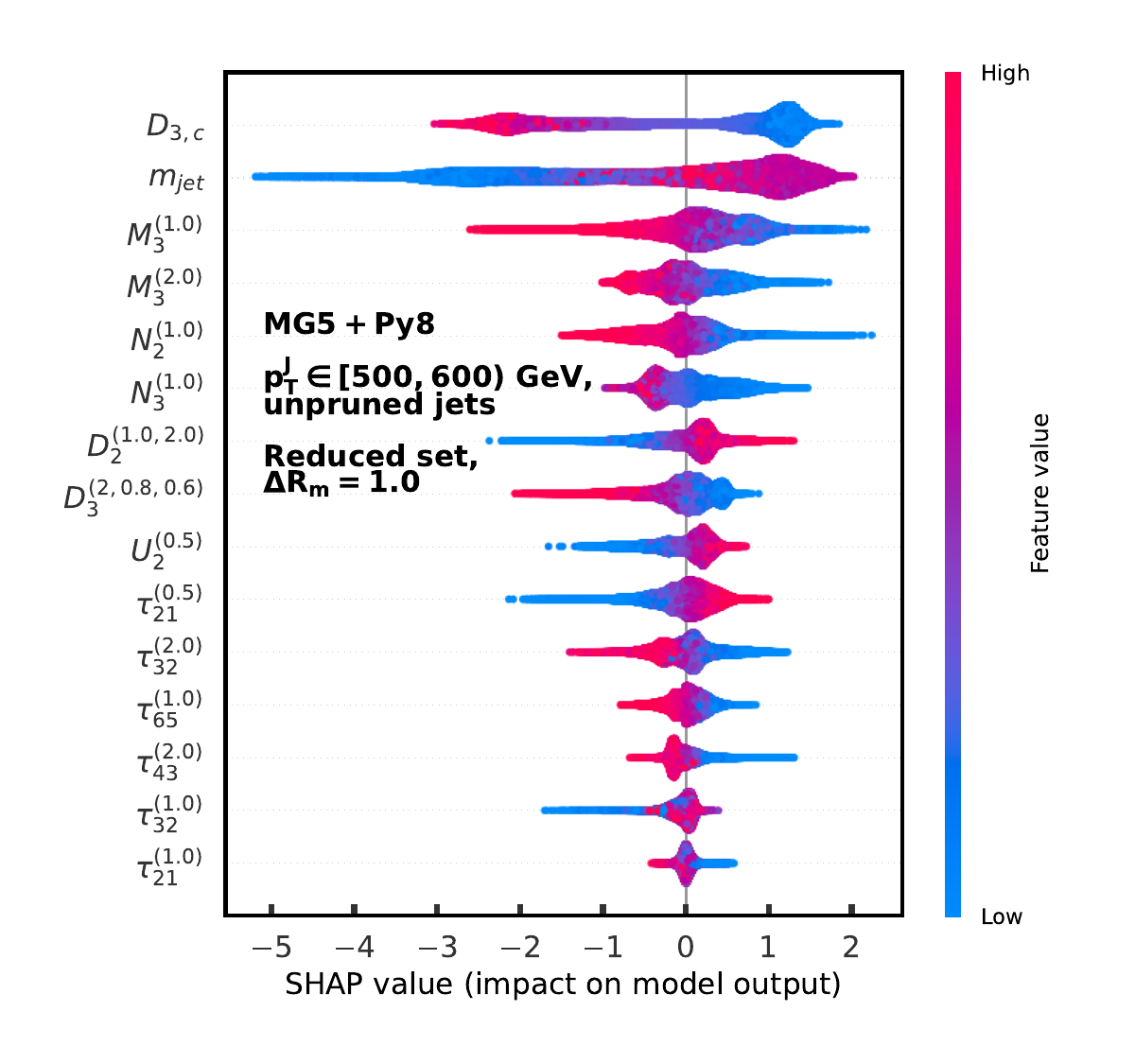}
    \caption{A summary plot showing the variation of SHAP values with the input feature values for the hybrid tagger consisting of the top 15 variables from all six sets combined. The dataset contains unpruned top and gluon jets with $\Delta R_m$ = 1.0 for top jets.}
    \label{fig:shap_bees}
\end{figure}

%SHAP_BEESWARM PLOT
Instead of looking at the feature importance as an average over all data points, we can plot each feature's SHAP value for individual data points. Figure \ref{fig:shap_bees} shows a summary plot. Each point on the plot represents an actual data point. The x-axis shows the SHAP value computed for each point and a particular feature on the y-axis. Positive SHAP values give the impact of a feature on classifying the data point as positive or 1 or, in our terms, a "top jet".
Similarly, negative SHAP values push the decision towards the negatively labeled, or 0, or "QCD jet" class. The higher the absolute value of SHAP, the more impact that variable has in classification. The colour gradient from blue to red indicates the variation in the feature's value from low to high. For example, in Figure \ref{fig:shap_bees}, the most important variable is $m_{jet}$. The high values of $m_{jet}$ push the decision towards "top-like," and the low values of $m_{jet}$ push the decision towards "QCD-like". The dense regions in each line show that most data points have similar SHAP values.

%DEPENDENCE PLOT
To understand how a feature value impacts the model prediction for each data point, we can construct a scatter plot between a feature value and the SHAP value. Figure \ref{fig:dependence_plot_mr10} shows such a dependence plot that shows the change in a feature's SHAP value with the feature's actual value. A dependence plot is a detailed picture of the beeswarm plot in Figure \ref{fig:shap_bees}. The variation of colour (red and blue) in each dependence plot shows the interaction of the feature with a second feature shown on the right. We can derive the following trends by looking at these plots:

\begin{figure}[htb!]
    \centering
    \includegraphics[width=0.45\textwidth]{./mr10_rank0_int.pdf}~
    \includegraphics[width=0.45\textwidth]{./mr10_rank1_int.pdf}\\
    \includegraphics[width=0.45\textwidth]{./mr10_rank2_int.pdf}~
    \includegraphics[width=0.45\textwidth]{./mr10_rank3_int.pdf}~
    \caption{Dependence plots of 5000 test events, of the four most important input features from Figure \ref{fig:shap_bar_top15}. The x-axis shows a variable's actual value, while the y-axis shows the SHAP values for each data point.  The dataset contains unpruned top and gluon jets with $\Delta R_m$ = 1.0 top jets.}
    \label{fig:dependence_plot_mr10}
\end{figure}

\begin{itemize}
    \item When $D_{3,c} \lesssim 1.5$, the events have a high SHAP value meaning they are predicted to be more top-like. These events have moderate to high $m_{jet}$ values while low $m_{jet}$ combined with high $D_{3,c}$ get a gluon-like prediction.
    \item SHAP value increases with an increase in $m_{jet}$, then reaches the highest $\sim 1$ near the top mass window and decreases for higher values of $m_{jet}$. Low values of $m_{jet}$ are also populated by events with high values of $M_3^{(2.0)}$. At higher values of jet mass, the points are more dispersed. Events with same $m_{jets}$ have different SHAP values ranging from -2 to +1 with increasing $M_3^{(2.0)}$.   
    \item SHAP values of $M_3^{(1.0)}$ increase from -2.0 to 1.5 with decreasing $M_3^{(1.0)}$. Points with $M_3^{(1.0)} \gtrsim 0.1 $ also have high values of $U_2^{(0.5)}$. Below $0.1$, $U_2^{(0.5)}$ varies from low to high, with higher values causing a rise in the SHAP value.
    \item Low $N_2^{(1.0)}$ result in positive SHAP values. When $N_2^{(1.0)} \gtrsim 0.35$, low $U_2^{(0.5)}$ pushes the prediction to be more QCD-like.
\end{itemize}

Figure \ref{fig:mjet_d3t3_mr10_500} shows the two-dimensional histograms of $m_{jet}$ and $D_{3,c}$ for both top and gluon jets. The top jets are mostly concentrated in a small region in the 2D plane, whereas the gluon jets have a wider spread along both axes. The reason $D_{3,c}$ works so well in distinguishing the two can be traced back to the fact that, combined with the jet mass, it provides a distinct boundary between them. This is also confirmed by the dependence plot in Figure \ref{fig:dependence_plot_mr10}.

\begin{figure}
    \centering
    \includegraphics[width=0.5\textwidth]{./mr10_mjet-d3t3-top.pdf}~
    \includegraphics[width=0.5\textwidth]{./mr10_mjet-d3t3-qcd.pdf}
    \caption{2-D histogram of $m_{jet}$ vs. $D_{3,c}$ for top jets \textit{(left)} and gluon jets \textit{(right)}. The red and the black ellipses highlight a simple boundary that selects and identifies top jets with  32$\%$ and 61$\%$ efficiency ($\epsilon_S$), respectively.}
    \label{fig:mjet_d3t3_mr10_500}
\end{figure}

This allows us to draw a two-dimensional boundary around the region populated by top jets and calculate the signal efficiency ($\epsilon_S$) and background rejection (1/$\epsilon_B$) for the two-dimensional cut. We highlight two simple ellipsoidal regions in Figure \ref{fig:mjet_d3t3_mr10_500} that correspond to 32$\%$ (red ellipse) and 61$\%$ (black ellipse) top selection efficiency. The background rejection for the two are 64 and 22, respectively. 

\begin{figure}[hbt!]
    \centering
    \includegraphics[width=0.48\textwidth]{./um_rank0_int.pdf}~
    \includegraphics[width=0.48\textwidth]{./um_rank1_int.pdf}\\
    \includegraphics[width=0.48\textwidth]{./um_rank2_int.pdf}~
    \includegraphics[width=0.48\textwidth]{./um_rank3_int.pdf}
    \caption{Dependence plots of 5000 test events, of the four most important input features. The x-axis shows a variable's actual value, while the y-axis shows the SHAP values for each data point.  The dataset contains unpruned top and gluon jets with no matching done for top jets.}
    \label{fig:dependence_plot_um}
\end{figure}

Figure \ref {fig:dependence_plot_um} shows the dependence plots for unmatched jets. When we remove the matching criterion, we observe that $m_{jet}$ has a prominent rise in SHAP value at $\lesssim$ 100 GeV, similar to the secondary peak at W mass ($\sim$ 80 GeV) in the jet mass distribution for unmatched top jets (see Figure \ref{fig:mjet_DRm}). The SHAP values in the W mass region turn positive for low $N_2^{(1.0)}$. Thus, if an unmatched top jet is partially reconstructed due to the loss of one hard subjet, it still gets tagged as top through a combination of features like $N_2^{(1.0)}$, which effectively captures the two-prong phase space. Similarly, $N_2^{(1.0)}$ takes positive SHAP values for low $D_2^{(1.0,2.0)}$. 

\subsection{Interaction effects in SHAP}
While computing SHAP values, the algorithm forms coalitions between the features. We can gain more knowledge about the data if we look into the pairwise interactions of features. Given two features $i, j$ out of N input features, SHAP can compute the interaction effect between the two using the Shapley interaction index \cite{DBLP:journals/corr/abs-1802-03888}.
\begin{equation}
    \Phi_{i,j} = \sum_{S\subseteq N\setminus \{i,j\}} \frac{|S|!(M-|S|-2)!}{2(M-1)!} \delta_{ij}(S),
\end{equation}
where $i\ne j$, and
\begin{equation}
    \delta_{ij}(S)=[f_x(S\cup \{i,j\})-f_x(S\cup \{i\})-f_x(S\cup \{j\})-f_x(S)]
\end{equation}
Thus, the interaction effect is calculated by subtracting the contribution of feature $i$ without $j$ present and feature $j$ without $i$ present from the contribution of the pair ${i,j}$ in the coalition. In other words, it can tell us how the effect of $i$ on the prediction might also be affected by $j$. The interaction effect is calculated as $\phi_{ij}=\phi_{ji}$.

\begin{figure}[hbt!]
    \centering
    \includegraphics[width=0.8\textwidth]{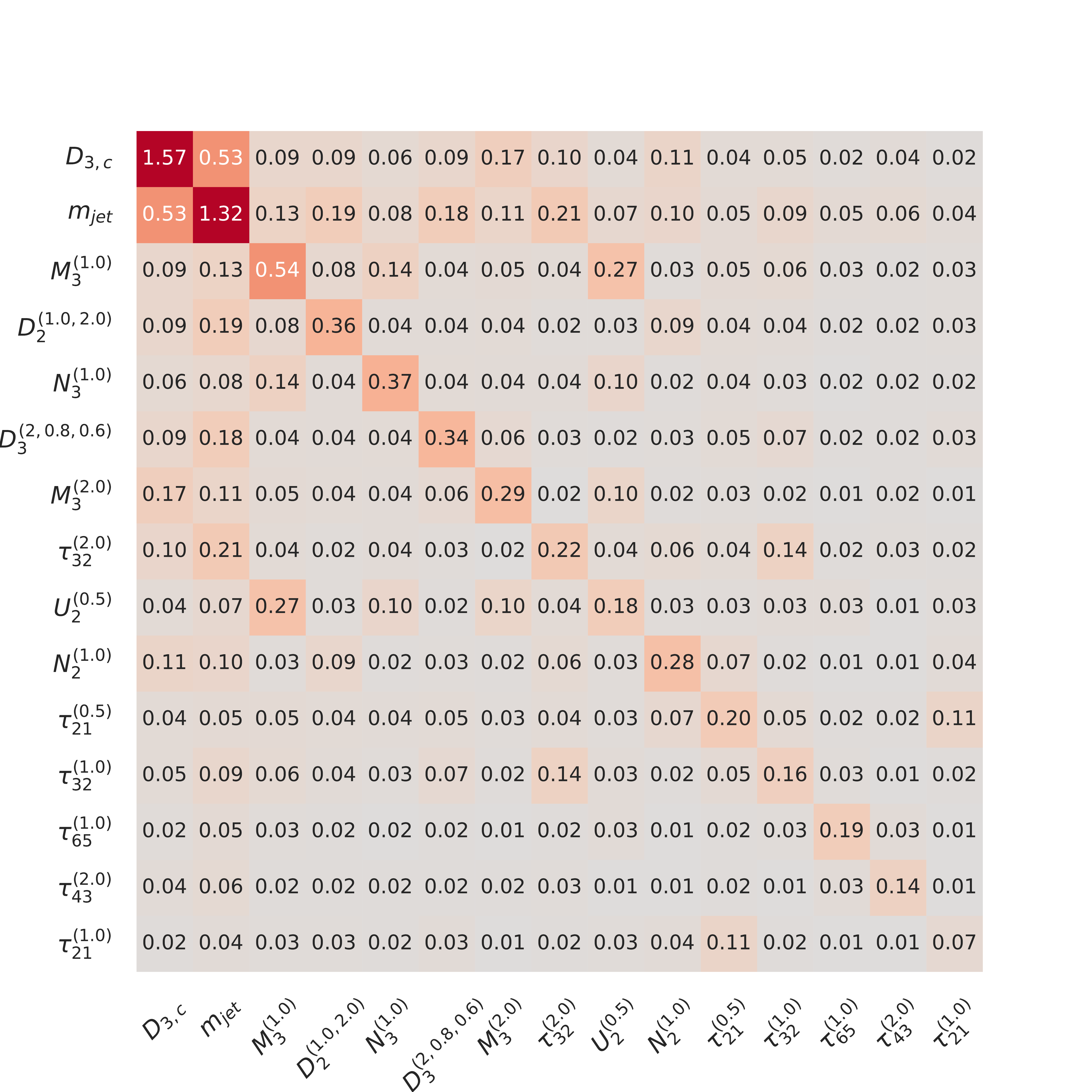}
    \caption{15 X 15 interaction matrix showing the interaction effects between features of the hybrid tagger. The dataset contains unpruned top and gluon jets with $\Delta R_m$ = 1.0 for top jets.}
    \label{fig:shap_bees_int}
\end{figure}

The main effect of a feature can be obtained if its interaction with all other features is subtracted from its total contribution (SHAP value).
\begin{equation}
    \Phi_{i,i}= \phi_i - \sum_{j\ne i} \Phi_{i,j}
\end{equation}
SHAP gives the interaction values for each data point as a matrix of N x N features. Figure \ref{fig:shap_bees_int} represents the interaction values between the features. 

\begin{figure}[hbt!]
    \centering
    \includegraphics[width=0.8\textwidth]{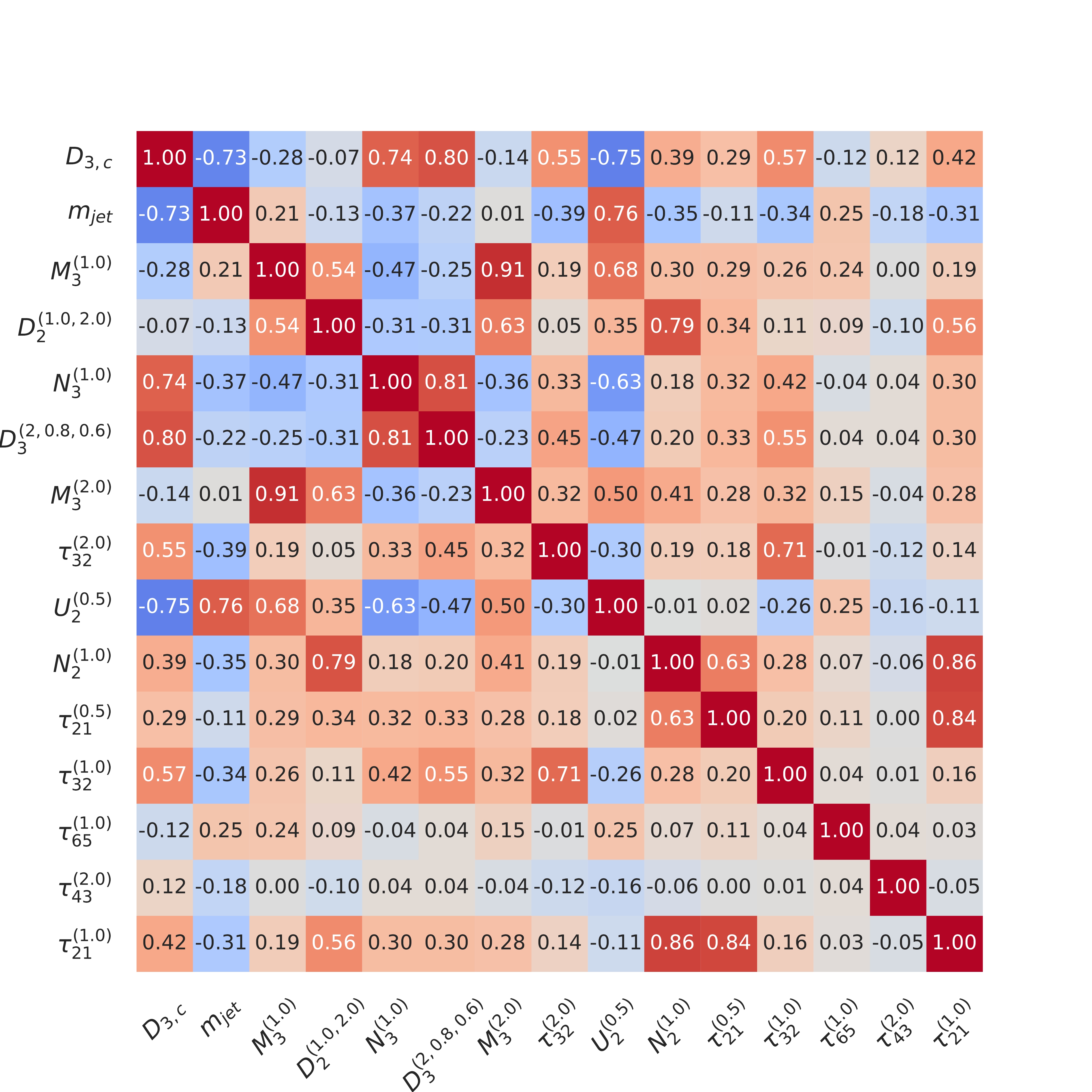}
    \caption{15 X 15 correlation matrix showing the correlation effects between features of the hybrid tagger. The dataset contains unpruned top and gluon jets with $\Delta R_m$ = 1.0 for top jets.}
    \label{fig:shap_bees_corr}
\end{figure}

The diagonal of the plot shows the mean of the absolute values of main effects, while the off-diagonal elements show the mean of the absolute values of interaction effects between two features. For example, $m_{jet}$ has the most interaction with $D_{3,c}$. Figure \ref{fig:mjet_d3t3_mr10_500} can also predict this effect. Interestingly, two variables having a high interaction need not be highly correlated. Two features may not be correlated but might populate different feature space regions for top and gluon jets. Comparing Figures \ref{fig:shap_bees_int} and \ref{fig:shap_bees_corr}, we see that $D_{3,c}$ and $D_3^{(2.0,0.8,0.6)}$ have an 80$\%$ correlation; however, they have a small interaction effect. The total SHAP value of a feature is the sum of the main and interaction effects.

\begin{figure}[htb!]
    \centering
    \includegraphics[width=0.5\textwidth]{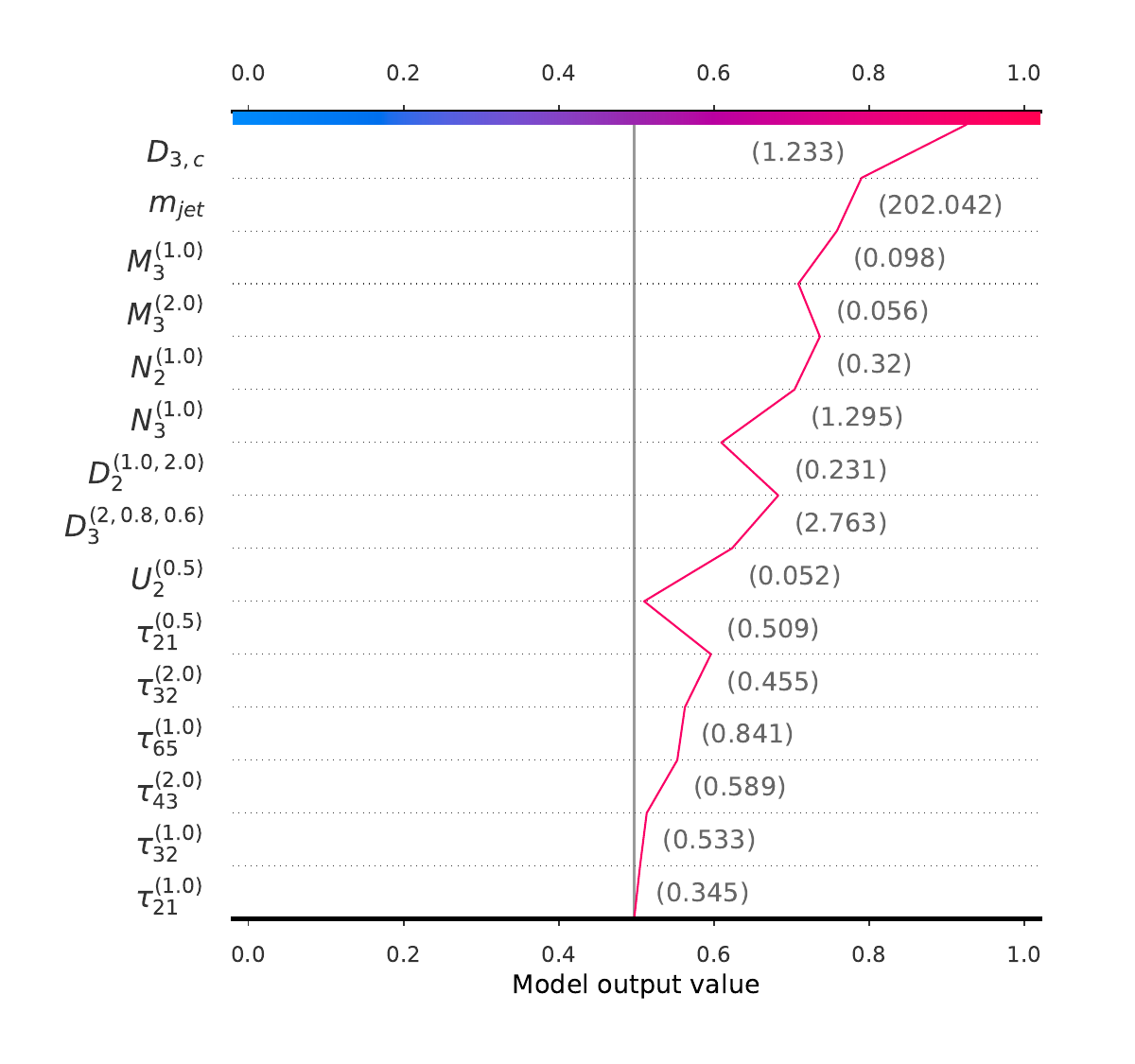}~
    \includegraphics[width=0.5\textwidth]{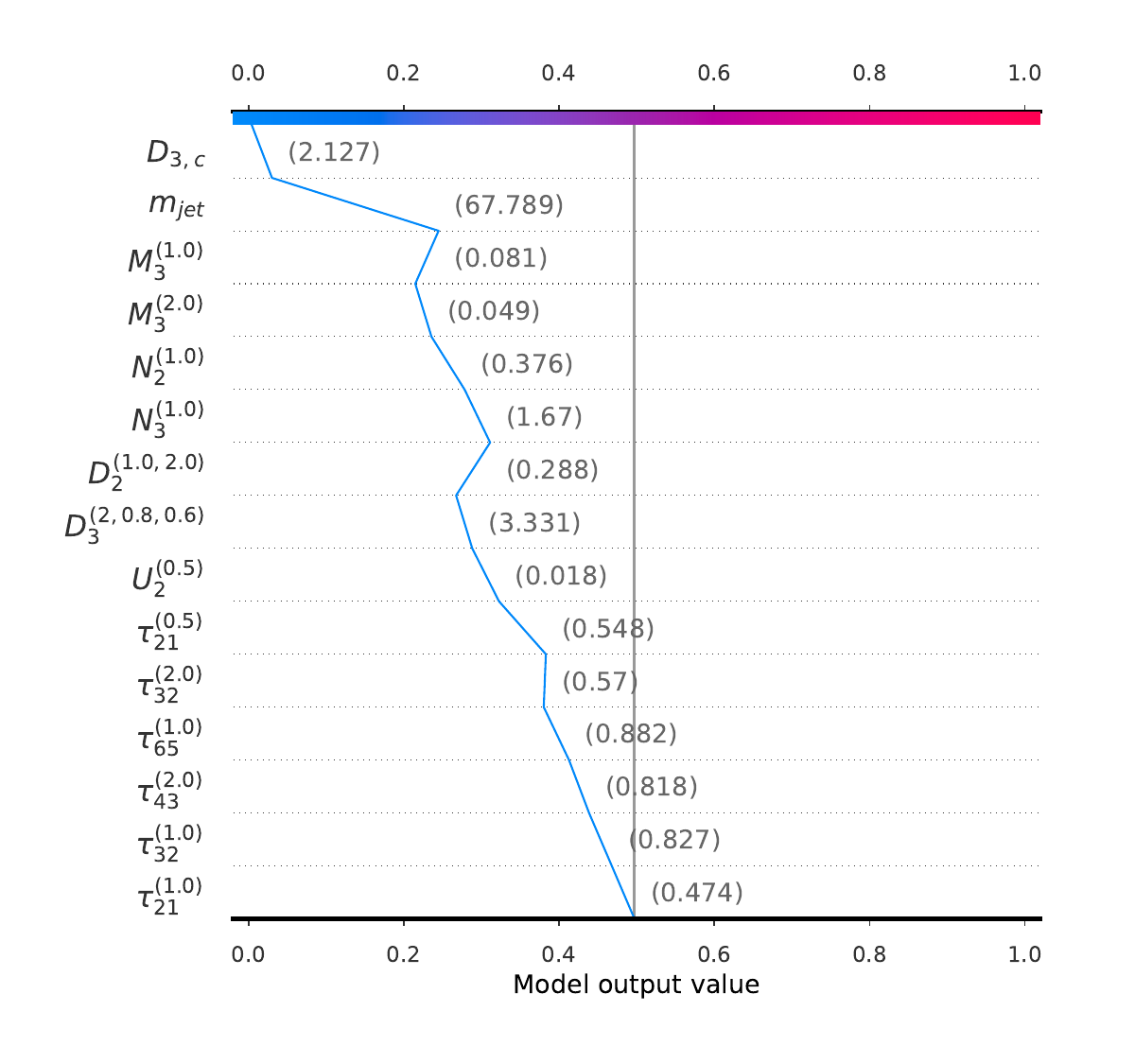}
    \caption{Decision plots of two correctly identified test events, one classified as top jet \textit{(left)} and the other as gluon jet \textit{(right)}. The number in brackets shows a variable's actual value, while the horizontal bar shows the probability score for that event. The line shows the change in the model's output with every feature.}
    \label{fig:dec_plot_um_py8}
\end{figure}

As stated earlier, SHAP is a local feature attribution method, meaning it can explain every event in the dataset. We can visualise a model's decision for every event through SHAP decision plots. Figure \ref{fig:dec_plot_um_py8} shows the decision plot for two different events, one classified as top jet and the other as gluon jet. The lines show how the model's decision is affected by every feature, moving through all the features.
The features in y-axis are ordered on the basis of importance. The numbers in brackets show the value of each feature for the specific event. The vertical line shows the base value, which is the average prediction of the model before training. In this case, the base value is 0.5 since we have an equal number of top and gluon jets. The SHAP value has been converted to probability, displayed in the horizontal colour bar on the top. The deviation of the red and the blue lines from the base value at each step represents the contribution of the feature to the model's output. Larger the distance, greater is the feature's impact on the model's decision.

\begin{figure}[htb!]
    \centering
    \includegraphics[width=0.5\textwidth]{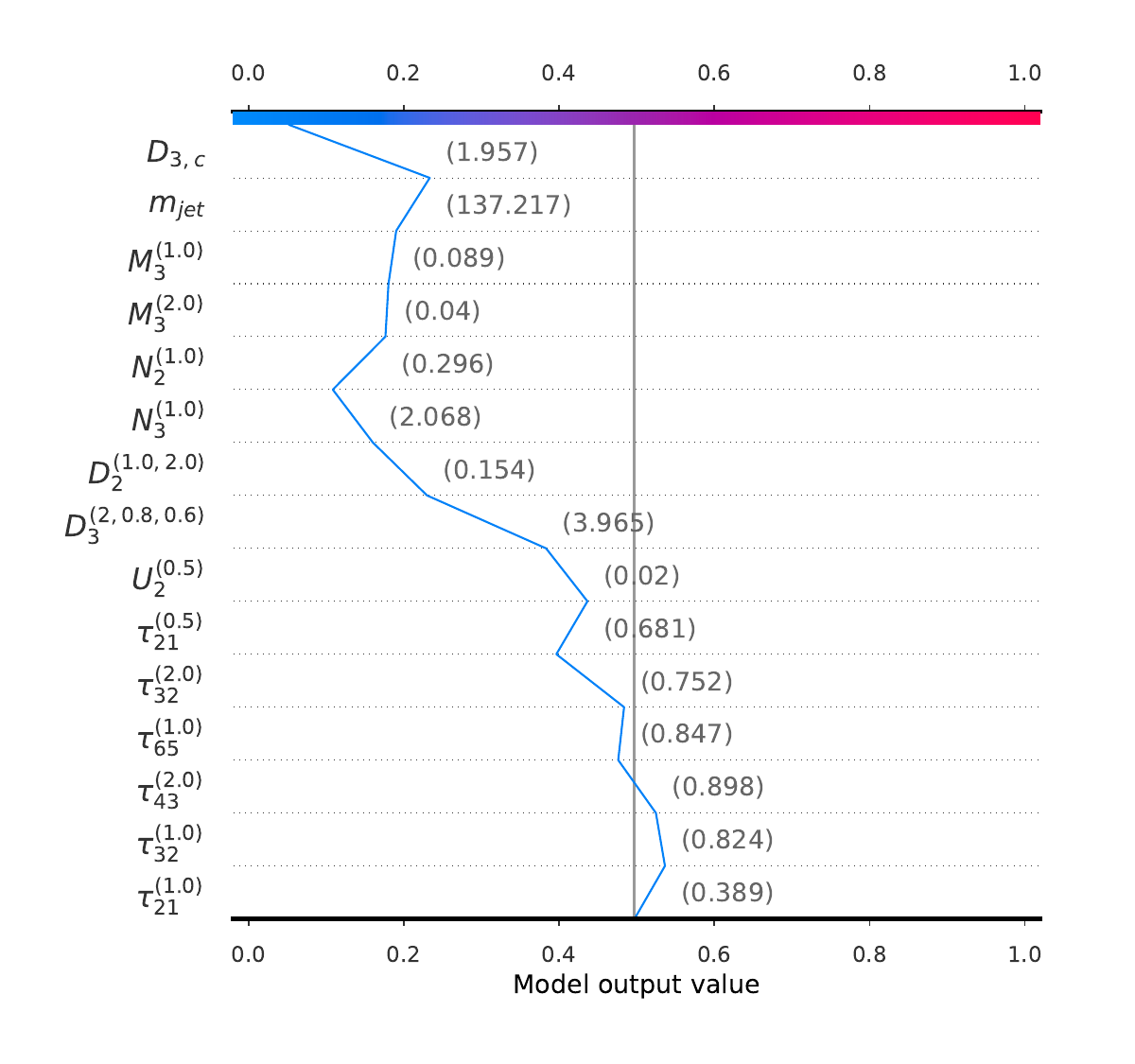}~
    \includegraphics[width=0.5\textwidth]{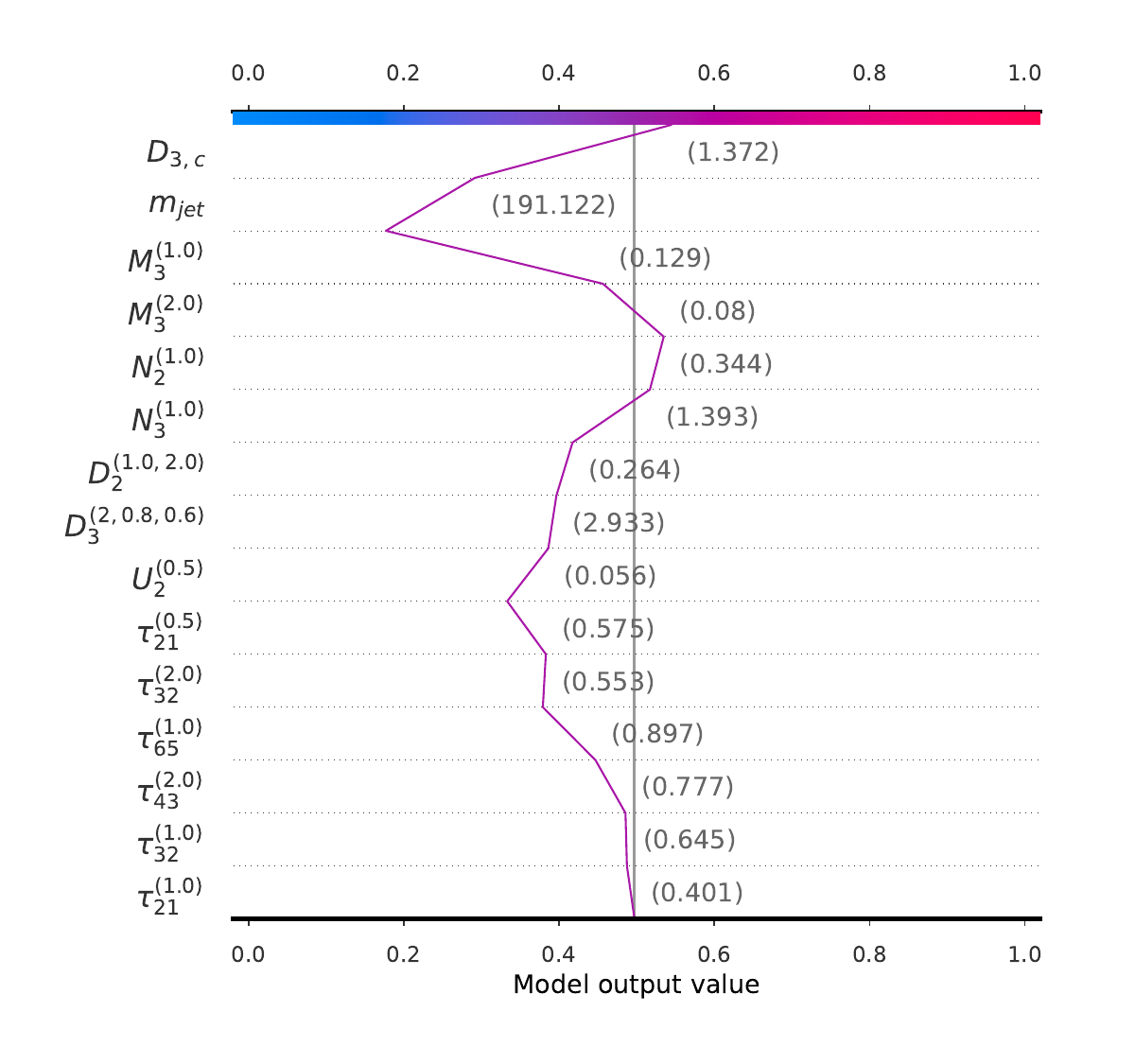}
    \caption{Decision plots of two test events, a top jet misclassified as gluon jet \textit{(left)} and a gluon jet mistagged as top jet \textit{(right)}. The number in brackets shows a variable's actual value, while the horizontal bar shows the probability score for that event. The line shows the change in the model's output with every feature.}
    \label{fig:dec_plot_um_mc_py8}
\end{figure}

Figure \ref{fig:dec_plot_um_mc_py8} presents the decision plots for two test events that were misclassified into the wrong category. We can see the pathway that the model takes to arrive at the decision. By focusing on the most influential feature, $D_{3,c}$, we observe that for the misclassified events, $D_{3,c}$ had an anomalously high value for a top jet and a low value for a gluon jet, contrary to the typical values expected for $D_{3,c}$ in top and gluon jets. This leads the model to make incorrect predictions.

\section{\texttt{XGBOOST} analysis with jet substructure observables for classification of top jets vs. quark jets}
\label{sec:xgb_quark}
Till now, we have performed \texttt{XGBOOST} classification to identify top jets from gluon jets. We have discovered that the performance of the taggers depends on the initial parton-level matching. The different matching conditions not only give rise to varying degrees of performance of the model, but the difference is also highlighted in the Shapley value based interpretation of the model. We have identified 15 most important features for all the sets and built a top tagger with the reduced set. We now replace gluon jets with quark jets and see how the performance and Shapley value interpretation of the reduced set tagger changes.

As mentioned in Section \ref{ssec:simulation} we generate quark jets using the process $pp\rightarrow Zq$ where Z decays invisibly and q, comprising u or d quarks, gives rise to quark jets. We construct the $N$-Subjettiness and Energy Correlation ratios on these jets. The unmatched top jets and quark jets are then used to train the reduced set taggers. The performance of this model is then tested on two datasets: top jets and quark jets, and top jets and gluon jets. Conversely, the reduced set tagger that we had built previously with gluon jets as background is tested on the aforementioned datasets. The purpose of this exercise is to check if the models trained on one category of QCD background jets could perform equally well while discriminating top jets from another category of QCD background. 

% \usepackage{multirow}
\begin{table}[htb!]
\centering
\begin{tabular}{|c|cc|}
\hline
\multirow{2}{*}{Train data} & \multicolumn{2}{c|}{Test data} \\ \cline{2-3} 
 & \multicolumn{1}{c|}{Top jets vs. Gluon jets} & Top jets vs. Quark jets \\ \hline
Top jets vs. Gluon jets & \multicolumn{1}{c|}{0.932} & 0.928 \\ \hline
Top jets vs. Quark jets & \multicolumn{1}{c|}{0.916} & 0.944 \\ \hline
\end{tabular}
\caption{AUC of XGBOOST model with the reduced set of features when training and testing data contain gluon jets versus light quark jets. The data samples contain unpruned and unmatched jets}
\label{tab:quark_auc}
\end{table}

The performance of the reduced set tagger is listed in Table \ref{tab:quark_auc}. The tagger that trains and tests on top and quark jets performs much better compared to the top vs. gluon jets taggers. It is more difficult to distinguish our signal from the background when the background consists of gluon-initiated jets. Moreover, we see that the models trained on gluon jets as background can yield similar performance, with a 0.4\% variation, when tested on quark jets which it has never seen before. However, the reverse is not true. This shows that the complex characteristics of gluon jets make them a sufficient candidate for signal vs. background classification as the model learns those characteristics well enough to be able to generalise over quark-initiated jets as well. We perform the rest of the study using gluon jets as background.

\begin{figure}[htb!]
    \centering
    \includegraphics[width=0.7\textwidth]{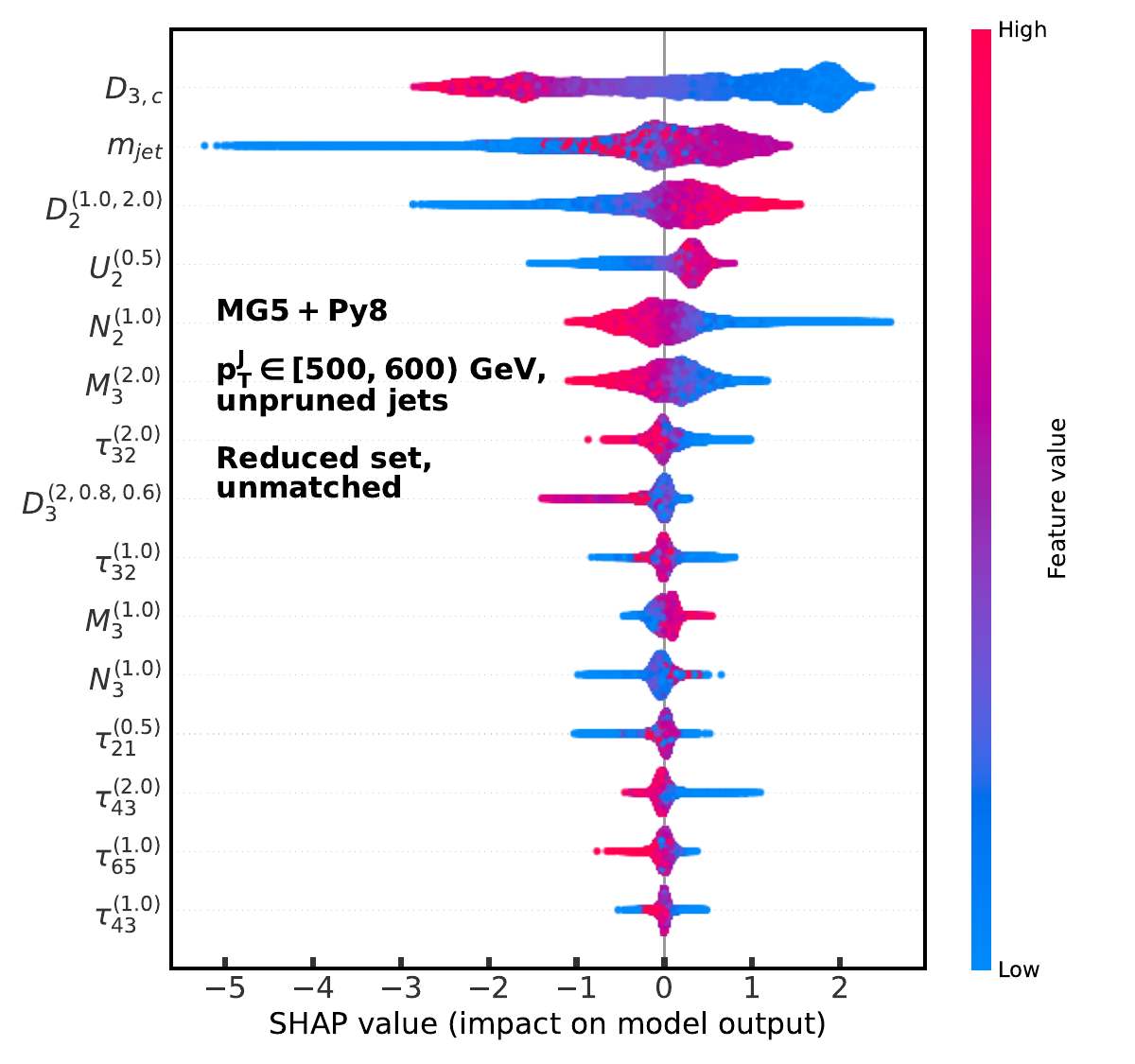}
    \caption{Summary plot depicting the hierarchy of feature importance for the hybrid tagger comprising the top 15 variables from all six sets combined. The dataset contains unpruned top and quark jets with unmatched (\textit{right}) top jets.}
    \label{fig:shap_bees_q}
\end{figure}

Now we explore further by plotting the Shapley values that the model assigns to the features in the case of top jets vs. quark jets classification. Figure \ref{fig:shap_bees_q}  shows the summary plot of the same. Comparing with Figure \ref{fig:shap_bar_top15}, we see that in addition to $D_{3,c}$ and $m_{jet}$, the features $D_2^{(1.0,2.0)}$, $U_2^{0.5}$, and $N_2^{1.0}$ also play significant roles. The two-point correlators are particularly effective at distinguishing light quark jets from top jets, followed by the three-prong identifiers.

\section{Generator dependence: PYTHIA vs. HERWIG}
\label{sec:herwig}
We have performed the showering and hadronisation of the top quark decay products using the \texttt{PYTHIA 8} generator.
To see if the output of our model has a generator dependence, we generate 1 million top jets and 1 million gluon jets using another showering and hadronisation generator, \texttt{HERWIG 7} \cite{Bahr:2008pv, Bellm:2015jjp}. We use $R_m$ = 1.0 matched top jets and unmatched top jets for the study. We then proceed to study the effects of pruning the jets generated by the two different shower algorithms. 

As stated previously in Section \ref{ssec:hybrid}, we calculate the correlation between all the variables and remove the ones with more than 95$\%$ correlation. In the case of \texttt{HERWIG 7} as well, the highly correlated variables are $C_1^{(1.0)},~C_1^{(2.0)}, ~U_1^{(0.5)}, ~U_3^{(0.5)}, ~U_1^{(1.0)}, ~U_2^{(1.0)}, ~U_3^{(1.0)}$, and $U_1^{(2.0)}$. We provide the combined set of 39 features as input, then select the 15 most important features given by SHAP as the reduced set. Figure \ref{fig:mjet_hw7} shows the $m_{jet}$ and $D_{3,c}$ distributions for jets generated with \texttt{PYTHIA 8} and \texttt{HERWIG 7}.  

\begin{figure}[htb!]
    \centering
    \includegraphics[width=0.5\textwidth]{./hw7_mr10_mjet_mr10.pdf}~
    \includegraphics[width=0.5\textwidth]{./hw7_um_mjet_um.pdf}\\
    \includegraphics[width=0.5\textwidth]{./hw7_mr10_d3_term3_mr10.pdf}~
    \includegraphics[width=0.5\textwidth]{./hw7_um_d3_term3_um.pdf}
    \caption{Normalised $m_{jet}$ distribution (\textit{top}) and normalised $D_{3,c}$ distribution (\textit{bottom}) of top jets with $\Delta R_m$ = 1.0 (\textit{left}) and unmatched top jets \textit{(right}), and gluon jets, generated by two different shower algorithms \texttt{PYTHIA 8} and \texttt{HERWIG 7}.}
    \label{fig:mjet_hw7}
\end{figure}

We train and test our decision tree with this reduced set of features using the same architecture as before; only this time, we also perform cross-generator training and testing. Table \ref{tab:py8_hw7_auc} shows the AUC for the cross-training and testing. The performance degrades when test events are generated with \texttt{HERWIG 7}, compared to when test events are generated with \texttt{PYTHIA 8}. When the model is trained with \texttt{PYTHIA 8} and tested with \texttt{HERWIG 7} the AUC lowers by 0.9 - 1.0 $\%$ but improves by 0.5 - 0.6$\%$ in the case where trained by \texttt{HERWIG 7} but tested by \texttt{PYTHIA 8}. 

\begin{table}[htb!]
\centering
\begin{tabular}{|cccccc|}
\hline
\multicolumn{6}{|c|}{Combined set tagger} \\ \hline
\multicolumn{3}{|c|}{$\Delta R_m$ = 1.0} & \multicolumn{3}{c|}{unmatched} \\ \hline
\multicolumn{1}{|c|}{\multirow{2}{*}{Train data}} & \multicolumn{2}{c|}{Test data} & \multicolumn{1}{c|}{\multirow{2}{*}{Train data}} & \multicolumn{2}{c|}{Test data} \\ \cline{2-3} \cline{5-6} 
\multicolumn{1}{|c|}{} & \multicolumn{1}{c|}{PYTHIA 8} & \multicolumn{1}{c|}{HERWIG 7} & \multicolumn{1}{c|}{} & \multicolumn{1}{c|}{PYTHIA 8} & HERWIG 7 \\ \hline
\multicolumn{1}{|c|}{PYTHIA 8} & \multicolumn{1}{c|}{0.934} & \multicolumn{1}{c|}{0.924} & \multicolumn{1}{c|}{PYTHIA 8} & \multicolumn{1}{c|}{0.970} & 0.960 \\ \hline
\multicolumn{1}{|c|}{HERWIG 7} & \multicolumn{1}{c|}{0.931} & \multicolumn{1}{c|}{0.926} & \multicolumn{1}{c|}{HERWIG 7} & \multicolumn{1}{c|}{0.969} & 0.961 \\ \hline
\multicolumn{6}{|c|}{Reduced set tagger} \\ \hline
\multicolumn{3}{|c|}{$\Delta R_m$ = 1.0} & \multicolumn{3}{c|}{unmatched} \\ \hline
\multicolumn{1}{|c|}{\multirow{2}{*}{Train data}} & \multicolumn{2}{c|}{Test data} & \multicolumn{1}{c|}{\multirow{2}{*}{Train data}} & \multicolumn{2}{c|}{Test data} \\ \cline{2-3} \cline{5-6} 
\multicolumn{1}{|c|}{} & \multicolumn{1}{c|}{PYTHIA 8} & \multicolumn{1}{c|}{HERWIG 7} & \multicolumn{1}{c|}{} & \multicolumn{1}{c|}{PYTHIA 8} & HERWIG 7 \\ \hline
\multicolumn{1}{|c|}{PYTHIA 8} & \multicolumn{1}{c|}{0.932} & \multicolumn{1}{c|}{0.923} & \multicolumn{1}{c|}{PYTHIA 8} & \multicolumn{1}{c|}{0.969} & 0.959 \\ \hline
\multicolumn{1}{|c|}{HERWIG 7} & \multicolumn{1}{c|}{0.930} & \multicolumn{1}{c|}{0.924} & \multicolumn{1}{c|}{HERWIG 7} & \multicolumn{1}{c|}{0.968} & 0.960 \\ \hline
\end{tabular}
\caption{AUC of \texttt{XGBOOST} models when training and testing data are generated with \texttt{PYTHIA 8} and \texttt{HERWIG 7}. The data samples contain unpruned $\Delta R_m$ = 1.0 matched and unmatched jets.}
\label{tab:py8_hw7_auc}
\end{table}

From Figure \ref{fig:mjet_hw7}, we see that the there is a difference in the distribution of gluon jets between the two generators. This causes a greater overlap between the distribution of top jets and gluon jets in the case of \texttt{HERWIG 7} than \texttt{PYTHIA 8}. This is also reflected in the $W_1(T,G)$ metric for both the generators. In the case of $m_{jet}$, going from \texttt{PYTHIA 8} to \texttt{HERWIG 7}, there a 14$\%$ decrease in $W_1(T, G)$ for $\Delta R_m$ = 1.0 jets, and a 17$\%$ decrease in $W_1(T, G)$ for unmatched jets. This can be due to the difference in parton shower modeling in the two shower algorithms. While \texttt{PYTHIA 8} uses $p_T$ ordering for shower and the Lund string model for hadronisation \cite{Andersson:1983jt}, \texttt{HERWIG 7} uses angular-ordering for shower and the cluster hadronisation model for hadronisation \cite{AMATI197987, WEBBER1984492}. This can lead to differences in how the radiation is distributed within jets, affecting the performance of jet substructure observables and also the ML models using those observables. Monte Carlo studies \cite{Larkoski:2013eya, Bhattacherjee:2015psa} have reported similar findings.  

The reduced sets for $\Delta R_m$ = 1.0 matched and unmatched jets generated with \texttt{HERWIG 7} share many features with the reduced set taggers for \texttt{PYTHIA 8} generated $\Delta R_m$ = 1.0 matched and unmatched jets. Table \ref{tab:best15_HW7} lists the common features between the two generators and features unique to each set. Specifically, variables $C_2^{2.0)}$ and $\tau_{21}^{(2.0)}$ replace $\tau_{43}^{(1.0)}$ and $\tau_{32}^{(1.0)}$ in the 15 highest ranked features for unmatched jets. Variables $D_2^{2.0,2.0)}$ and $\tau_{21}^{(2.0)}$ replace $N_2^{(1.0)}$ and $\tau_{21}^{(1.0)}$ in the 15 highest ranked features for $\Delta R_m$ = 1.0 matched jets. 

% \usepackage{multirow}
\begin{table}[hbt!]
\centering
\begin{tabular}{|ccccc|}
\hline
\multicolumn{5}{|c|}{$p_T^J\in[500,600)$ GeV} \\ \hline
\multicolumn{5}{|c|}{Features in the reduced set taggers} \\ \hline
\multicolumn{1}{|c|}{12 common features} & \multicolumn{4}{c|}{\begin{tabular}[c]{@{}c@{}}$m_{jet}$, $\tau_{21}^{(0.5)}$, $\tau_{32}^{(2.0)}$, $\tau_{43}^{(2.0)}$, $D_2^{(1.0,2.0)}$,$D_{3,c}$,\\  $D_3^{(2,0.8,0.6)}$, $N_3^{(1.0)}$, $M_3^{(1.0)}$, $M_3^{(2.0)}$, $U_2^{(0.5)}$, $\tau_{65}^{(1.0)}$\end{tabular}} \\ \hline
\multicolumn{1}{|c|}{\multirow{2}{*}{Features unique to each set}} & \multicolumn{2}{c|}{$\Delta R_m$ = 1.0} & \multicolumn{2}{c|}{unmatched} \\ \cline{2-5} 
\multicolumn{1}{|c|}{} & \texttt{HERWIG 7} & \multicolumn{1}{c|}{\texttt{PYTHIA 8}} & \texttt{HERWIG 7} & \texttt{PYTHIA 8} \\ \hline
\multicolumn{1}{|c|}{$\tau_{21}^{(2.0)}$} & $\times$ & \multicolumn{1}{c|}{$\checkmark$} & $\times$ & $\times$ \\
\multicolumn{1}{|c|}{$\tau_{21}^{(2.0)}$} & $\checkmark$ & \multicolumn{1}{c|}{$\times$} & $\checkmark$ & $\times$ \\
\multicolumn{1}{|c|}{$\tau_{32}^{(1.0)}$} & $\checkmark$ & \multicolumn{1}{c|}{$\checkmark$} & $\times$ & $\checkmark$ \\
\multicolumn{1}{|c|}{$\tau_{43}^{(1.0)}$} & $\times$ & \multicolumn{1}{c|}{$\times$} & $\times$ & $\checkmark$ \\
\multicolumn{1}{|c|}{$C_2^{(2.0)}$} & $\times$ & \multicolumn{1}{c|}{$\times$} & $\checkmark$ & $\times$ \\
\multicolumn{1}{|c|}{$D_2^{(2.0, 2.0)}$} & $\checkmark$ & \multicolumn{1}{c|}{$\times$} & $\times$ & $\times$ \\
\multicolumn{1}{|c|}{$N_2^{(1.0)}$} & $\times$ & \multicolumn{1}{c|}{$\checkmark$} & $\checkmark$ & $\checkmark$ \\ \hline
\end{tabular}
\caption{Features appearing in the reduced set with 15 highest ranked features for unpruned jets with $\Delta R_m = ~1.0$ and unmatched matching criteria for \texttt{HERWIG 7} generated and \texttt{PYTHIA 8} generated jets.} 
\label{tab:best15_HW7}
\end{table}

\begin{figure}[htb!]
    \centering
    \includegraphics[width=0.45\textwidth]{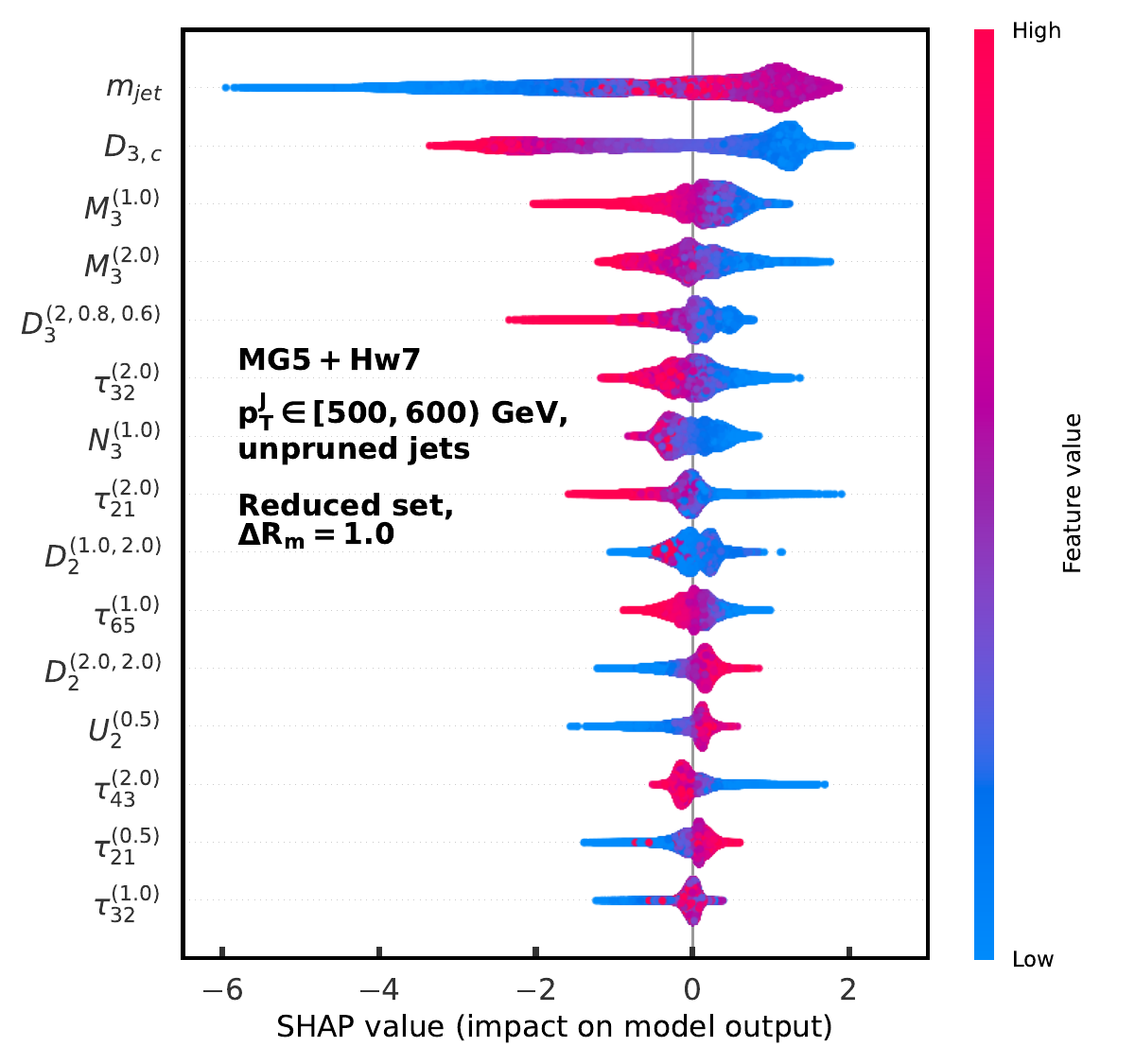}~
    \includegraphics[width=0.45\textwidth]{./mr10_shap_bees_15_mr10_test_py8.pdf}\\
    \includegraphics[width=0.45\textwidth]{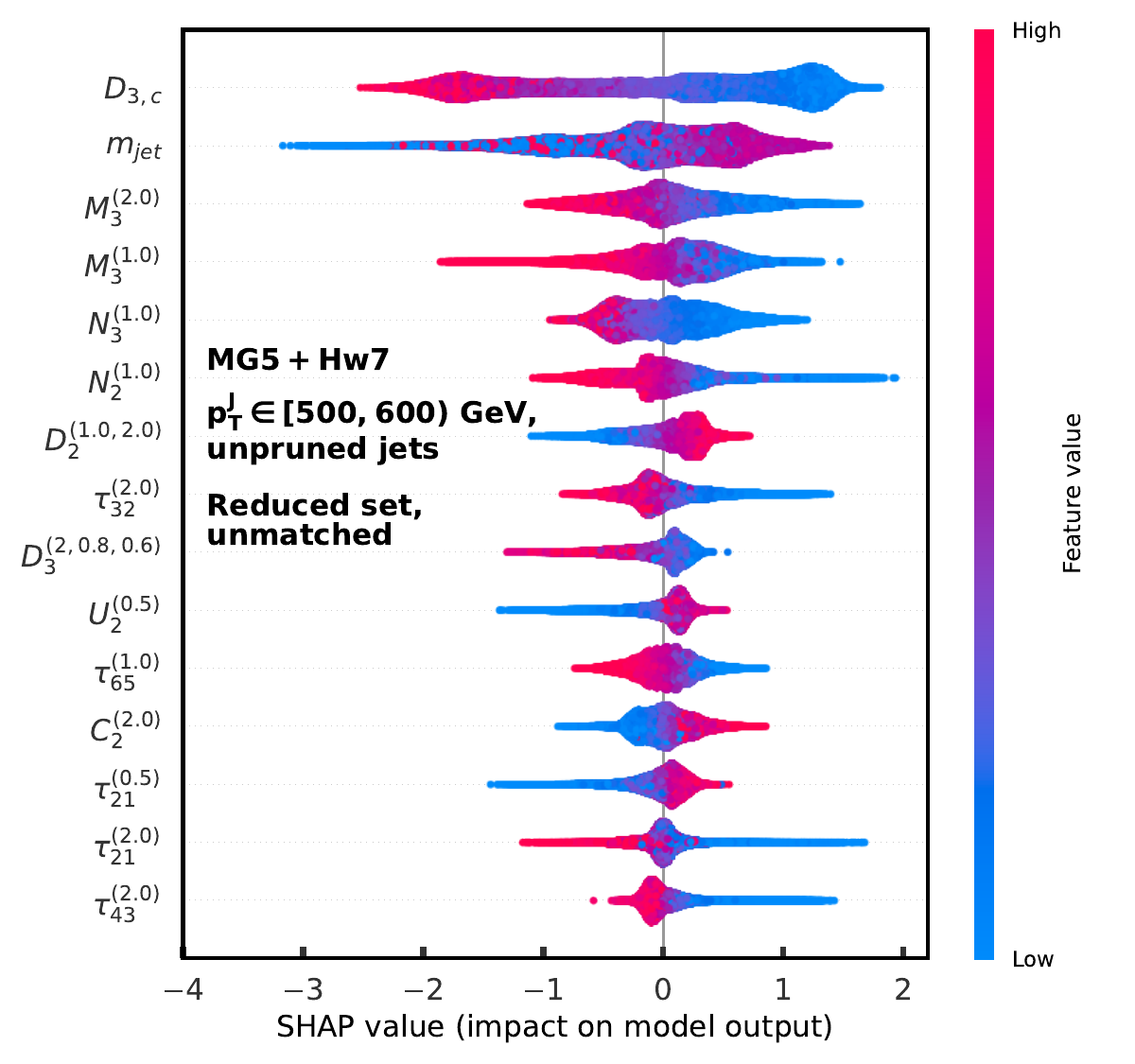}~
    \includegraphics[width=0.45\textwidth]{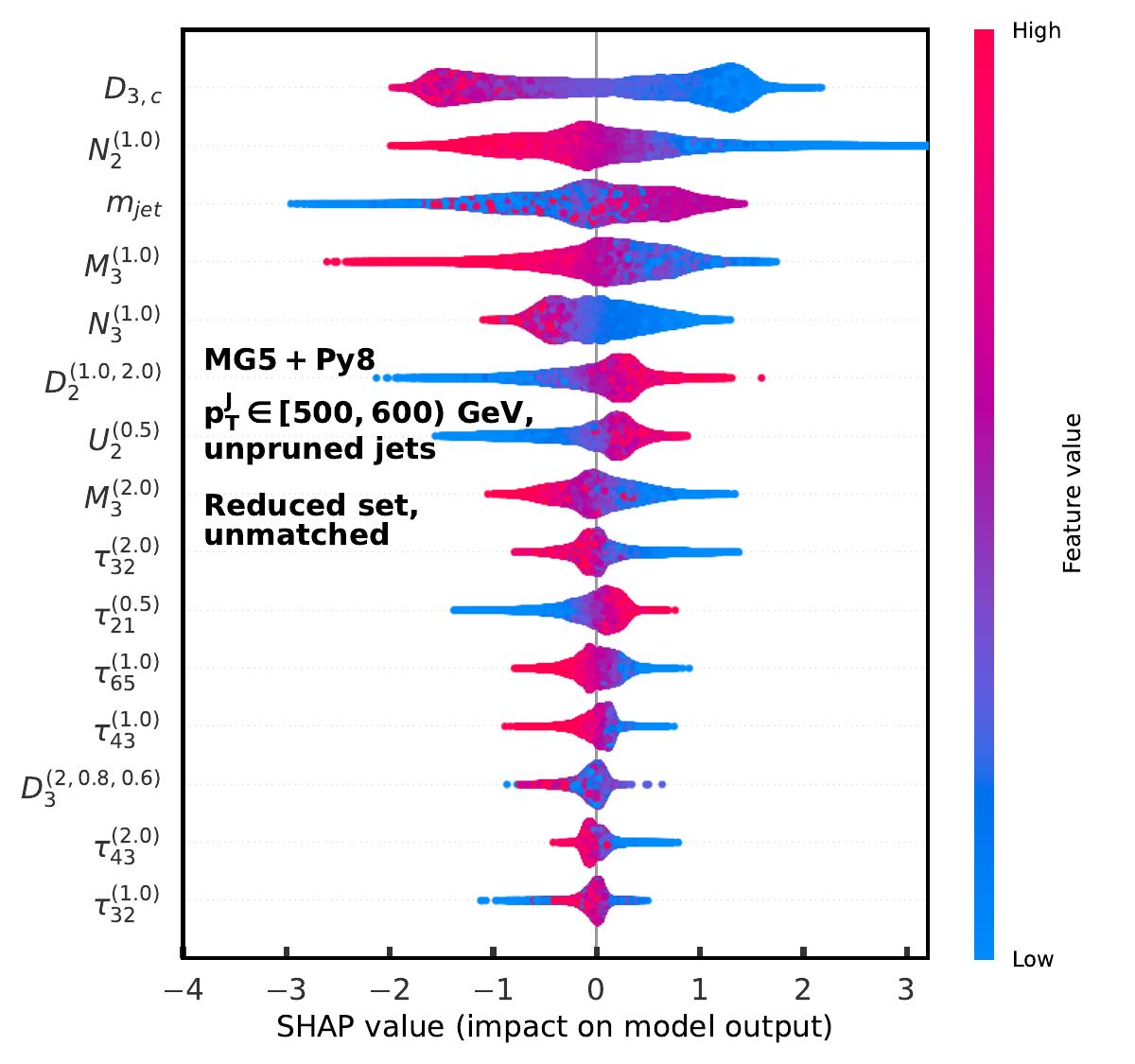}
    \caption{A summary plot showing the variation of SHAP values with the input feature values for the four hybrid taggers consisting of the the reduced set of features. \textit{Top row:} Top jets were generated with \texttt{HERWIG 7} (\textit{left}), \texttt{PYTHIA 8} (\textit{right}) and partons were matched to $\Delta R_m$ = 1.0. \textit{Bottom row:} Top jets were generated with \texttt{HERWIG 7} (\textit{left}), \texttt{PYTHIA 8} (\textit{right}) and partons were not matched.}  
    \label{fig:shap_bees_um_hw7}
\end{figure}

Figure \ref{fig:shap_bees_um_hw7} shows the SHAP feature importance for \texttt{PYTHIA 8} and \texttt{HERWIG 7} when trained and tested on $\Delta R_m$ = 1.0 and unmatched jets. In the case of unmatched jets, the variables that are unique to both the generators rank lower and contribute less to the model output compared to the 9 common features.  $N_2^{(1.0)}$ has a greater contribution in the case of \texttt{PYTHIA 8} jets compared to \texttt{HERWIG 7} while $M_3^{(2.0)}$ ranks higher for \texttt{HERWIG 7}. 

We perform cross-generator training and testing on pruned jets. Table \ref{tab:py8_hw7_auc_p1} lists the AUC. While the model trained with \texttt{HERWIG 7} jets gives similar results when tested on \texttt{PYTHIA 8} jets, the model trained with \texttt{PYTHIA 8} jets performs poorly when cross-tested. In fact, \texttt{PYTHIA 8} trained model drops off by 5$\%$ when tested with \texttt{HERWIG 7} jets. 

\begin{table}[htb!]
\centering
\begin{tabular}{|l|ll|}
\hline
\multicolumn{3}{|c|}{Combined set tagger} \\ \hline
\multicolumn{1}{|c|}{\multirow{2}{*}{Train data}} & \multicolumn{2}{c|}{Test data} \\ \cline{2-3} 
\multicolumn{1}{|c|}{}                                    & \multicolumn{1}{l|}{PYTHIA 8} & HERWIG 7 \\ \hline
PYTHIA 8                                                  & \multicolumn{1}{l|}{0.923}    & 0.874  \\ \hline
HERWIG 7                                                    & \multicolumn{1}{l|}{0.895}    & 0.894  \\ \hline
\end{tabular}
\caption{AUC of \texttt{XGBOOST} model when different training and testing samples are used. The samples contain unmatched and pruned jets with the parameters $z_{cut}$ = 0.1 and $r_{cut}$ = $\frac{m_{jet}}{p_{T^J}}$.}
\label{tab:py8_hw7_auc_p1}
\end{table}

As discussed earlier, pruning removes soft and wide-angled radiations from jets, and in doing so, it might affect the performance of the observables that use angular correlations between jet constituents to differentiate between top and QCD. We saw that \texttt{PYTHIA 8} jets exhibit better separation in the 1D distributions of jet substructure observables that \texttt{HERWIG 7} jets. As a result, models trained on \texttt{HERWIG 7} jets learn to identify features in a more conservative and generalised manner that results in better performance with \texttt{PYTHIA 8}. For pruned \texttt{HERWIG 7} jets, the improvement in performance is smaller compared to the unpruned case.  Models trained on \texttt{PYTHIA 8} jets may learn details that fit the well separated distributions of the \texttt{PYTHIA 8} simulation, failing to recognise the features of \texttt{HERWIG 7} data. The degradation gets amplified after pruning the jets due to greater mismatch in substructure features.

\textbf{Other test cases.} We generate top jets and gluon jets by varying certain collider conditions to check the dependence of the performance of the taggers on them. We use unmatched top jets for the purpose as it presents the most realistic scenario. We use the reduced set tagger trained on \texttt{PYTHIA 8} generated jets to find out its performance on the different test cases.

\begin{enumerate}
    \item \textbf{Center-Of-Mass (COM) energy.} We had mentioned earlier that the events were generated at $\sqrt{s}$ = 14 TeV. We increase the COM energy to $\sqrt{s}$ = 27 TeV and generate the same signal and background processes. The rest of the simulation details are the same as earlier. When the trained decision tree with the reduced set of features is tested on this new dataset, the performance decreases by 1$\%$ from a test accuracy of 0.860 to 0.851.

    \item \textbf{Parton Distribution Function (PDF).} Instead of the \texttt{cteq6l1} PDF, we use the \texttt{NNPDF2.3NNLO} PDF. By changing the PDF, we find that the AUC of the model remains unchanged, indicating that the model does not depend on PDF. The accuracy sees an increase by only 0.2\%.
    
    \item \textbf{Pileup (PU).} We have not considered pileup throughout our study. To check the dependence on pileup, we generate events with 140 PU by using the \texttt{Delphes CMS Phase II} detector card for simulation. After simulating pileup events, the PileUp Per Particle Identification (PUPPI) algorithm \cite{Bertolini:2014bba} is used to mitigate the effects of pileup from the jets. Pileup subtracted jets were used as the third test case for the tagger that was initially trained on 0 PU events. We obtain an accuracy of 0.852. It is a 0.9$\%$ decrease from the zero pileup scenario. 
\end{enumerate}

The performance of the reduced set tagger varies up to 1$\%$ only with the varying simulation conditions. We also study the dependence of our results on the $p_T$ of the jets produced in the colliders, which is discussed in detail in the next section.

\section{\texttt{XGBOOST} analysis with jet substructure observables for 1 TeV jets}
\label{sec:1000GeV}

We generate top and gluon jet samples with $p_T^J\in[1.0,1.1)$ TeV. The simulation procedure remains the same as in Section \ref{ssec:simulation}.

Figure \ref{fig:mjet_d3t3_1TeV} show the distribution of the two variables, $m_{jet}$ and $D_{3,c}$ for signal and background in both $p_T$ ranges. The jet mass distribution has a better resolution for higher $p_T$ jets. The small peak near W mass is reduced because even without matching, the higher the boost, the more collimated the partons are within the jet radius. However, the jet mass distributions have a longer tail, due to increased soft radiation like ISR and MPI being captured within the jet.

\begin{figure}[htb!]
    \centering
    \includegraphics[width=0.52\textwidth]{./PT_1000_mjet.pdf}~
    \includegraphics[width=0.52\textwidth]{./PT_1000_d3_term3.pdf}
    \caption{Normalised $m_{jet}$ distribution \textit{(left)} and $D_{3,c}$ distribution \textit{(right)} of unmatched top jets and gluon jets for the two $p_T$ bins of the jets.}
    \label{fig:mjet_d3t3_1TeV}
\end{figure}

% \usepackage{multirow}
\begin{table}[htb!]
\centering
\begin{tabular}{|cclll|}
\hline
\multicolumn{5}{|c|}{Features in the reduced set taggers} \\ \hline
\multicolumn{1}{|c|}{9 common features} & \multicolumn{4}{c|}{\begin{tabular}[c]{@{}c@{}}$m_{jet}$, $\tau_{21}^{(0.5)}$, $D_2^{(1.0,2.0)}$, $D_{3,c}$,\\  $M_3^{(1.0)}$, $M_3^{(2.0)}$, $U_2^{(0.5)}$, $\tau_{65}^{(1.0)}$\end{tabular}} \\ \hline
\multicolumn{1}{|c|}{\multirow{2}{*}{\begin{tabular}[c]{@{}c@{}}Features unique to $p_T^J\in[500,600)$ GeV\\ jets\end{tabular}}} & \multicolumn{4}{c|}{\multirow{2}{*}{$N_2^{(1.0)}$, $N_3^{(1.0)}$, $D_3^{(2,0.8,0.6)}$, $\tau_{32}^{(1.0)}$, $\tau_{43}^{(1.0)}$}} \\
\multicolumn{1}{|c|}{} & \multicolumn{4}{c|}{} \\ \hline
\multicolumn{1}{|c|}{\multirow{2}{*}{\begin{tabular}[c]{@{}c@{}}Features unique to $p_T^J\in[1000,1100)$ GeV\\ jets\end{tabular}}} & \multicolumn{4}{c|}{\multirow{2}{*}{$M_2^{(2.0)}$, $N_2^{(2.0)}$, $D_{3,a}$, $D_{3,b}$, $\tau_{21}^{(2.0)}$}} \\
\multicolumn{1}{|c|}{} & \multicolumn{4}{c|}{} \\ \hline
\end{tabular}
\caption{Features appearing in the reduced set with 15 highest ranked features for unmatched jets with $p_T\in[500,600)$ GeV and $p_T\in[1000,1100)$ GeV jets.}
\label{tab:best15_1000}
\end{table}

We find that in the case of $p_T^J\in[1.0,1.1)$ TeV jets, the features that are more than 95$\%$ correlated are $C_1^{(1.0)},~C_1^{(2.0)},~D_3^{(2.0,0.8,0.6)} ~U_1^{(0.5)}, ~U_3^{(0.5)}, ~U_1^{(1.0)}, ~U_2^{(1.0)}, ~U_3^{(1.0)}$, and $U_1^{(2.0)}$. In Figure \ref{fig:corr_d3} , we show the correlation between the D-Series of Energy Correlation observables in the case of $p_T^J\in[1.0,1.1)$ TeV and $p_T^J\in[500,600)$ GeV jets. Removing these features and using 38 combined set of features as input to the \texttt{XGBOOST}, we obtain a classification accuracy of 0.844 and an AUC of 0.924, which is a $\sim$1$\%$ degradation from the $p_T^J\in[500,600)$ GeV case.

\begin{figure}[htb!]
    \centering
    \includegraphics[width=0.52\textwidth]{./PT_500_corr_um_D3_500GeV.pdf}~
    \includegraphics[width=0.52\textwidth]{./PT_1000_corr_um_D3_1000GeV.pdf}
    \caption{Matrix showing the correlations between the D-Series of observables for the two $p_T$ bins of the jets.}
    \label{fig:corr_d3}
\end{figure}

\begin{figure}[htb!]
    \centering
    \includegraphics[width=0.5\textwidth]{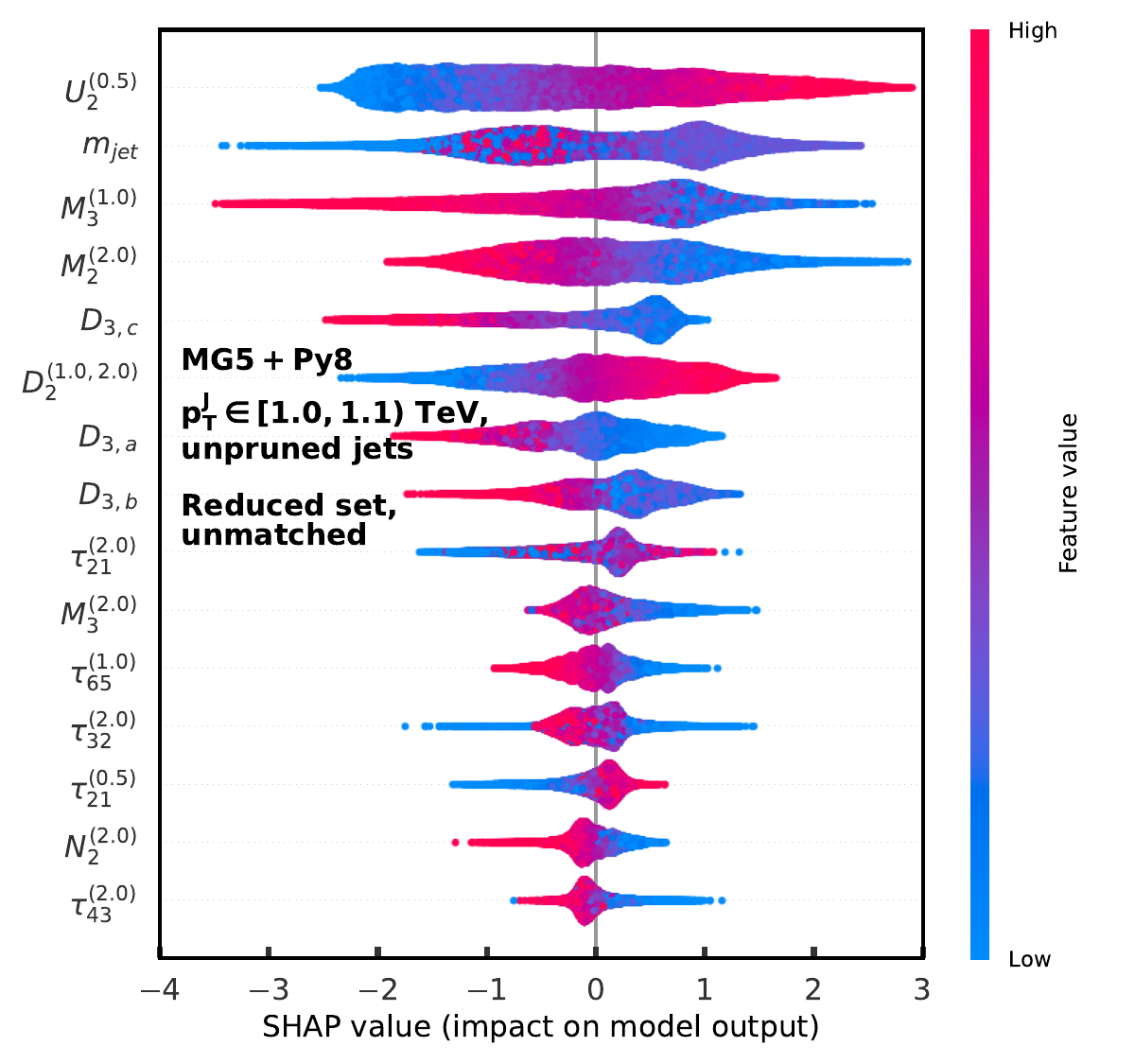}~
    \includegraphics[width=0.5\textwidth]{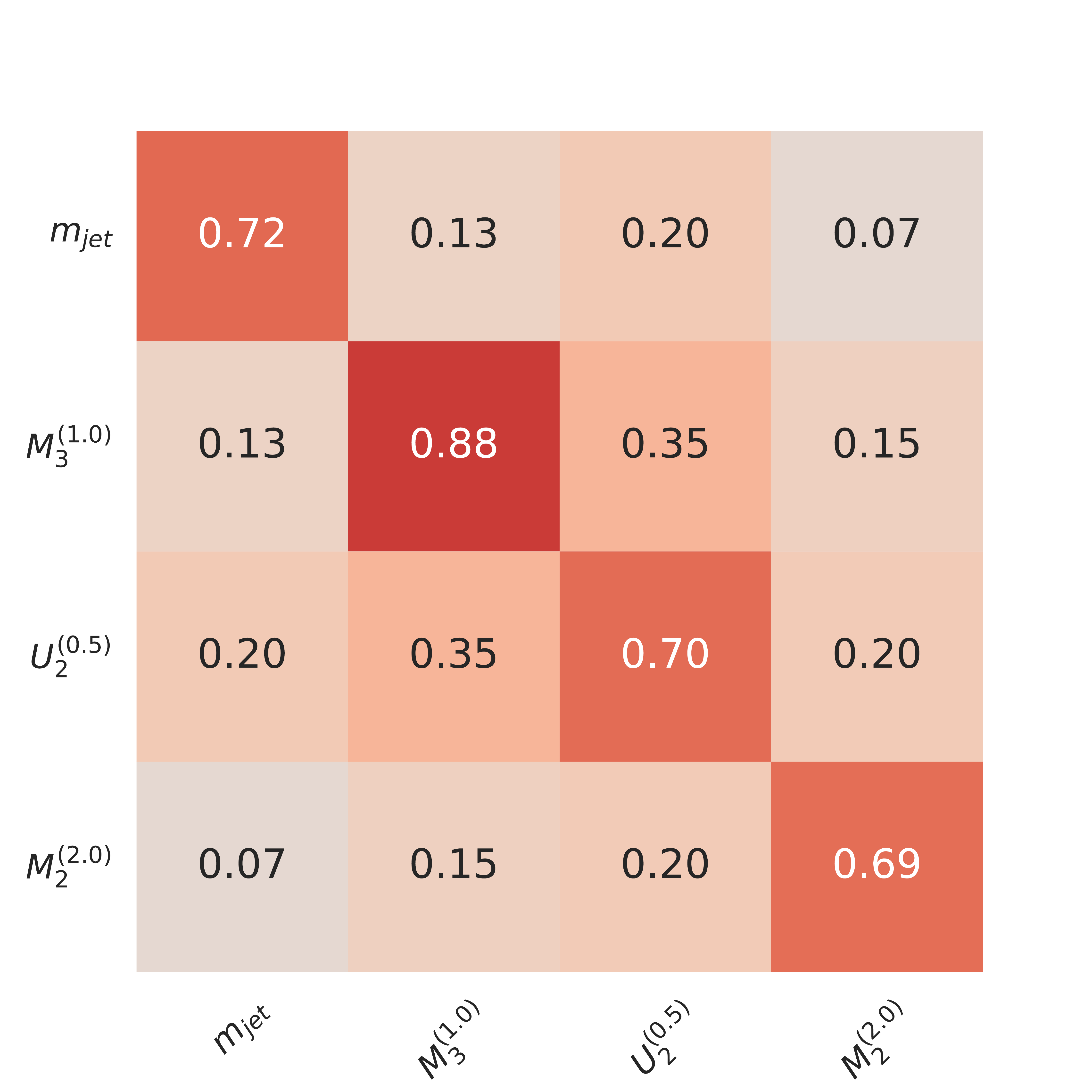}
    \caption{\textit{Left: }A summary plot showing the variation of SHAP values with the input feature values for the hybrid tagger consisting of 15 features. The dataset contains unpruned top and gluon jets having $p_T^J\in[1.0,1.1)$ TeV, generated using \texttt{PYTHIA 8}. The top jets were not matched to partons. \textit{Right: }4$\times$ 4 interaction matrix showing the interaction effects between four most important features of the hybrid tagger. The dataset contains unmatched top and gluon jets with $p_T^J\in[1.0,1.1)$ TeV.}
    \label{fig:shap_bees_15_1TeV}
\end{figure}

With the reduced set of features we obtain by choosing the 15 most important features as input, the accuracy and AUC becomes 0.841 and 0.921. Table \ref{tab:best15_1000} lists the common features in the reduced taggers of the two $p_T$ ranges as well as the features unique to both of them. We see that the observables for $p_T^J\in[1.0,1.1)$ TeV jets include $M_2^{(2.0)}$, $N_2^{(2.0)}$ and $\tau_{21}^{(2.0)}$ which are typically used to identify two-prong structure in jets. Figure \ref{fig:shap_bees_15_1TeV} shows the SHAP feature importance and the interaction matrix between the four most important features from the SHAP analysis. The interaction values have been calculated for 5000 test events.

$U_2^{(0.5)}$ becomes the most important feature. Figure \ref{fig:u2b5_m3b1_1TeV} shows the distributions of $U_2^{(0.5)}$ and $M_3^{(1.0)}$ for $p_T^J\in[1.0,1.1)$ TeV and $p_T^J\in[500,600)$ GeV jets. $M_3^{(1.0)}$ shows a greater separation for the higher $p_T$ jets. $U_2^{(0.5)}$ distribution changes considerably from $p_T^J\in[500,600)$ GeV to $p_T^J\in[1.0,1.1)$ TeV. The angular factor of 0.5 allows the variable to probe small angle emissions within the jet. 

\begin{figure}[htb!]
    \centering
    \includegraphics[width=0.52\textwidth]{./PT_1TeV_U2b5.pdf}~
    \includegraphics[width=0.52\textwidth]{./PT_1TeV_M3b1.pdf}
    \caption{Normalised $U2^{(0.5)}$ distribution \textit{(left)} and $m_{jet}$ distribution \textit{(right)} of unmatched top jets and gluon jets for the two $p_T$ bins of the jets.}
    \label{fig:u2b5_m3b1_1TeV}
\end{figure}

$U_2^{0.5}$ being the most important feature maybe attributed to its high interaction between $M_3^{(1.0)}$ and $m_{jet}$. Figure \ref{fig:2d_1TeV} shows the two dimensional distributions of $U_2^{0.5}$-$M_3^{(1.0)}$ and $U_2^{0.5}$-$m_{jet}$. 
\begin{figure}[htb!]
    \centering
    \includegraphics[width=0.45\textwidth]{./PT_1TeV_M3-U2-top.pdf}~
    \includegraphics[width=0.45\textwidth]{./PT_1TeV_M3-U2-gluon.pdf}\\
    \includegraphics[width=0.45\textwidth]{./PT_1TeV_mjet-U2-top.pdf}~
    \includegraphics[width=0.45\textwidth]{./PT_1TeV_mjet-U2-gluon.pdf}
    \caption{2-D histograms of $M_3^{(1.0)}$ vs. $U_2^{(0.5)}$ (\textit{top}) and $m_{jet}$ vs. $U_2^{(0.5)}$ (\textit{bottom}) for top jets \textit{(left)} and gluon jets \textit{(right)}. }
    \label{fig:2d_1TeV}
\end{figure}

Unlike in the case of $p_T\in[500,600)$ GeV jets, the $m_{jet}$ distributions of top and gluon jets have a long overlapping tail. This shows up in the SHAP plot of Figure \ref{fig:shap_bees_15_1TeV} where such events result in negative SHAP values. $D_{3,a}$ and $D_{3,b}$ have significant contribution to the model's output despite $D_3^{2.0,0.8,0.6}$ being removed, possibly due to the decreased correlation among them. 

\section{Background from jets with a broad range of transverse momentum}
\label{sec:high_pt_gluon}
In our work, we have performed top vs. gluon jet classification for two $p_T$ ranges of the jets - 500-600 GeV, and 1.0-1.1 TeV. It might so happen that jets from the background processes that originated with much higher boost loose their energy and fake as jets having $p_T$ in the range of our interest. It would be interesting to note the effect this has on the tagger's ability to discriminate between signal and background.

To study this phenomenon of migrated background jets, we generate gluon jets with $p_T$ greater than 550 GeV, 650 GeV, 750 GeV, 850 GeV, and 950 GeV and check the fraction that  they contribute in the $p_T\in$ [500,600) GeV bin. Similarly, for the $p_T\in$[1.0,1.1) TeV bin, we generate five sets of gluon jets demanding a minimum $p_T$ of 1.050 TeV to 1.45 TeV with an increment 100 GeV for each set. We then calculate the fraction of events from each of the higher $p_T$ that lie in the $p_T\in$ [500,600) GeV and $p_T\in$[1.0,1.1) TeV range. With these jets, we test the two reduced set taggers for the two $p_T$ cases on the migrated gluon jets. The accuracy with which the model can identify these boosted gluon jets varies between 2-4\% from the original accuracies.

% \usepackage{multirow}
\begin{table}[htb!]
\centering
\begin{tabular}{|c|c|c|c|}
\hline
\multirow{2}{*}{\begin{tabular}[c]{@{}c@{}}Generation cut $p_T>X$\\ on $pp\rightarrow Zg$\end{tabular}} & \multirow{2}{*}{\begin{tabular}[c]{@{}c@{}}Percentage of jets with \\ $p_T\in[500,600)$ GeV\end{tabular}} & \multirow{2}{*}{\begin{tabular}[c]{@{}c@{}}Relative cross-section\\ $\sigma_{p_T>X}/\sigma_{p_T>450 ~GeV}$\end{tabular}} & \multirow{2}{*}{\begin{tabular}[c]{@{}c@{}}Gluon\\accuracy\end{tabular}} \\
 &  &  &  \\ \hline
550 GeV & 34\% & 0.40 & 0.855 \\ \hline
650 GeV & 10\% & 0.18 & 0.833 \\ \hline
750 GeV & 4\% & 0.08 & 0.829 \\ \hline
850 GeV & 2\% & 0.04 & 0.822 \\ \hline
950 GeV & 1\% & 0.02 & 0.819 \\ \hline
\end{tabular}
\caption{The percentage of jets that lie in the $p_T^J\in$[500,600) GeV range after being generated with higher boosts with generation cut of $p_T^g>X$, where X is the minimum $p_T$ of the jets. The table also lists the cross-sections of the boosted gluon jets relative to the original $p_T^g>450$ GeV jets, and the accuracy with which our reduced set tagger identifies the gluon jets correctly.}
\label{tab:500_gbins}
\end{table}

% \usepackage{multirow}
\begin{table}[htb!]
\centering
\begin{tabular}{|c|c|c|c|}
\hline
\multirow{2}{*}{\begin{tabular}[c]{@{}c@{}}Generation cut $p_T>X$\\ on $pp\rightarrow Zg$\end{tabular}} & \multirow{2}{*}{\begin{tabular}[c]{@{}c@{}}Percentage of jets with \\ $p_T\in[1.0,1.1)$ TeV\end{tabular}} & \multirow{2}{*}{\begin{tabular}[c]{@{}c@{}}Relative cross-section\\ $\sigma_{p_T>X}/\sigma_{p_T>950 ~GeV}$\end{tabular}} & \multirow{2}{*}{\begin{tabular}[c]{@{}c@{}}Gluon\\accuracy\end{tabular}} \\
&  &  &  \\ \hline
1050 GeV & 25\% & 0.55 & 0.830 \\ \hline
1150 GeV & 11\% & 0.31 & 0.820 \\ \hline
1250 GeV & 6\% & 0.18 & 0.815 \\ \hline
1350 GeV & 3\% & 0.10 & 0.820 \\ \hline
1450 GeV & 2\% & 0.06 & 0.825 \\ \hline
\end{tabular}
\caption{The percentage of jets that lie in the $p_T^J\in$[1.0,1.1) TeV range after being generated with higher boosts with generation cut of $p_T^g>X$, where X is the minimum $p_T$ of the jets. The table also lists the cross-sections of the boosted gluon jets relative to the original $p_T^g>950$ GeV jets, and the accuracy with which our reduced set tagger identifies the gluon jets correctly.}
\label{tab:1000_gbins}
\end{table}

Tables \ref{tab:500_gbins} and \ref{tab:1000_gbins} show these percentages along with the generation cross-section relative to the cross-section of the actual $p_T^J \in$ [500,600) GeV and [1.0,1.1) TeV jets. As we can see, the percentage of gluon jets faking the $p_T^J$ of top jets in $p_T^J \in$ [500,600) GeV and [1.0,1.1) TeV ranges decrease with increasing $p_T^g$, and so do their cross-sections. When the reduced set tagger is used to classify these jets, the gluon identification accuracy decreases with increasing $p_T^g$ in the case of $p_T^J \in$ [500,600) GeV. This means that background jets that originate with a higher transverse momentum have more chances of being mistagged as top jets. However, this effect  shall not be very pronounced at the LHC, because of the lower cross-sections and fake percentages as highlighted in the tables. A similar trend is observed in [1.0,1.1) TeV jets too, however, the tagging gets better at $p_T^g$>1.35 TeV. 
\clearpage

\section{Conclusion}
\label{sec:conclusion}

This study has provided a thorough investigation into the performance of decision tree based top tagging algorithm using jet substructure observables, with emphasis on interpretability and robustness under realistic collider conditions. The key findings from the study, which also answer the questions that were asked at the beginning, are as follows:

\begin{enumerate}
\item 
The parton that carries a significant fraction of the $p_T$ of its parent top lies closer to it. Jets matched with smaller radii ($\Delta R_m$ = 0.6) showed the best classification performance, indicating that a tighter matching condition leads to more accurate reconstruction and improved tagger effectiveness. Moreover, we see that pruning the top and the gluon jets negatively impacts the performance of top and gluon jet classifiers, making them less accurate compared to their unpruned counterparts. However, combining pruned and unpruned jet features led to a notable improvement in the tagger's accuracy.

\item 
To find how well the \texttt{XGBOOST} model distinguishes between top jets and two types of background jets, we treat the gluon jets and quark jets separately. When gluon jets are used as the background, the tagging accuracy decreases compared to using quark jets, indicating the greater difficulty of identifying top jets in the presence of gluon jets. We lose only 0.4\% in accuracy when the tagger trained on gluon jets is tested on quark jets. This suggests that gluon jets serve as a more robust background for developing generalised top jet taggers. 

\item 
With the gluon jets as background, we proceed to study the dependence of tagger performance on simulation conditions. The \texttt{XGBOOST} tagger demonstrated stability when subjected to variations in center-of-mass energy, PDFs, and pile-up effects, with performance deviations limited to around 1\%. This robustness signifies that the tagger's efficiency is largely unaffected by these changes, making it suitable for practical applications in collider experiments.

\item 
We further observe that the performance degrades by $\sim 1\%$ with jets having a higher boost of $p_T^J\in [1.0,1.1)$ TeV. Jets that migrate across different $p_T$ ranges due to energy loss or gain showed a slight reduction in classification accuracy, ranging from 2-4\%. However, the impact of these migrated jets on the overall performance of the tagger was minimal, given their lower cross-sections and occurrence rates.

We have also performed cross-generator training and testing between \texttt{PYTHIA 8} and \texttt{HERWIG 7}. We found that when \texttt{PYTHIA 8} generated events are used for testing, we have better results irrespective of the generator used for the training events. The differences in model performance between \texttt{PYTHIA 8} and \texttt{HERWIG 7} can be considered as a source of systematic uncertainty in jet substructure analyses. These differences arise due to differences in the parton shower algorithms used by the two generators. Such variations can affect the jet characteristics and, consequently, the performance of the taggers trained on these simulations. To mitigate these effects, further Monte Carlo (MC) tuning and matching with experimental data are necessary.

\item 
The SHAP framework helped us interpret the machine learning model's decisions by showing which features contributed the most to the tagger's predictions. The third term of the generalised $D_3$ variable has a very high contribution in the top vs. gluon classification.

\end{enumerate}

From the SHAP analysis in all the scenarios discussed, the 15 highest-ranked features share many variables in common across all datasets. Tables \ref{tab:best15_deltaRm}, \ref{tab:best15_HW7} and \ref{tab:best15_1000} list the common features across datasets that \texttt{XGBOOST} learns consistently. Interestingly, they belong to not one but different categories of the variables. Beyond just feature importance, SHAP allowed us to understand how different features interact during the decision-making process. We used this information to create a simpler analysis based on the two most important variables, \(m_{\text{jet}}\) and \(D_{3,c}\). Extending this approach to include more interacting features could lead to even more effective tagging strategies in future studies.

\newpage 
\appendix
\section{Matching the top parton to jets}
\label{sec:t_match}
We set the matching radius $\Delta R_m$ to 0.6 and compare two matching methods, one where each individual parton $u$, $d$, and $b$ is matched to within $\Delta R_m$ from of the jet axis, and the other where only the $t$ is matched to the jet. Table \ref{tab:match_500_600_t} compares the two methods and the case of unmatched jets. Only 34$\%$ of top jets matched with the $t$ parton have all three partons inside $\Delta R_m$ = 0.6.

\begin{table}[htb!]
\centering
    \begin{tabular}{|c|c|c|}
    \hline
    \multicolumn{3}{|c|}{$p_T^{J} \in [500,600)$ GeV}  \\ \hline
     \multicolumn{1}{|c|}{\begin{tabular}[c]{@{}c@{}}Matching criteria\end{tabular}} & \multicolumn{1}{c|}{\begin{tabular}[c]{@{}c@{}}Fraction of top jets \\ where all three quarks \\ lie inside $\Delta R_m$ = 0.6\end{tabular}} & \begin{tabular}[c]{@{}c@{}}Fraction of top jets \\ where the top quark \\ lies inside $\Delta R_m$ = 0.6\end{tabular} \\ \hline
     \multicolumn{1}{|c|}{$u$,$d$,$b$ matched to $\Delta R_m$ = 0.6}   & 1.0 & 1.0 \\ \hline
     \multicolumn{1}{|c|}{only $t$ matched to $\Delta R_m$ = 0.6}  & 0.34 & 1.0 \\ \hline
     \multicolumn{1}{|c|}{unmatched}    & 0.32 & 0.94 \\ \hline
    \end{tabular}
    \caption{Comparison between matching individual partons to top jets, matching only top partons to top jets, and unmatched top jets $p_T^J\in[500,600)$ GeV.}
\label{tab:match_500_600_t}
\end{table}

When all three decay products of the top quark, $u$, $d$, and $b$, are matched to a certain radius within the jet, the $t$ parton lies close to the three partons. However, the reverse is not true. This means that in the cases where only the $t$ parton is matched to the jet, it does not guarantee that the $u$, $d$, and $b$ partons are matched. Jets with only $t$ parton matched are similar to the case of unmatched jets. 

\section{Definitions of observables}
\label{app:standard_top_taggers}

In this section, we briefly introduce the jet substructure observables, namely 
$N$-Subjettiness \cite{Thaler:2010tr} and Energy Correlation Functions (ECFs) \cite{Larkoski:2013eya} and their ratios.

\subsection{$N$-Subjettiness}
$N$-Subjettiness \cite{Thaler:2010tr}\footnote{It is an adaption of the $N$-jettiness variable, which is an event-shape variable describing the number of isolated jets in an event.} is a variable to count subjets inside a fat jet. It is defined as

\begin{equation}
\tau_N^{(\beta)} = \frac{1}{\sum_{\alpha\in\text{jet}} p_{T,\alpha}R_0^{\beta}} \sum_{\alpha\in\text{jet}}p_{T,\alpha} \min_{k=1,...,N}(\Delta R_{k,\alpha})^{\beta}
\label{eq:Nsub}
\end{equation}
\noindent
relative to $N$ subjet directions $\hat{n}_{j}$ and where $\beta>0$ is an arbitrary weighting exponent to ensure infrared safety. 
For a jet having $N$ subjets, $\tau_N<\tau_{N-1}$ and the ratio $\tau_{N}/\tau_{N-1}$ is small. 
Therefore, for top decays producing three separated subjets, the ratio $\tau_3/\tau_2$ is expected to peak at lower values, compared to the QCD case. However, other $N$-Subjettiness ratios like $\tau_{21}^{(\beta)}=\tau_2^{(\beta)}/\tau_1^{(\beta)}$ might also have some distinguishing power between the top and QCD jets.

We combine the following $N$-Subjettiness variables as inputs to our \texttt{XGBOOST} model.
\begin{equation}
    \tau_{21}^{(\beta)},~\tau_{32}^{(\beta)},~\tau_{43}^{(\beta)},~\tau_{54}^{(\beta)},~\tau_{65}^{(\beta)}, ~and ~m_{jet}
\end{equation}
with $\beta =(0.5,~1.0,~2.0)$ for a total of 16 input features. 

\subsection{Energy Correlation Functions}

The normalised $n$-point Energy Correlation Function (ECF) is defined in \cite{Larkoski:2013eya} as,

\begin{equation}
e_n^{(\beta)} = \frac{1}{(p_T^J)^n}\sum_{i_{1}<i_{2}<...<i_{n}\in J}\left(\prod_{a=1}^{n} p_T^{i_{a}}\right)\left(\prod_{b=1}^{n-1} \prod_{c=b+1}^{n} R_{i_{b}i_{c}}\right)^{\beta}
\end{equation}
$J$ denotes a jet, and $p_T^J$ is the transverse momenta of that jet. The sum runs over all the particles in $J$ with $R_{ij}=\sqrt{(\phi_{i}-\phi_{j})^{2}+(y_{i}-y_{j})^2}$ where $\phi$ and $y$ are the azimuthal angle and rapidity of a particle respectively. The first term in brackets is just the product of the transverse momenta of \textit{n} particles. The second term is the product of separation variables between $\binom{n}{2}$ pairs of particles out of the n particles. The angular exponent $\beta$ is a free parameter which we set $\beta$ > 0 to make it infrared and collinear safe. From the definition, it is clear that an $n$-pronged jet will yield a smaller value of $e_{n+1}^{\beta}$ compared to $e_{n}^{\beta}$. It has been shown that ratios constructed out of these functions by power counting analysis \cite{Larkoski:2013eya, Larkoski:2014gra, Moult:2016cvt} provide good discrimination between top and QCD jets.

In the following sections, we list 5 series of variables formulated using ECFs and combine some of these variables from each series to build a top tagger. 
%In each section, we list the performance of 5 such top taggers on our test data sets while using unpruned, pruned, and combined sets of variables.

\subsubsection{The C-Series}

The $C_i$ series is defined in \cite{Larkoski:2013eya} as: 
\begin{equation}
C_{i}^{(\beta)} = \frac{e_{n+1}^{(\beta)}e_{n-1}^{(\beta)}}{(e_{n}^{(\beta)})^2}
\end{equation} According to Ref. \cite{Larkoski:2013eya}, $C_3^{(\beta)}$, along with a mass cut and a cut on $C_2^{(\beta)}$ to ensure its IRC safety, has considerable discriminating power for 3-prong jets. 
\begin{equation}
C_{3}^{(\beta)} = \frac{e_{4}^{(\beta)}e_{2}^{(\beta)}}{(e_{3}^{(\beta)})^2}
\end{equation}

Although $C_3$ is typically used in boosted top tagging, an ML model might capture more correlations from $C_1$ and $C_2$ variables with different values of $\beta$.  The variables from the C-Series used in the \texttt{XGBOOST} analysis are listed below in Equation \ref{eq:C_Series}.
\begin{equation}
\begin{aligned}
    & C_1^{(1.0)},C_2^{(1.0)},~C_3^{(1.0)},
    & C_1^{(2.0)},C_2^{(2.0)},~C_3^{(2.0)},~m_{jet}
    \label{eq:C_Series}
\end{aligned}
\end{equation}

\subsubsection{The D-Series}

The $D_{3}^{(\alpha,\beta,\gamma)}$ variable defined in terms of the ratios of the ECFs shows a significant discriminating power between three-prong and one/two-prong phase space \cite{Larkoski:2014zma}. It is defined as,
\begin{equation}
D_{3}^{(\alpha,\beta,\gamma)}=\frac{e_{4}^{(\gamma)}(e_{2}^{(\alpha)})^{\frac{3\gamma}{\alpha}}}{e_{3}^{(\beta)})^{\frac{3\gamma}{\beta}}}+x\frac{e_{4}^{(\gamma)}(e_{2}^{(\alpha)})^{\frac{2\gamma}{\beta}-1}}{e_{3}^{(\beta)})^{\frac{2\gamma}{\beta}}}+y\frac{e_{4}^{(\gamma)}(e_{2}^{(\alpha)})^{\frac{2\beta}{\gamma}-\frac{\gamma}{\alpha}}}{e_{3}^{(\beta)})^{2}},
\end{equation}
where x and y are constants whose value depends on top jet kinematics. The above expression is the linear combination of three phase space regions that contain triple splitting, strongly ordered splitting, and soft emissions. Following \cite{Larkoski:2014zma}, 
%one can define two quantities x and y using the method of power counting, which are given below, 
%\begin{equation}
%x=\kappa_{1}\left(\frac{(p_{T}^{cut})^2}{m_{top}^2}\right)^{\frac{\alpha\gamma}{\beta}-\frac{\alpha}{2}}
%\end{equation}
%\begin{equation}
%y=\kappa_{2}\left(\frac{(p_{T}^{cut})^2}{m_{top}^2}\right)^{\frac{5\gamma}{2}-2\beta} .
%\end{equation}
we use the following values of the constants: $\alpha=2 , \beta=0.8, \gamma=0.6$, $x=5, y=0.35$, and the scaling parameters $\kappa_{1} = \kappa_2 = 1$. The reason for choosing $\alpha=2$ is because 
a cut on the jet mass restricts the study to a certain region of the phase space, as given by,
\begin{equation}
e_2^{(2)}\sim \frac{m_{J}^2}{p_{TJ}^2}
\end{equation}
In other words, a mass cut is typically applied on the jet, to restrict $e_2$ and simplify the phase space with only $e_3^{(\beta)}$ and $e_{4}^{(\gamma)}$ remaining. We label the three terms of $D_{3}^{(\alpha,\beta,\gamma)}$ without the x and y coefficients as $D_{3,a},~D_{3,b},$ and $D_{3,c}$. We use all the terms along with the standard $D_3$ variable as input to \texttt{XGBOOST}.  We also consider the two-prong discriminant $D_2^{(\alpha,\beta)}$ \cite{Larkoski:2014gra,Larkoski:2015kga} defined as,
\begin{equation}
    D_2^{(\alpha,\beta)} = \frac{e_3^{(\alpha)}}{(e_2^{(\beta)})^{\frac{3\alpha}{\beta}}}
\end{equation}

 The variables used in this series are listed below. 
\begin{equation}
\begin{aligned}
   & D_2^{(1.0,1.0)}, D_2^{(1.0,2.0)}, D_2^{(2.0,1.0)}, D_2^{(2.0,2.0)}, \\
   & D_{3,a},~D_{3,b},~D_{3,c},~D_3^{(2,0.8,0.6)},~m_{jet} 
\end{aligned}
\end{equation}

\subsubsection{The U-Series}
One can also define the generalised correlator functions by replacing the angular part by \textit{v} factors out of the $\binom{n}{2}$ pairs of angles \cite{Moult:2016cvt},
\begin{equation}
{}_{v}e_{n}^{(\beta)}=\sum_{i_{1}<i_{2}<...<i_{n}\in J}z_{i_{1}}z_{i_{2}}..z_{i_{n}} \prod_{m=1}^{v}\min^{(m)}_{s<t\in \{i_{1},i_{2}..i_{n}\}}\left\lbrace\theta^{(\beta)}_{st}\right\rbrace
\end{equation}
where
\begin{equation}
z\equiv\frac{p_{T,i}}{\sum_{i\in J} p_{T,i}}.
\end{equation}
For a particular value of \textit{v}, the function will consist of the product of the $m^{th}$ smallest angles, with \textit{m} running from 1 to \textit{v}, meaning that the expression will contain only \textit{v} factors of pairwise angles. This simplifies the angular part to a great extent and also increases the flexibility of angular scales. With the new definition, a larger number of boost invariant ratios have been constructed using different combinations. Still, the complexity of computing the angular part is now reduced due to selecting only the minimum of the pairs. The simplest variable is the $U_i$ series, proposed in \cite{Moult:2016cvt} as quark vs. gluon discriminants. 
\begin{equation}
U_i^{(\beta)}={}_{1}e_{i+1}^{(\beta)}
\end{equation} 

We set $\theta_{ij}$ to $R_{ij}$ for our work, where $R_{ij} = \sqrt{\Delta \eta_{ij}^2 + \Delta \phi_{ij}^2}$ is the distance between two particles $i$ and $j$ in the $\eta-\phi$ plane. We use the following variables as the input features of the U-Series tagger,
\begin{equation}
\begin{aligned}
   & U_1^{(0.5)}, ~U_2^{(0.5)},  ~U_3^{(0.5)},\\
   & U_1^{(1.0)}, ~U_2^{(1.0)},  ~U_3^{(1.0)}, \\
   & U_1^{(2.0)}, ~U_2^{(2.0)},  ~U_3^{(2.0)}, ~m_{jet} 
\end{aligned}
\end{equation}

\subsubsection{The M-Series}
The $M_{i}$ series of observables are dimensionless, boost invariant ratios of the generalised ECFs with $v=1$ \cite{Moult:2016cvt}.
\begin{equation}
M_{i}^{(\beta)}=\frac{{}_{1}e_{i+1}^{(\beta)}}{{}_{1}e_{i}^{(\beta)}}
\end{equation}
 
In the case of boosted top tagging, i.e., tagging of 3-prong objects, the $M_{3}$ discriminant is used, defined as, 
\begin{equation}
M_{3}=\frac{{}_{1}e_{4}^{(\beta)}}{{}_{1}e_{3}^{(\beta)}}
\end{equation}

We use the following variables as the input features of the M-Series tagger,
\begin{equation}
M_2^{(1.0)},~M_3^{(1.0)},~M_2^{(2.0)},~M_3^{(2.0)},~m_{jet}
\end{equation}

\subsubsection{The N-Series}

Similarly, the $N_{i}$ series is defined in \cite{Moult:2016cvt} as,
\begin{equation}
N_{i}=\frac{{}_{2}e_{i+1}^{(\beta)}}{({}_{1}e_{i}^{(\beta)})^{2}}
\end{equation}
Consisting of 2 angles in the numerator and the denominator, this is also a boost invariant quantity. The $N_{3}$ observable is used for boosted top tagging.
\begin{equation}
N_{3}=\frac{{}_{2}e_{4}^{(\beta)}}{({}_{1}e_{3}^{(\beta)})^{2}}
\end{equation}

The N-Series tagger uses the following variables as input features,

\begin{equation}
N_2^{(1.0)},~N_3^{(1.0)},~N_2^{(2.0)},~N_3^{(2.0)},~m_{jet}
\end{equation}

\subsection{Wasserstein distance}
\label{app:wass}
The Wasserstein distance, also known as the Earth Mover's Distance (EMD), is a metric used to quantify the distance between two probability distributions. It measures the minimum "cost" of transforming one distribution into the other, where the cost is defined in terms of the distance one would need to "move" probability mass.

The Wasserstein distance $W_p$ between two probability distributions $\mu$ and $\nu$ is defined for $p\geq 1$ as:
\begin{equation}
W_p(\mu, \nu) = \left( \inf_{\gamma \in \Gamma(\mu, \nu)} \int_{X \times Y} d(x, y)^p \, d\gamma(x, y) \right)^{\frac{1}{p}}
\end{equation}
where:
$d(x, y)$ is the distance between points $x$ and $y$ in the metric space, and
$\Gamma(\mu,\nu)$ is the set of all joint distributions (couplings) with marginals $\mu$ and $\nu$.

When comparing two one-dimensional probability distributions, the Wasserstein distance can be related to their cumulative distribution functions (CDFs). For distributions $F$ and $G$ with CDFs $F(x)$ and $G(x)$, the first Wasserstein distance $W_1$ is given by:
\begin{equation}
W_1(F, G) = \int_{-\infty}^{\infty} |F(x) - G(x)| \, dx
\end{equation}
This integral represents the area between the two CDFs, providing a direct measure of the difference between the distributions.

In summary, the Wasserstein distance offers a robust way to measure the distance between probability distributions by considering the minimum transportation cost, and for one-dimensional cases, it can be directly calculated from the CDFs of the distributions.

\subsection{Kinematic observables constructed on light quark jets}
\label{app:quark_dist}

Figure \ref{fig:quark_dis} shows the distribution of the four kinematic observables $\tau_{32}^{(1.0)}$, $C_3^{(1.0)}$ , $D_3^{(2.0,0.8,0.6)}$, and $N_3^{(1.0)}$ without the cut on jet mass. The top and quark jet distributions are more separated than top and gluon jet distributions in terms of the observables $D_3^{(2.0,0.8,0.6)}$, and $N_3^{(1.0)}$. 

\begin{figure}[hbt!]
    \centering
    \includegraphics[width=0.45\textwidth]{./PT_500_tau32b1m_q.pdf}~
    \includegraphics[width=0.45\textwidth]{./PT_500_C3b1mc2_q.pdf}\\
    \includegraphics[width=0.45\textwidth]{./PT_500_D3m_q.pdf}~
    \includegraphics[width=0.45\textwidth]{./PT_500_N3b1m_q.pdf}
    \caption{Normalised distributions of jet substructure observables $\tau_{32}^{(1.0)}$ (\textit{top left}), $C_3^{1.0)}$ (\textit{top right}), $D_3^{(2.0,0.8,0.6)}$ (\textit{bottom left}), and $N_3^{(1.0)}$ (\textit{bottom right}) of top jets with $\Delta R_m$ = 0.6, 0.8, 1.0, unmatched top jets, and quark jets.}
    \label{fig:quark_dis}
\end{figure}

\subsection{Effect of a jet mass cut}
\label{app:mass_cut}
Figure \ref{fig:no_mass_cut} shows the distribution of the four kinematic observables $\tau_{32}^{(1.0)}$, $C_3^{(1.0)}$ , $D_3^{(2.0,0.8,0.6)}$, and $N_3^{(1.0)}$ without the cut on jet mass. Comparing with Figure \ref{fig:tau32b1_DRm}, the top and gluon jet distributions of $\tau_{32}^{(1.0)}$ and $C_3^{(1.0)}$ increasingly overlap with each other with increasing $\Delta R_m$ while $D_3^{(2.0,0.8,0.6)}$, and $N_3^{(1.0)}$ show less change on the removal of the cut.

\begin{figure}[hbt!]
    \centering
    \includegraphics[width=0.45\textwidth]{./PT_500_tau32b1.pdf}~
    \includegraphics[width=0.45\textwidth]{./PT_500_C3b1.pdf}\\
    \includegraphics[width=0.45\textwidth]{./PT_500_D3.pdf}~
    \includegraphics[width=0.45\textwidth]{./PT_500_N3b1.pdf}
    \caption{Normalised distributions of jet substructure observables $\tau_{32}^{(1.0)}$ (\textit{top left}), $C_3^{1.0)}$ (\textit{top right}), $D_3^{(2.0,0.8,0.6)}$ (\textit{bottom left}), and $N_3^{(1.0)}$ (\textit{bottom right}) of top jets with $\Delta R_m$ = 0.6, 0.8, 1.0, unmatched top jets, and gluon jets without any additional cut.}
    \label{fig:no_mass_cut}
\end{figure}

Figure \ref{fig:no_mass_cut_p} shows similar distributions of the above observables constructed on pruned jets. Without a cut on the jet mass, distributions of pruned gluon jets shift towards the top jet distributions.

\begin{figure}[hbt!]
    \centering
    \includegraphics[width=0.45\textwidth]{./PT_500_tau32b1_p.pdf}~
    \includegraphics[width=0.45\textwidth]{./PT_500_C3b1_p.pdf}\\
    \includegraphics[width=0.45\textwidth]{./PT_500_D3_p.pdf}~
    \includegraphics[width=0.45\textwidth]{./PT_500_N3b1_p.pdf}
    \caption{Normalised distributions of jet substructure observables $\tau_{32}^{(1.0)}$ (\textit{top left}), $C_3^{1.0)}$ (\textit{top right}), $D_3^{(2.0,0.8,0.6)}$ (\textit{bottom left}), and $N_3^{(1.0)}$ (\textit{bottom right}) of top jets with $\Delta R_m$ = 0.6, 0.8, 1.0, unmatched top jets, and gluon jets after pruning but without any additional cut.}
    \label{fig:no_mass_cut_p}
\end{figure}

\section{Hyperparameter optimisation of \texttt{XGBOOST}}
\label{app:hyperopt}

Optimization was performed on the hyperparameters listed in Table \ref{tab:hyperparams} using the \texttt{HYPEROPT} package \cite{bergstra2013making}. 
%It uses the Tree-structured Parzen Estimators (TPE) algorithm to minimise the objective function by efficiently selecting hyperparameters from the hyperparameter space

\begin{table}[hbt!]
    \centering
    \begin{tabular}{|c|c|}
    \hline
        Hyperparameter & Definition\\ \hline
        \texttt{eta} & Learning rate\\ \hline
%        \texttt{lambda} & L2 regularization on weights\\ \hline
        \texttt{gamma} & Minimum loss reduction required to split a node\\ \hline
        \texttt{max\_depth} & Maximum depth of the tree\\ \hline
        \texttt{min\_child\_weight}& Minimum sum of weights in a child to split a node\\ \hline
        \texttt{colsample\_by\_tree} & Fraction of features selected for each boosting round \\\hline
        \texttt{subsample} & Fraction of training data to be subsampled\\ \hline
        
    \end{tabular}
    \caption{List of hyperparameters used to train the \texttt{XGBOOST} and their definition.}
    \label{tab:hyperparams}
\end{table}

The values of the hyperparameters after optimisation are listed in Table \ref{tab:optimised_hyperparameters_six_models}. 
\begin{table}[ht]
\centering
\small
\begin{tabular}{|l|c|c|c|c|c|c|c|}
\hline
Model & \texttt{colsample\_bytree} & \texttt{eta} & \texttt{gamma} & \texttt{max\_depth} & \texttt{min\_child\_weight} & \texttt{subsample} \\
\hline
$N$-Subjettiness & 1.0 & 0.175 & 0.5 & 5 & 1.0 & 0.95 \\
C-Series         & 1.0 & 0.075 & 0.75 & 5 & 1.0 & 0.7 \\
D-Series         & 1.0 & 0.075 & 0.75 & 5 & 1.0 & 0.9 \\
U-Series         & 1.0 & 0.15 & 1.0 & 5 & 1.0 & 0.8 \\
M-Series         & 1.0 & 0.075 & 0.6 & 5 & 1.0 & 0.75 \\
N-Series         & 1.0 & 0.075 & 0.65 & 5 & 1.0 & 0.8  \\
Combined set     & 1.0 & 0.175 & 0.5 & 5 & 1.0 & 0.95  \\
Reduced set      & 1.0 & 0.175 & 0.5 & 5 & 1.0 & 0.95  \\
\hline
\end{tabular}
\caption{Optimised Hyperparameter Values for Six Models and two models with combined and reduced set of features.}
\label{tab:optimised_hyperparameters_six_models}
\end{table}
\clearpage
\newpage

%%%%%%%%%%%%%%%%%%%%%%%%%%%%%%%%%%%%%%%%%%%%%%%
%% Bibliography 
%%%%%%%%%%%%%%%%%%%%%%%%%%%%%%%%%%%%%%%%%%%%%%%

\providecommand{\href}[2]{#2}\begingroup\raggedright\endgroup

\end{document}